\documentclass{article}
\usepackage{graphics}

\begin{document}

\title{Cell-Centric Heuristics for the Classification of Cellular Automata}

\author{Valmir C. Barbosa\thanks{Corresponding author
({\tt valmir@cos.ufrj.br}).}\\
Fernando M. N. Miranda\\
Matheus C. M. Agostini\\
\\
Universidade Federal do Rio de Janeiro\\
Programa de Engenharia de Sistemas e Computa\c c\~ao, COPPE\\
Caixa Postal 68511\\
21941-972 Rio de Janeiro - RJ, Brazil}

\maketitle

\begin{abstract}
We study the classification of cellular-automaton update rules into Wolfram's
four classes. We start with the notion of the input entropy of a spatiotemporal
block in the evolution of a cellular automaton, and build on it by introducing
two novel entropy measures, one that is also based on inputs to the cells, the
other based on state transitions by the cells. Our two new entropies are both
targeted at the classification of update rules by parallel machines, being
therefore mindful of the necessary communications requirements; we call them
cell-centric input entropy and cell-centric transition entropy to reflect
this fact. We report on extensive computational experiments on both one- and
two-dimensional cellular automata. These experiments allow us to conclude that
the two new entropies possess strong discriminatory capabilities, therefore
providing valuable aid in the classification process.

\end{abstract}

\section{Introduction}\label{intr}

Cellular automata are discrete-time dynamical systems comprising finite-state
units, called cells, whose states evolve in time as a result of the interactions
with other cells. Since their introduction nearly five decades ago by
von Neumann \cite{vn66}, cellular automata have acquired an ever more prominent
status as a modeling tool in several research areas (cf., e.g.,
\cite{kvw97,bms01} and the references therein), and have even come to be
regarded by some as a central abstraction in the modeling of nature's
fundamental processes \cite{w02}.

For $S=\{0,\ldots,s-1\}$ the set of possible states, and for $t\ge 0$ an
integer, a cellular automaton with $n$ cells evolves from time $t$ to time
$t+1$ by synchronously updating all $n$ states by the application of a
deterministic mapping $F_f$ from $S^n$ to $S^n$. This mapping $F_f$ is global in
nature and depends on the local update rule $f$, which dictates how each
individual state is to be updated given the cell's state at time $t$ as well as
the states of those cells that lie within a neighborhood of size $\delta$. The
update rule $f$ is then a mapping from $S^{1+\delta}$ to $S$.

Normally a cell's neighborhood in a cellular automaton is determined by an
underlying multidimensional lattice according to several possible criteria.
For example, a cell's neighbors relative to a certain dimension of the lattice
may be taken to be those cells that are $r>0$ edges away along that dimension
but no edges away along any other dimension, $r$ being usually referred to as
the radius of the update rule in that dimension. For unit radii in all
dimensions, this characterizes what is known as the von Neumann neighborhood,
but in this paper we employ the same denomination also for greater radii.
Another example neighborhood comes from letting two cells be neighbors of each
other whenever one can be reached from the other by treading no more edges along
a certain dimension than the update rule's radius along that dimension. For unit
radii this is the Moore neighborhood, but once again we generalize and in this
paper employ the same denomination under greater radii as well. When $n$ is
finite, it is customary to regard the lattice as having cylindrical boundaries,
that is, as allowing every cell to have exactly two nearest neighbors along each
dimension.

Finite cellular automata, those for which $n$ is finite, are necessarily such
that $F_f$ eventually leads to a fixed point, or a limit cycle, of
configurations in $S^n$, that is, either $x$ such that $F_f(x)=x$ or
$x_0,\ldots,x_{p-1}$, with $p>0$, such that $x_0=F_f(x_{p-1})$, $x_1=F_f(x_0)$,
and so on \cite{wl92}. The case of infinite cellular automata, on the other
hand, is far more complicated and intriguing, since now $n$ is formally infinite
and no periodicity is guaranteed to emerge from the successive application of
$F_f$.

Both in the finite and in the infinite cases, cellular automata have along the
years been the subject of theoretical and experimental analyses. For a summary
of key results, the reader is referred, for example, to \cite{w94,i01} and to
their many references. One of the most appealing topics of investigation has
been the classification of the update rule $f$, and consequently of the cellular
automata based on it, regarding its ``complexity.''

Interest in this question received its initial impetus from the study by Wolfram
of infinite one-dimensional cellular automata \cite{w84}, which resulted in the
empirical finding that, nearly regardless of initial states, $f$ consistently
falls within one of four possible qualitative categories: (i) evolution leads to
a homogeneous configuration, i.e., a configuration in which all cells have the
same state; (ii) evolution leads to an inhomogeneous fixed point or to a limit
cycle; (iii) evolution leads to a chaotic succession of configurations; or (iv)
evolution leads to complex localized spatiotemporal structures that are
``sometimes long-lived.'' Although initially conceived for the one-dimensional
case, there is in principle no reason why such a qualitative classification
should not also be applicable to higher-dimensional cases. In fact, similar
studies for the two-dimensional case have appeared as early as in \cite{pw85}.

Naturally, class-(iv) update rules are intuitively associated with the
realization of ``complex'' computations by the cellular automata that are built
on them, that is, precisely those computations that underlie so much of the
interest in cellular automata as modeling tools. Not surprisingly, then,
considerable effort has been channeled into finding approaches to automatically
categorize update rules into the classes (i)--(iv). Formally, all such efforts
hover around the so-called limit set of an update rule $f$ in the infinite case,
which is the set of configurations that result from all possible initial
configurations after the passage of arbitrarily long time. As it turns out,
every nontrivial property of a limit set (i.e., a property that holds for at
least one cellular automaton and does not hold for at least one other) can be
proven undecidable through a reduction from the problem of whether a limit set
is a singleton \cite{k94}, itself known to be undecidable \cite{k92}.

As a consequence of this inherent undecidability, every effective strategy for
categorizing update rules must necessarily be of a heuristic nature or else
eventually boil down to a heuristic if it is to have any practical use. Our
interest in this paper is the study of heuristics that can be coupled with the
parallel simulation of cellular automata in order to analyze the spatiotemporal
patterns that emerge, aiming at categorizing the underlying update rule within
Wolfram's four classes. Efficiency in the form of minimal communications needs
is then an essential requirement, leading to what we term cell-centric
heuristics, that is, heuristics that depend as minimally as possible on the
exchange of information among processors during the simulation of a cellular
automaton.

We start in Section~\ref{back} with a review of some of the prominent heuristics
that have been proposed for automatically classifying update rules, and proceed
in Section~\ref{ientr} to a discussion of the so-called input-entropy measures.
Our cell-centric heuristics are presented in Section~\ref{heur}, with results
from computational experiments on one- and two-dimensional cellular automata
given in Section~\ref{exp}. Further considerations on the computational results
appear in Section~\ref{disc} and concluding remarks come in Section~\ref{concl}.

\section{Background}\label{back}

In broad terms, we distinguish two essential kernel classes of strategies for
the categorization of update rules. The first class comprises those techniques
that aim at extracting the update rules' computational capabilities by solely
considering the update rule itself, not simulations of cellular automata for
examination of the resulting spatiotemporal patterns. Approaches of this type
have concentrated on one-dimensional cellular automata, so a cell's neighborhood
size is in fact $\delta=2r$.

The pioneering step within this class of approaches was taken by Langton
\cite{l90}, who proposed to classify an update rule $f$ into the four Wolfram
classes by examining a single parameter, denoted by $\lambda_f$ and given by
\begin{equation}
\label{langton}
\lambda_f=1-\frac{q}{s^{1+2r}}.
\end{equation}
In (\ref{langton}), $q$ is the number of distinct $(1+2r)$-tuples on which $f$
outputs $\sigma$, where $\sigma\in S$ is any of the so-called quiescent states,
that is, $\sigma$ is such that $f(\sigma,\ldots,\sigma)=\sigma$.

The initial report on the use of the $\lambda_f$ parameter indicated that it
behaves as an order parameter with respect to which a phase transition occurs:
on generating update rules $f$ with increasing $\lambda_f$ one first encounters
class-(i) behavior, then class (ii), then class (iv) around $\lambda_f=0.5$,
then finally class-(iii) behavior. This seemed to suggest that complexity was to
be found at the region in the $\lambda_f$ space that became known as the ``edge
of chaos.'' But, in addition to the obvious difficulties regarding the existence
and choice of a quiescent state, criticism regarding the existence and nature of
the purported phase transition soon came from several sources (cf., e.g.,
\cite{mhc93,mch94,mhk97}). In particular, it now appears that update rules
belonging to several classes, not just class (iv), are to be found near
$\lambda_f=0.5$.

Two other interesting approaches have been introduced that are also of the same
nature in that they also dispense with the need for computer simulations of
cellular automata. The two approaches share the goal of investigating how the
information contained in an update rule $f$ affects the overall behavior of
cellular automata built on $f$. One of the approaches is topological in nature,
that is, it seeks to analyze a cellular automaton's global behavior by
identifying finite-size spatial patterns and characterizing their appearance and
evolution \cite{bcfqv95}. By contrast, the other one \cite{ddf01} is algorithmic
and aims at characterizing update rule $f$ from the perspective of Kolmogorov
complexity \cite{lv97}, that is, the perspective of the shortest possible
description of $f$. Both approaches relate clearly to the Wolfram
classification, while at the same time shedding new light on it, each from its
particular perspective. The latter approach, in addition, may also hold a key to
some of the incongruities that are inherent to Langton's parameter $\lambda_f$.
It is worth remarking, however, that because each of the two approaches induces
its own class system, neither one is found to relate clearly to class (iv). The
reader is referred to the original sources for details.

A wholly distinct class of strategies to categorize cellular-automaton update
rules concentrates on the examination of space-time patterns as they appear
during the evolution of cellular automata from as representative a sample of
initial configurations as possible. Now, of course, the fact that the cellular
automata under examination are formally infinite has to be reckoned with; we
will come to this later, and will for now ignore any difficulties that such
infinities may cause in practice. We do mention, however, that some successful
approaches are built from the start on the assumption that $n$ is finite. One
example is the ``computational mechanics'' exemplified in \cite{hc97}, which for
one-dimensional cellular automata draws heavily on finite-state machines
\cite{s98} derived from patterns in the cellular automaton's evolution to
characterize the fundamental spatiotemporal features that are inherent to each
update rule.

The approaches in this class that are central to our interest are those that
rely on some form of entropy measure as the basis of the categorization effort.
The initial approach along these lines appeared in the same paper that
introduced the four-class Wolfram classification \cite{w84}. Essentially, what
it does is to consider the probability distribution of space-time blocks as
they occur during the evolution of cellular automata for a fixed update rule $f$
and then use this distribution to define the desired entropies.

For a more precise characterization, let $d>0$ be the number of dimensions of
the cellular automata in question, and let $X_1,\ldots,X_d$ denote numbers of
contiguous cells along each dimension. For $T>0$ a number of successive time
steps during an evolution of the cellular automaton that employs update rule
$f$, we need the probability, given $f$, that an
$X_1\times\cdots\times X_d\times T$ block of states appears somewhere in the
spatiotemporal trace of the cellular automaton's evolution. Clearly, the number
of possible blocks is $s^{X_1\ldots X_dT}$. We denote by $P_i$ the probability
of the $i$th block, $1\le i\le s^{X_1\ldots X_dT}$.

Two basic entropies can now be defined. These are the set entropy
\begin{equation}
E_f(X_1,\ldots,X_d,T)=\frac{1}{T}\log_s
\left(\sum_{i=1}^{s^{X_1\ldots X_dT}}\theta(P_i)\right),
\end{equation}
where $\theta(p)=1$ for $p>0$ and $\theta(0)=0$, and the measure entropy
\begin{equation}
E_f^\mu(X_1,\ldots,X_d,T)=-\frac{1}{T}
\sum_{i=1}^{s^{X_1\ldots X_dT}}P_i\log_sP_i.
\end{equation}
From them, we obtain the limiting quantities
\begin{equation}
\label{setentropy}
H_f=\lim_
{{X_1,\ldots,X_d\to\infty}\atop{{T\to\infty}\atop{T/X_1,\ldots,T/X_d\to\infty}}}
E_f(X_1,\ldots,X_d,T)
\end{equation}
and
\begin{equation}
\label{measureentropy}
H_f^\mu=\lim_
{{X_1,\ldots,X_d\to\infty}\atop{{T\to\infty}\atop{T/X_1,\ldots,T/X_d\to\infty}}}
E_f^\mu(X_1,\ldots,X_d,T),
\end{equation}
respectively.

The quantity in (\ref{setentropy}) gives the asymptotic rate at which the
diversity of spatiotemporal patterns increases with time, and the one in
(\ref{measureentropy}) represents the average amount of ``new information''
that each new configuration of the cellular automaton contributes as time
elapses. As it turns out, these quantities (or variations thereof obtained by
taking the limit exclusively as $X_1,\ldots,X_d\to\infty$ while $T$ is kept
constant, or as $T\to\infty$ while $X_1,\ldots,X_d$ are kept constant) yield
insight into how to categorize $f$. Coarsely, all indicators vanish for
class-(i) cellular automata and are nonzero for class-(iii) cellular automata.
Discriminating class (ii), in turn, requires decoupling space and time:
indicators resulting from letting $T\to\infty$ alone are zero, while those
related to letting $X_1,\ldots,X_d\to\infty$ for fixed $T$ are nonzero. As for
class (iv), once again the attempt at identification is eluded.

\section{Input entropy}\label{ientr}

The concept of input entropy is due to Wuensche \cite{w99} and constitutes an
attempt to merge together some of the key features of the two classes of
strategies discussed in Section~\ref{back}, those that seek to base update-rule
classification on examining the update rule solely and those that rely on
space-time signatures of evolving cellular automata. Given one of the
$X_1\times\cdots\times X_d\times T$ state blocks of that section, we start by
considering the probability that, inside the block, each of the possible
$s^{1+\delta}$ inputs to a cell occurs. Denoting the probability of the $i$th
input by $Q_i$, $1\le i\le s^{1+\delta}$, the input entropy is defined by
\begin{equation}
\label{inputentropy1}
I_f(X_1,\ldots,X_d,T)=-\sum_{i=1}^{s^{1+\delta}}Q_i\log_sQ_i.
\end{equation}

The use of the input entropy in practice starts by fixing the values of
$X_1,\ldots,X_d$ and of $T$ and then choosing the $X_1\ldots X_d$ cells to be
observed during simulations of the cellular automaton built on $f$. Each
simulation is conducted from a randomly selected initial configuration and runs
for $t_+$ time steps, for some $t_+\ge T-1$, generating a new configuration at
each time step. For each time $t$ in the interval $[t_0,t_+]$ with
$T-1\le t_0\le t_+$, the probability $Q_i$ that the $i$th input occurs within a
$X_1\times\cdots\times X_d\times T$ block can be approximated by
$q_i^t/X_1\ldots X_dT$, where $q_i^t$ is the number of occurrences of the $i$th
input within the block that ends at time $t$. The practical entropy figure that
stems from (\ref{inputentropy1}) is then, for the block that ends at time $t$,
\begin{equation}
\label{inputentropy2}
I_f^t(X_1,\ldots,X_d,T)=-\sum_{i=1}^{s^{1+\delta}}
\left(\frac{q_i^t}{X_1\ldots X_dT}\right)
\log_s\left(\frac{q_i^t}{X_1\ldots X_dT}\right).
\end{equation}

The mean and variance of the quantity in (\ref{inputentropy2}) for
$t=t_0,\ldots,t_+$, that is,
\begin{equation}
\label{iemean}
\overline{I_f}(X_1,\ldots,X_d,T)=\frac{1}{t_+-t_0+1}
\sum_{t=t_0}^{t_+}I_f^t(X_1,\ldots,X_d,T)
\end{equation}
and
\begin{eqnarray}
\label{ievariance}
\lefteqn{\sigma^2\left(I_f(X_1,\ldots,X_d,T)\right)}
\nonumber\\
&&\hspace{0.25in}
=\frac{1}{t_+-t_0+1}\sum_{t=t_0}^{t_+}
\left[I_f^t(X_1,\ldots,X_d,T)-\overline{I_f}(X_1,\ldots,X_d,T)\right]^2,
\end{eqnarray}
respectively, can be used to reveal the Wolfram class to which update rule $f$
belongs, after having themselves been averaged over some number of simulations
for randomly chosen initial configurations. Roughly, for $d=1$ it has been found
that low means and variances bespeak class-(i) or (ii) behavior, while high
means and low variances indicate a class-(iii) update rule in action. Class-(iv)
behavior is characterized by medium-valued means and high variances \cite{w99}.

\section{Cell-centric heuristics}\label{heur}

Computing the mean and variance of the input entropy as indicated respectively
in (\ref{iemean}) and (\ref{ievariance}) requires simulating the cellular
automaton that is based on $f$ for $t_+$ time steps and accumulating the
quantity given in (\ref{inputentropy2}) while an
$X_1\times\cdots\times X_d\times T$ block ``window'' is slid from an initial
position that makes the block end at time $t_0$ through a final position at
which the block ends at time $t_+$. When the simulation is performed in
parallel, the $X_1\ldots X_d$ cells do not all reside at the same processor,
so computing the input entropy as given by (\ref{inputentropy2}) for all values
of $t$ requires a considerable amount of communication involving the processors
that lodge the cells.

Given that one processor, call it $P$, has been singled out for coalescing all
the information required for computing the mean and the variance, in essence the
number of integers that needs to be communicated to $P$ is $O(Xt_+\delta)$,
where $X$ is the number of cells allocated outside $P$. In the case of true
massive parallelism (one cell per processor), this becomes
$O(X_1\ldots X_dt_+\delta)$. In general, for each cell and each time $t$, the
integers to be communicated are the $1+\delta$ integers needed for specifying
the input to that cell at time $t$. If that input is the $i$th possible input,
then communicating the $1+\delta$ integers contributes one unit to each of
$q_i^t,\ldots,q_i^{t+T-1}$; that is, it contributes to the calculation of the
input entropy of (\ref{inputentropy2}) for $T$ blocks (the one ending at time
$t$ through the one ending at time $t+T-1$).

\subsection{Cell-centric input entropy}

The crux of the $O(Xt_+\delta)$ communications requirement is that the
logarithm appearing in (\ref{inputentropy2}) can only be assessed after the
contributions to $q_i^t$ have been taken into account for all $X_1\ldots X_d$
cells, in particular for all the $X$ cells lodged outside processor $P$. The
first step towards obtaining a cell-centric approximation to
(\ref{inputentropy2}), one that allows communications requirements to be reduced
dramatically, is to examine more closely the argument to that logarithm and to
notice that $q_i^t/X_1\ldots X_d$ is the average number of occurrences of the
$i$th input per cell within the $X_1\times\cdots\times X_d\times T$ block.

Let $q_i^{c,t}$ denote the number of occurrences of the $i$th input for the
$c$th cell inside the block that ends at time $t$; clearly,
$q_i^t=\sum_{c=1}^{X_1\ldots X_d}q_i^{c,t}$ and (\ref{inputentropy2}) can be
rewritten as
\pagebreak
\begin{eqnarray}
\label{inputentropy3}
\lefteqn{I_f^t(X_1,\ldots,X_d,T)}
\nonumber\\
&&\hspace{0.25in}=-\sum_{i=1}^{s^{1+\delta}}
\frac{\sum_{c=1}^{X_1\ldots X_d}q_i^{c,t}}{X_1\ldots X_dT}
\log_s\left(\frac{\sum_{c=1}^{X_1\ldots X_d}q_i^{c,t}}{X_1\ldots X_dT}\right)
\nonumber\\
&&\hspace{0.25in}=-\frac{1}{X_1\ldots X_d}
\sum_{c=1}^{X_1\ldots X_d}\sum_{i=1}^{s^{1+\delta}}
\left(\frac{q_i^{c,t}}{T}\right)
\log_s\left(\frac{\sum_{c'=1}^{X_1\ldots X_d}q_i^{c',t}}{X_1\ldots X_dT}\right).
\end{eqnarray}

Our cell-centric input entropy is defined by approximating $q_i^{c,t}$ by its
average over all cells in the block whenever convenient. That is, we use the
approximation
\begin{equation}
\label{approximation}
\frac{\sum_{c'=1}^{X_1\ldots X_d}q_i^{c',t}}{X_1\ldots X_d}
\approx
q_i^{c,t}
\end{equation}
in the argument to the logarithm in (\ref{inputentropy3}) for all $c$ such that
$1\le c\le X_1\ldots X_d$. We then obtain
\begin{equation}
\label{ccentropy}
C_f^t(X_1,\ldots,X_d,T)
=-\frac{1}{X_1\ldots X_d}
\sum_{c=1}^{X_1\ldots X_d}\sum_{i=1}^{s^{1+\delta}}
\left(\frac{q_i^{c,t}}{T}\right)\log_s\left(\frac{q_i^{c,t}}{T}\right),
\end{equation}
which is the cell-centric input entropy of the
$X_1\times\cdots\times X_d\times T$ block that ends at time $t$.

The essential question, of course, is whether the cell-centric input entropy
defined in (\ref{ccentropy}) still has discriminatory capabilities analogous to
those of the input entropy, given that the two are related by the approximation
in (\ref{approximation}). The answer to this question is affirmative and is
explored in Section \ref{exp} through the mean
$\overline{C_f}(X_1,\ldots,X_d,T)$ and the variance
$\sigma^2\left(C_f(X_1,\ldots,X_d,T)\right)$, defined analogously to
(\ref{iemean}) and (\ref{ievariance}), respectively.

Let us then consider how much communication must be directed towards the special
processor $P$ during a parallel simulation of a cellular automaton. Clearly, a
processor $Q\neq P$ lodging $X_Q$ cells can calculate its portion of the double
summation in (\ref{ccentropy}) for each $t$ (i.e., let $c$ range over its $X_Q$
cells) completely locally. If $N$ denotes the number of processors, then $P$
has to receive $O(Nt_+)$ floating-point numbers for the entire simulation. In
the limit of true massive parallelism, this becomes $O(X_1\ldots X_dt_+)$,
which relates to the communications requirements of the original input entropy
by a factor of $O(\delta)$ if we disregard any differences between sending
integers and sending floating-point numbers. So using the cell-centric
approximation to input entropy saves considerable amounts of communication in
the current technological reality of $N\ll X_1\ldots X_d$ but makes little sense
in the limit of true massive parallelism.

\subsection{Cell-centric transition entropy}

We perceive the functional form of (\ref{ccentropy}) as being suggestive of a
host of possible different criteria that may be experimented with when
attempting to classify cellular-automaton update rules. One possibility that
we have considered is the following. For a fixed cell $c$ inside the
$X_1\times\cdots\times X_d\times T$ block that ends at time $t$, let
$\tau^{c,t}$ denote the number of state transitions within the block that cause
the state of the cell to change during the simulation. This quantity does not
depend explicitly on the inputs to the cell, so when deriving the corresponding
cell-centric entropy measure from (\ref{ccentropy}), and considering that there
are $T-1$ state transitions per cell within the block, we obtain
\begin{equation}
\label{cctentropy}
T_f^t(X_1,\ldots,X_d,T)
=-\frac{1}{X_1\ldots X_d}
\sum_{c=1}^{X_1\ldots X_d}
\left(\frac{\tau^{c,t}}{T-1}\right)\log_s\left(\frac{\tau^{c,t}}{T-1}\right).
\end{equation}

We call the quantity in (\ref{cctentropy}) the cell-centric transition entropy
relative to the block that ends at time $t$. Naturally, it shares with the
cell-centric input entropy all the relevant characteristics that have to do with
the parallel simulation of cellular automata. In addition, as will become
apparent in Section~\ref{exp}, it also offers interesting glimpses into the
categorization of the underlying update rule when analyzed from the perspective
of its mean $\overline{T_f}(X_1,\ldots,X_d,T)$ and its variance
$\sigma^2\left(T_f(X_1,\ldots,X_d,T)\right)$, defined in analogy to
(\ref{iemean}) and (\ref{ievariance}), respectively.

\subsection{Upper bounds}

Upper bounds on the cell-centric input entropy of (\ref{ccentropy}) and the
cell-centric transition entropy of (\ref{cctentropy}) can be established easily
if we recall that entropies are maximized when all the mutually exclusive events
at hand are ascribed the same probability. In the case of (\ref{ccentropy}) this
amounts to setting $q_i^{c,t}/T$ to $s^{-(1+\delta)}$ for all appropriate values
of $i$ and $c$, which yields
\begin{equation}
\label{ccebound}
C_f^t(X_1,\ldots,X_d,T)\le 1+\delta.
\end{equation}
The case of (\ref{cctentropy}) is even simpler, as it suffices to recognize that
$x\log_sx$ is maximized for $x=e^{-1}$ and to set $\tau^{c,t}/(T-1)$ to this
value for all appropriate values of $c$, thus yielding
\begin{equation}
\label{cctebound}
T_f^t(X_1,\ldots,X_d,T)\le\frac{e^{-1}}{\ln s}.
\end{equation}

But some of the results described in Section~\ref{exp} are based on the
so-called outer-totalistic update rules \cite{pw85}, that is, update rules whose
outcomes depend not on the cell's individual state and those of its neighbors,
but rather on the cell's state and the sum of its neighbors' states. For such
update rules, and considering the $s=2$ case only, while the bound given by
(\ref{cctebound}) is still correct and gives a value slightly above $0.53$, in
the case of the cell-centric input entropy it no longer makes sense to assume a
uniform probability distribution on all inputs for entropy maximization, and
consequently (\ref{ccebound}) has to be revised. The correct level at which to
assume the uniform distribution for entropy to be maximized is now the level at
which inputs are grouped with one another according to the sum of the neighbor
states that they comprise.

For $s=2$ the number of distinct such groups is $1+\delta$, each corresponding
to one of the possible sum values, from $0$ to $\delta$. The probability
associated with each group is then $(1+\delta)^{-1}$, so each individual input
is assumed to occur with probability $[(1+\delta)n(\sigma)]^{-1}$, where
$n(\sigma)$ is the number of inputs whose states sum up to $\sigma$ with
$0\le\sigma\le\delta$. But $n(\sigma)={\delta\choose\sigma}$, so from
(\ref{ccentropy}) we obtain
\begin{equation}
\label{ccebound2}
C_f^t(X_1,\ldots,X_d,T)\le\log_2(1+\delta)+
\frac{1}{1+\delta}\sum_{\sigma=0}^\delta\log_2{\delta\choose\sigma}.
\end{equation}

\section{Computer experiments}\label{exp}

An infinite cellular automaton cannot be simulated in its entirety, nor can a
portion of it be simulated for an indefinitely long number of time steps. One
crucial first decision when planning such a simulation is which contiguous cells
to observe along each of the $d$ dimensions and also the number of time steps
$t_+$ during which to perform the simulation. Choosing a finite number of cells
to observe poses the question of how to handle the boundaries of the observed
region, since those boundaries affect the simulation but cannot be extended
indefinitely. Adopting artificial cylindrical boundaries or feeding randomly
picked values to the boundary cells at each time step of the simulation will not
in principle do, since this would have direct impact on the assumed infinite and
deterministic nature of the cellular automaton.

The solution, naturally, comes from first setting the value $t_+$ of the number
of steps during which the cellular automaton is to be observed in the
simulation, as well as the number $\ell_k$ of contiguous cells to be observed
along the $k$th dimension, $1\le k\le d$, and then working backwards from them.
We start by assuming a von Neumann neighborhood and then split a cell's
neighborhood $\delta$ into its dimension-wise constituents; that is, we write
$\delta=2(r_1+\cdots+r_d)$, where each $r_k$ is the update rule's radius along
the $k$th dimension. In order to output the state at time $t_+$ of a cell that
lies at a boundary along the $k$th dimension as if it were indeed embedded in an
infinite cellular automaton, the states of additional $r_k$ off-boundary cells
are needed along that dimension at time $t_+-1$. The number of boundary cells
along the $k$th dimension at time $t_+$ is
\begin{equation}
2\prod_{{l=1}\atop{l\neq k}}^d\ell_l,
\end{equation}
so the total number of cells for which states are needed at time $t_+-1$ is
\begin{equation}
\label{sizeatzero1}
\ell_1\ldots\ell_d+2\sum_{k=1}^dr_k\prod_{{l=1}\atop{l\neq k}}^d\ell_l
\le\prod_{k=1}^d(\ell_k+2r_k).
\end{equation}
Note that, for $d=1$, equality holds in (\ref{sizeatzero1}). The upper bound is
useful, though, because it generalizes easily as we work backwards through time
$t=0$, revealing that initial states are needed for cells that number no more
than
\begin{equation}
\label{sizeatzero2}
\prod_{k=1}^d(\ell_k+2r_kt_+).
\end{equation}

The upper bound in (\ref{sizeatzero2}) is clearly an exaggeration for $d>1$
under a von Neumann neighborhood, since it corresponds to an
$(\ell_1+2r_1t_+)\times\cdots\times(\ell_d+2r_dt_+)$ patch of cells. Here we
only point out that specifying the precise cells whose initial states are needed
is totally feasible, however cumbersome their determination or the inner
mechanics of a parallel simulation involving exactly those cells and no others.
In any event, we now have a set of cells that can be simulated through time
$t_+$ with the certainty that the observed behavior of the core
$\ell_1\ldots\ell_d$ cells is fully compatible with the assumption of an
infinite cellular automaton and of the deterministic character of its update
rule. Boundaries still exist with respect to the extended set of cells, but the
way they are handled is now immaterial. Either cylindrical boundaries may be
assumed or randomly picked states may be used to fill up the inputs needed by
the boundary cells. The effects of either choice can only affect the states of
the core cells after time $t_+$.

When we assume a Moore neighborhood to start with, we write
$\delta=(1+2r_1)\ldots(1+2r_d)-1$ instead, so that the number of cells for which
states are needed at time $t_+-1$ is exactly the upper bound appearing in
(\ref{sizeatzero1}). In this case, clearly the number of cells given by
(\ref{sizeatzero2}) is no longer an exaggeration, but expresses precisely what
is needed.

\begin{figure}
\centering
\scalebox{1.0}{\includegraphics{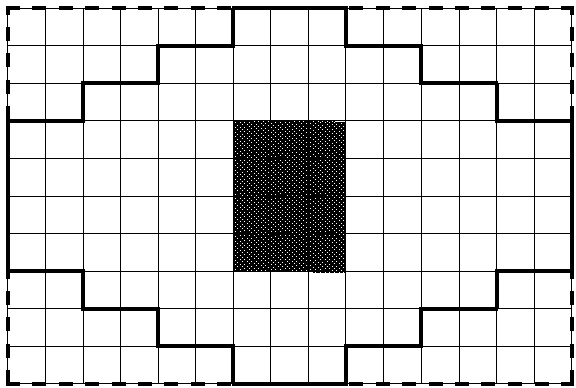}}
\caption{The $15\times 10$ patch of cells for which initial states are needed in
the two-dimensional case of $\ell_1=3$, $\ell_2=4$, $r_1=2$, and $r_2=1$, so
that the shaded cells may be observed correctly for $t_+=3$. The thick-line
enclosures refer to the minimal sets of cells that are needed under a von
Neumann (solid line) or a Moore (dashed line) neighborhood.}
\label{infinited2}
\vspace{0.45in}
\end{figure}

In Figure~\ref{infinited2}, we provide an illustration of these issues in the
two-dimensional case when $\ell_1=3$ and $\ell_2=4$ with $r_1=2$ and $r_2=1$.
What is shown is the set of cells for which initial states are needed if the
states of the shaded cells are to be observed for $t_+=3$ further time steps as
if those cells were part of an infinite, deterministic cellular automaton. Cells
enclosed within the thick solid contour are those for which initial states are
needed in the case of a von Neumann neighborhood. Those enclosed with the thick
dashed contour must have initial states specified if a Moore neighborhood is
used. Our practice henceforth is to employ sets of cells of the latter type
regardless of the neighborhood type in use.

\subsection{The value of $T$}

When a cellular automaton is simulated with the goal of computing the input
entropy of Section~\ref{ientr} or one of the cell-centric quantities of
Section~\ref{heur} inside an $X_1\times\cdots\times X_d\times T$ state block,
the core set of observed cells is such that $\ell_1=X_1,\ldots,\ell_d=X_d$. In
this section, we discuss the choice of $T$ that maximizes the discriminatory
capabilities of our cell-centric heuristics in the context of the Wolfram
classes. We henceforth assume $S=\{0,1\}$, i.e., $s=2$.

Our approach has been to perform a set of initial experiments with $t_+=500$ on
a single processor and to analyze their outcomes aiming at finding a suitable
$T$ value for use in the main experiments. We ran four sets of initial
experiments: one for $d=1$ and $r_1=2$, one for $d=1$ and $r_1=3$, one for $d=2$
under a von Neumann neighborhood with $r_1=r_2=1$, and one last for $d=2$ under
a Moore neighborhood with $r_1=r_2=1$. For the one-dimensional cases, each set
comprised four runs for $X_1=150$ and four runs for $X_1=300$; for the
two-dimensional cases, we did four runs for each of $X_1=X_2=15$ and
$X_1=X_2=30$. Within each set, the first run corresponds to a known class-(i)
update rule, the second to a known class-(ii) update rule, and so on. The known
update rules we used are detailed in Table~\ref{updtrules}.

In Table~\ref{updtrules}, the update rules are specified according to the
following conventions. For the one-dimensional experiments, each update rule is
the hexadecimal form of the binary number whose most significant bit is the
update rule's output to the input $11\ldots 1$, read left to right, the next bit
corresponds to $11\ldots 1-1$, and so on (cf., e.g., \cite{w83}). The
two-dimensional von Neumann case is similar, except that the most significant
bit corresponds to $00\ldots 0$, the next one to $00\ldots 0+1$, and so on,
inputs being read in the self-north-east-south-west order \cite{mcell-lex-url}.
The two-dimensional Moore case comprises outer-totalistic update rules only, for
which we adopt Conway's Life \cite{g70,bcg82} usual notation style: {\tt b}$x$
indicates that the cell's state moves from $0$ to $1$ (it ``is born'') if the
cell has $x$ neighbors in the $1$ state; {\tt s}$x$ means that the cell's state
remains $1$ (it ``survives'') if the cell has $x$ neighbors in the $1$ state; in
all cases not listed explicitly, the cell's state becomes $0$
\cite{mcell-lex-url}.

\begin{table}
\centering
\caption{Update rules used to generate the plots in
Figures~\ref{ie-initiald1}--\ref{te-initiald2}.}
\begin{tabular}{lll}
\hline
Experiment	&Class	&Update rule					\\
\hline
$d=1$,		&(i)	&{\tt 1d000a20}					\\
$r_1=2$,	&(ii)	&{\tt 01dc3610}					\\
Figures~\ref{ie-initiald1}(a--d) and \ref{te-initiald1}(a--d)
		&(iii)	&{\tt 994a6a65}					\\
		&(iv)	&{\tt 6c1e53a8}					\\
\hline
$d=1$,		&(i)	&{\tt 1df00000000f00000000000000000020}		\\
$r_1=3$,	&(ii)	&{\tt 7fdc3610fc48472c01dc361001dc3660}		\\
Figures~\ref{ie-initiald1}(e--h) and \ref{te-initiald1}(e--h)
		&(iii)	&{\tt 994f6a65994a6a65a94a6a65994a6a99}		\\
		&(iv)	&{\tt 3b469c0ee4f7fa96f93b4d32b09ed0e0}		\\
\hline
$d=2$,		&(i)	&{\tt 00000601}					\\
$r_1=r_2=1$,	&(ii)	&{\tt 06900600}					\\
von Neumann,	&(iii)	&{\tt 69969669}, Fredkin2 \cite{mcell-lex-url}	\\
Figures~\ref{ie-initiald2}(a--d) and \ref{te-initiald2}(a--d)
		&(iv)	&{\tt 6db6fac8}, Crystal2 \cite{mcell-lex-url}	\\
\hline
$d=2$,		&(i)	&{\tt b3b6b7 s3s6s7s8}				\\
$r_1=r_2=1$,	&(ii)	&{\tt b3 s2s5s6}				\\
Moore,		&(iii)	&{\tt b1b3b5 s1s3s5}				\\
Figures~\ref{ie-initiald2}(e--h) and \ref{te-initiald2}(e--h)
		&(iv)	&{\tt b3 s2s3}					\\
\hline
\end{tabular}

\label{updtrules}
\vspace{0.45in}
\end{table}

Within each run, the cellular automaton is simulated for $T=5,10,\ldots,250$
and for each simulation the mean and variance indicators of Section~\ref{heur}
are computed. Simplifying the notation in the obvious way, these are
$\overline{C_f}$, $\sigma^2(C_f)$, $\overline{T_f}$, and $\sigma^2(T_f)$, $f$
being the update rule under consideration. All simulations sharing the same
value of $d$, $X_1,\ldots,X_d$, and $r_1,\ldots,r_d$ start at the same initial
configuration, itself generated at the beginning for that group of simulations
by randomly choosing initial states for the $(X_1+2r_1t_+)\ldots(X_d+2r_dt_+)$
cells involved.

The results of these initial experiments are given in Figures~\ref{ie-initiald1}
through \ref{te-initiald2}, and also in Appendix~\ref{patterns}, where plots of
spatiotemporal patterns are given for selected runs. Figures~\ref{ie-initiald1}
and \ref{ie-initiald2} refer, respectively, to the behavior of the cell-centric
input entropy for the one- and two-dimensional cases as $T$ varies.
Figures~\ref{te-initiald1} and \ref{te-initiald2}, in turn, refer to the
behavior of the cell-centric transition entropy for the one- and two-dimensional
cases, respectively, as $T$ varies. Notice that, but virtue of our experiments'
setup, the plots in Figures~\ref{ie-initiald1} and \ref{te-initiald1} for which
$r_1$ and $X_1$ have the same values correspond to the same initial
configuration, and similarly for Figures~\ref{ie-initiald2} and
\ref{te-initiald2}.

\begin{figure}
\centering
\begin{tabular}{c@{\hspace{0.00in}}c}
\scalebox{0.300}{\includegraphics{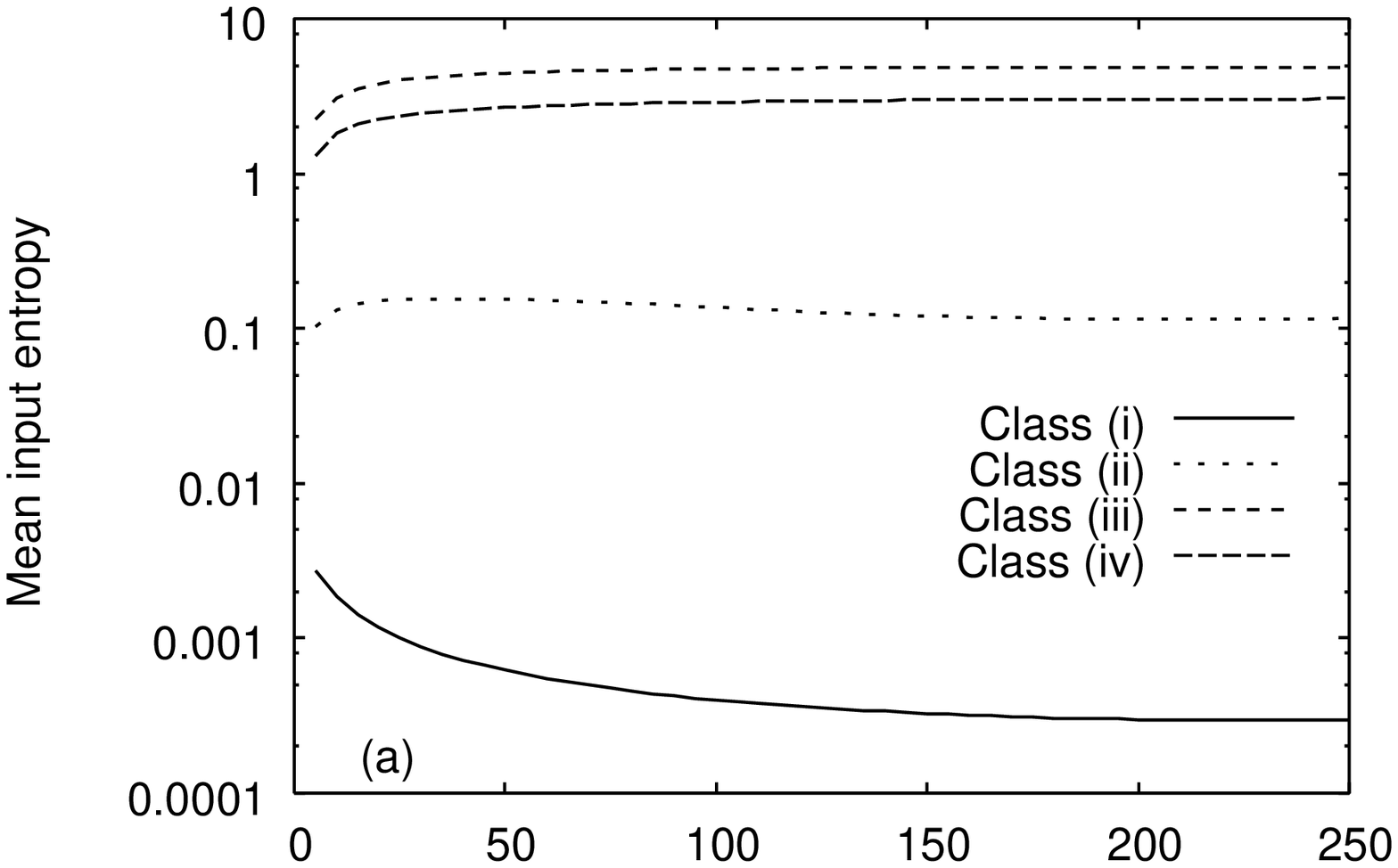}}&
\scalebox{0.300}{\includegraphics{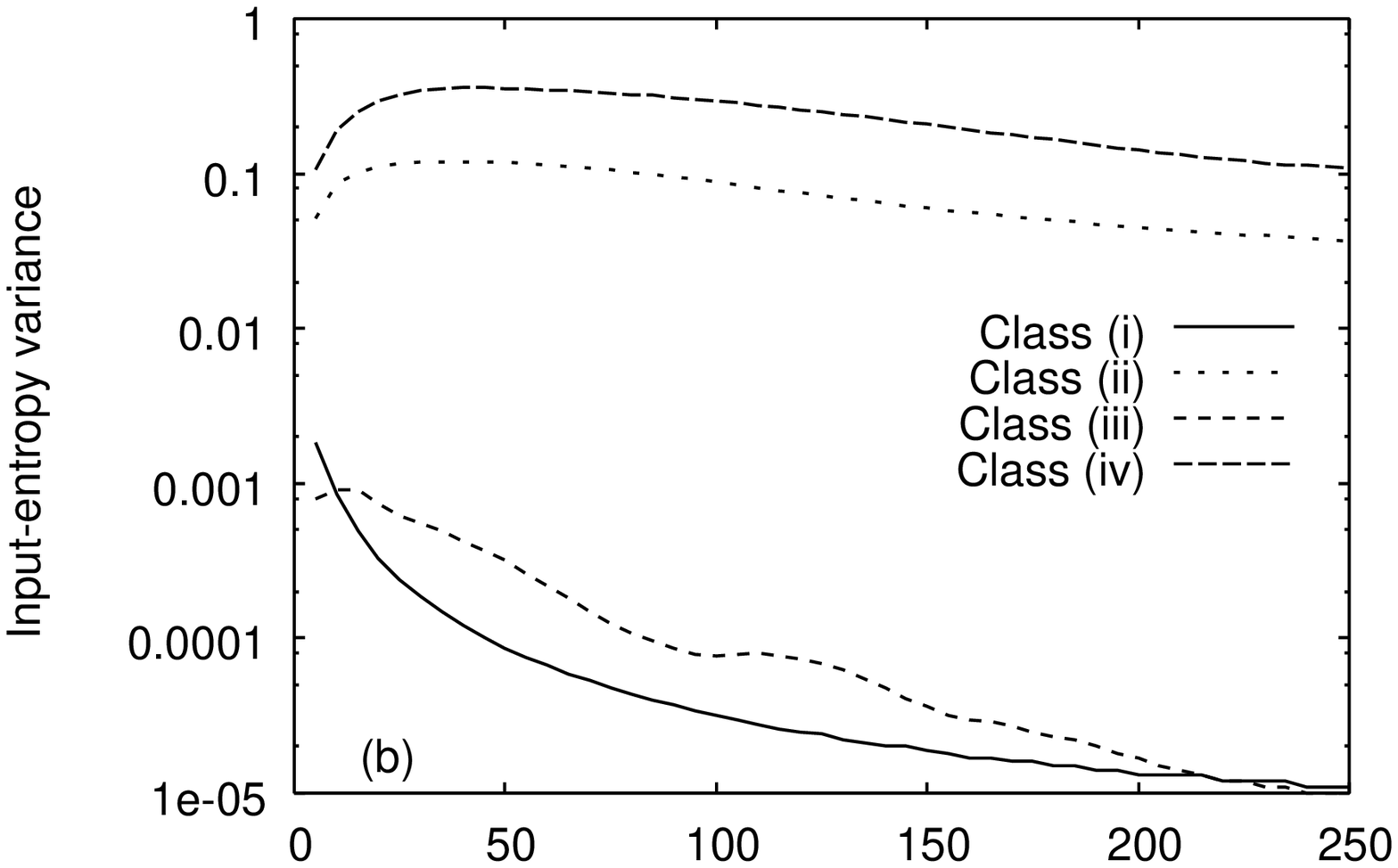}}\\
\scalebox{0.300}{\includegraphics{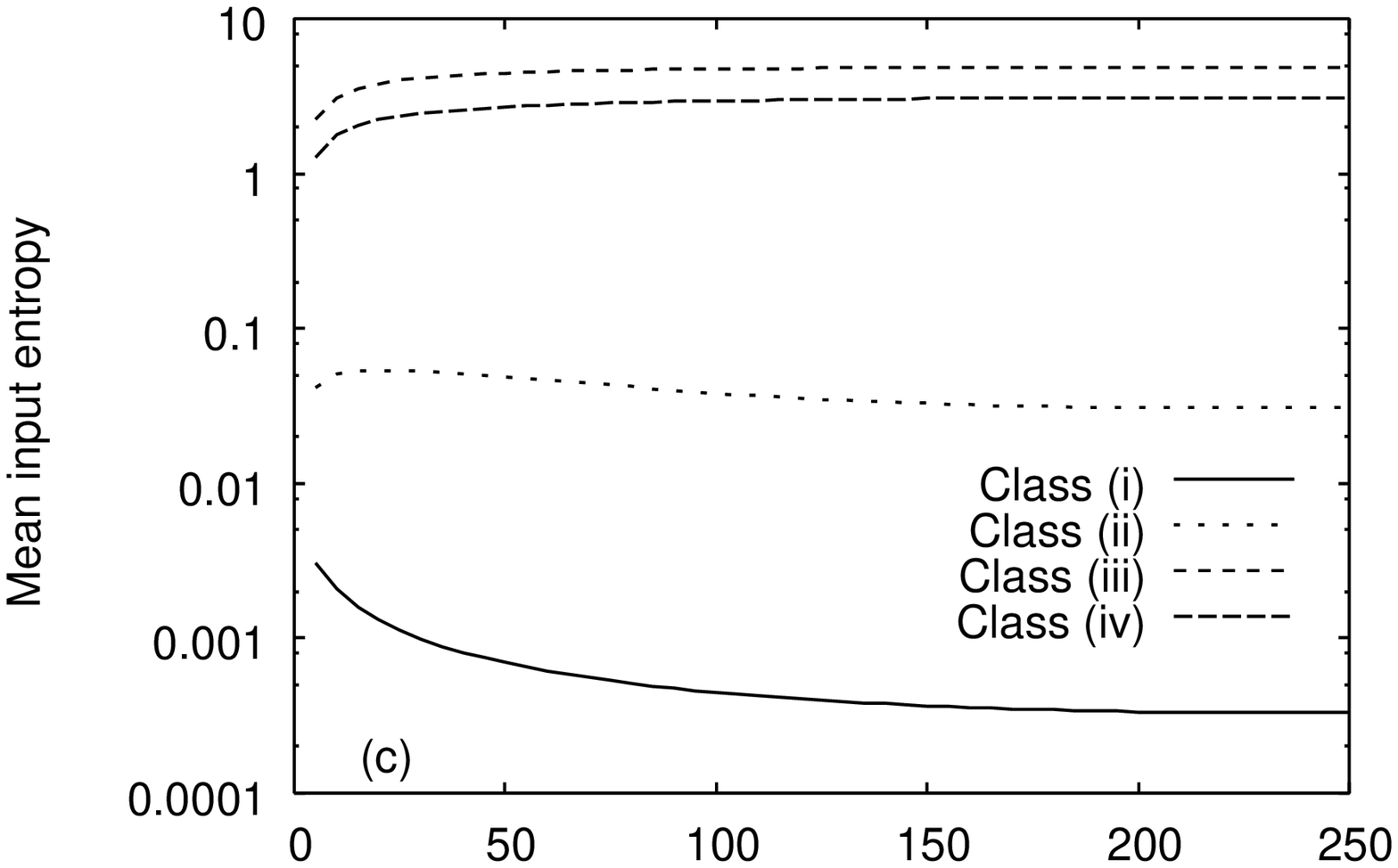}}&
\scalebox{0.300}{\includegraphics{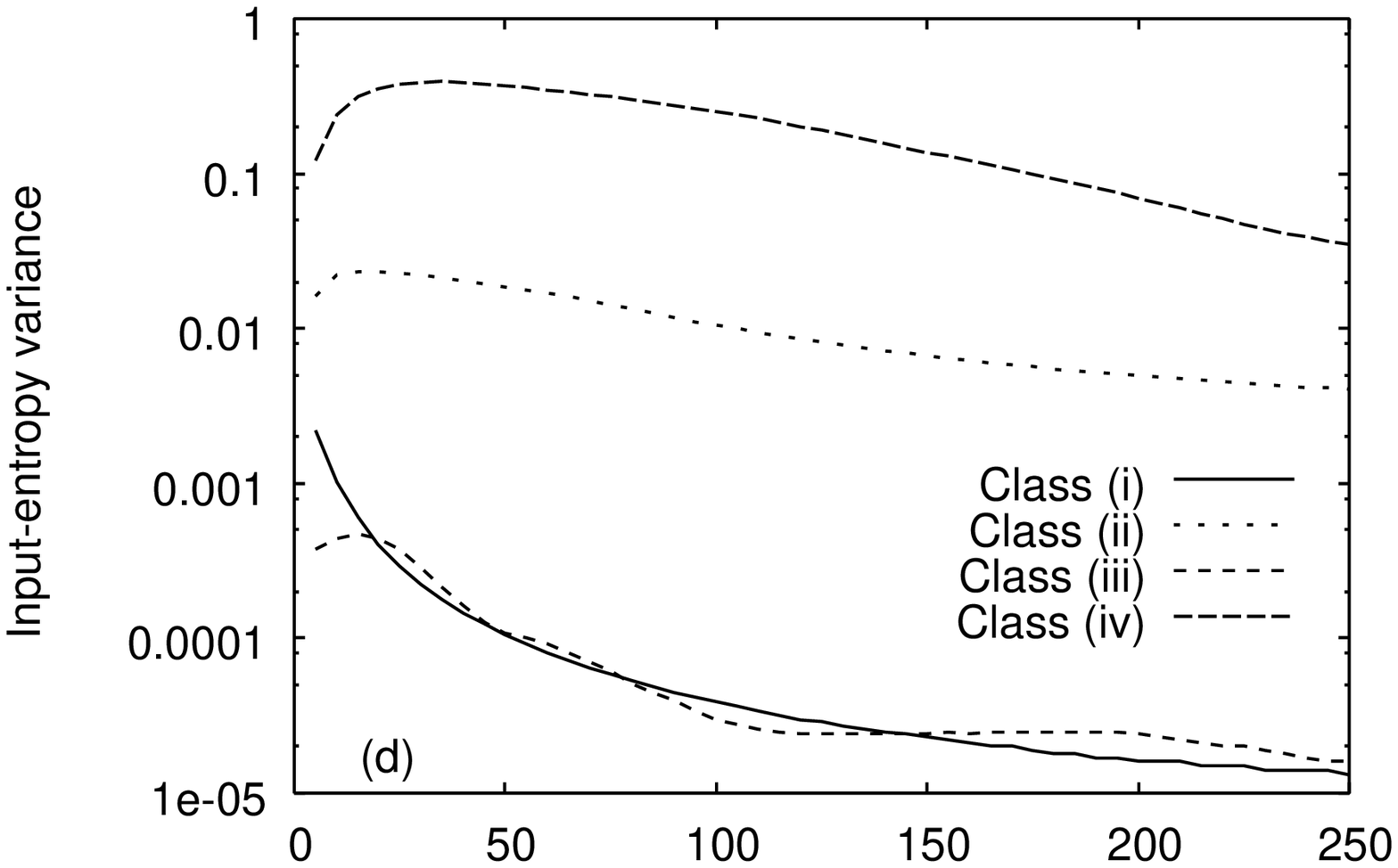}}\\
\scalebox{0.300}{\includegraphics{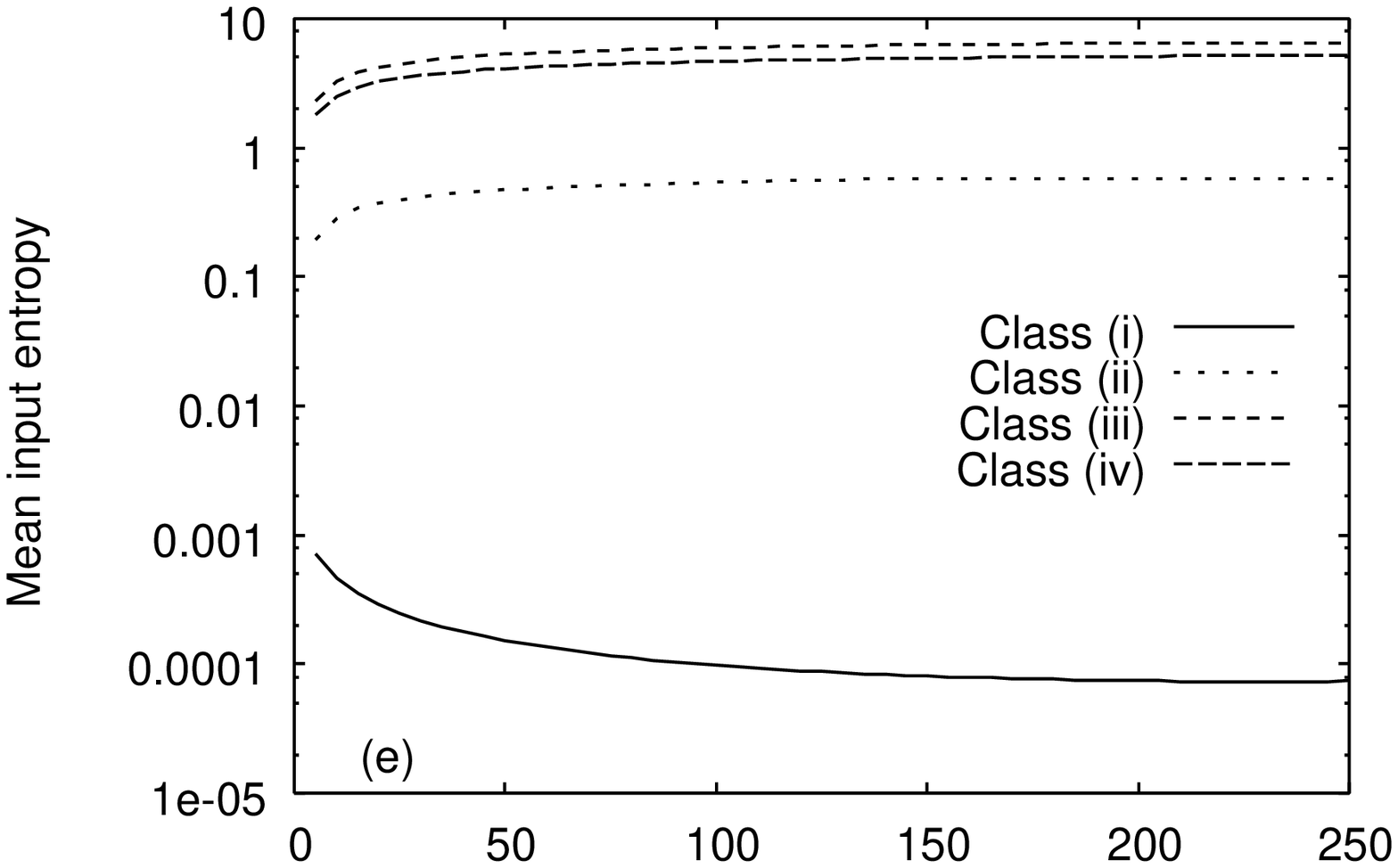}}&
\scalebox{0.300}{\includegraphics{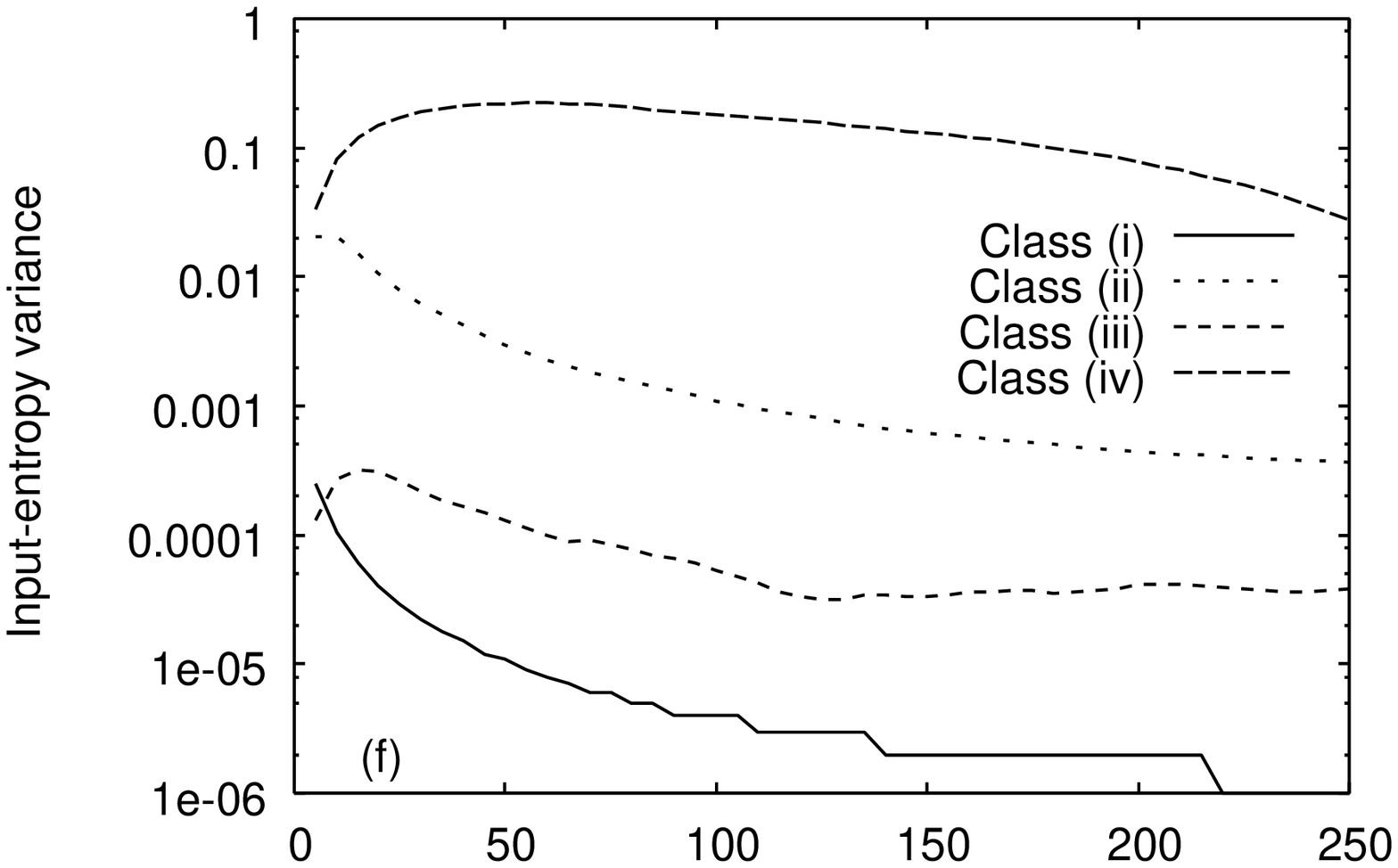}}\\
\scalebox{0.300}{\includegraphics{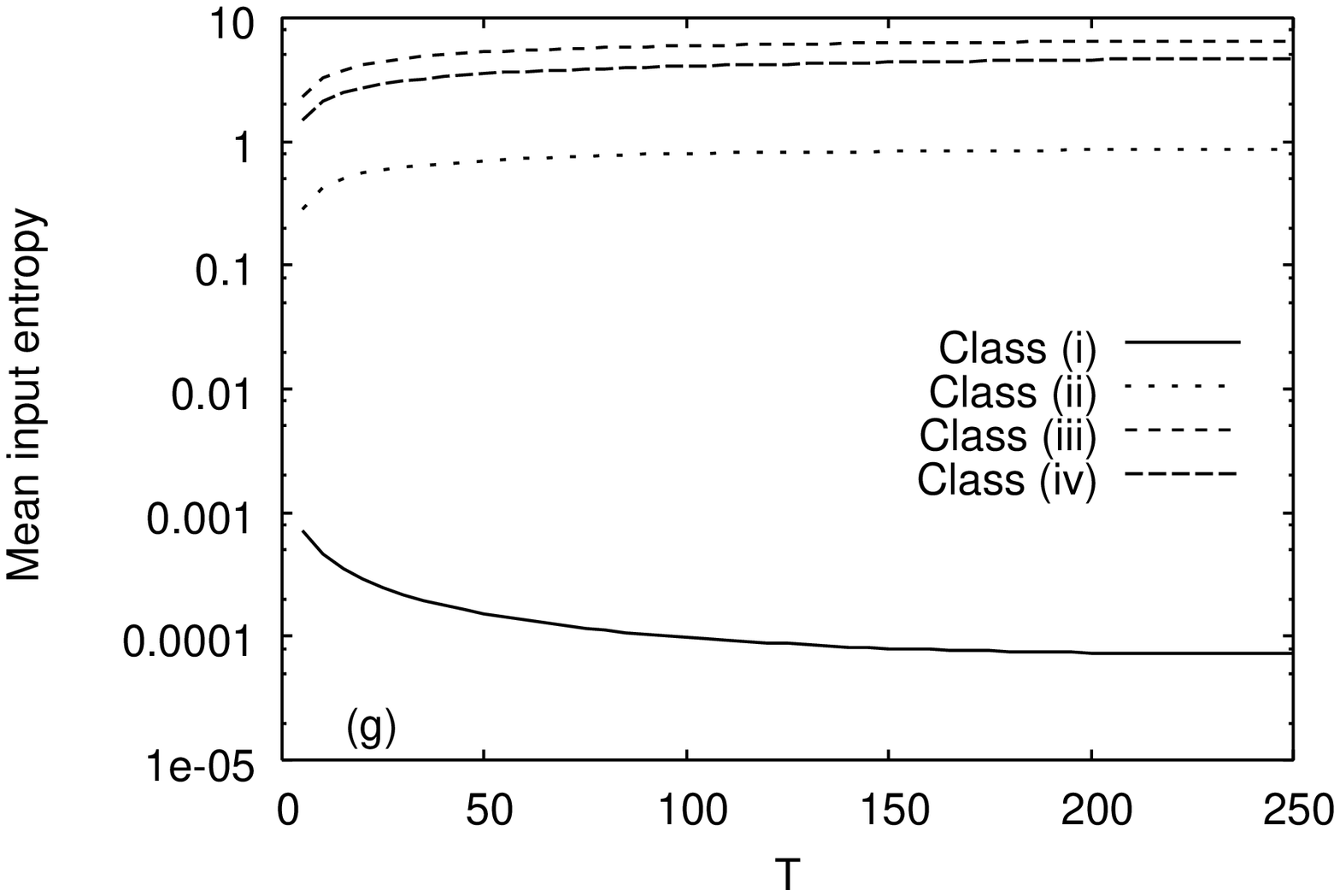}}&
\scalebox{0.300}{\includegraphics{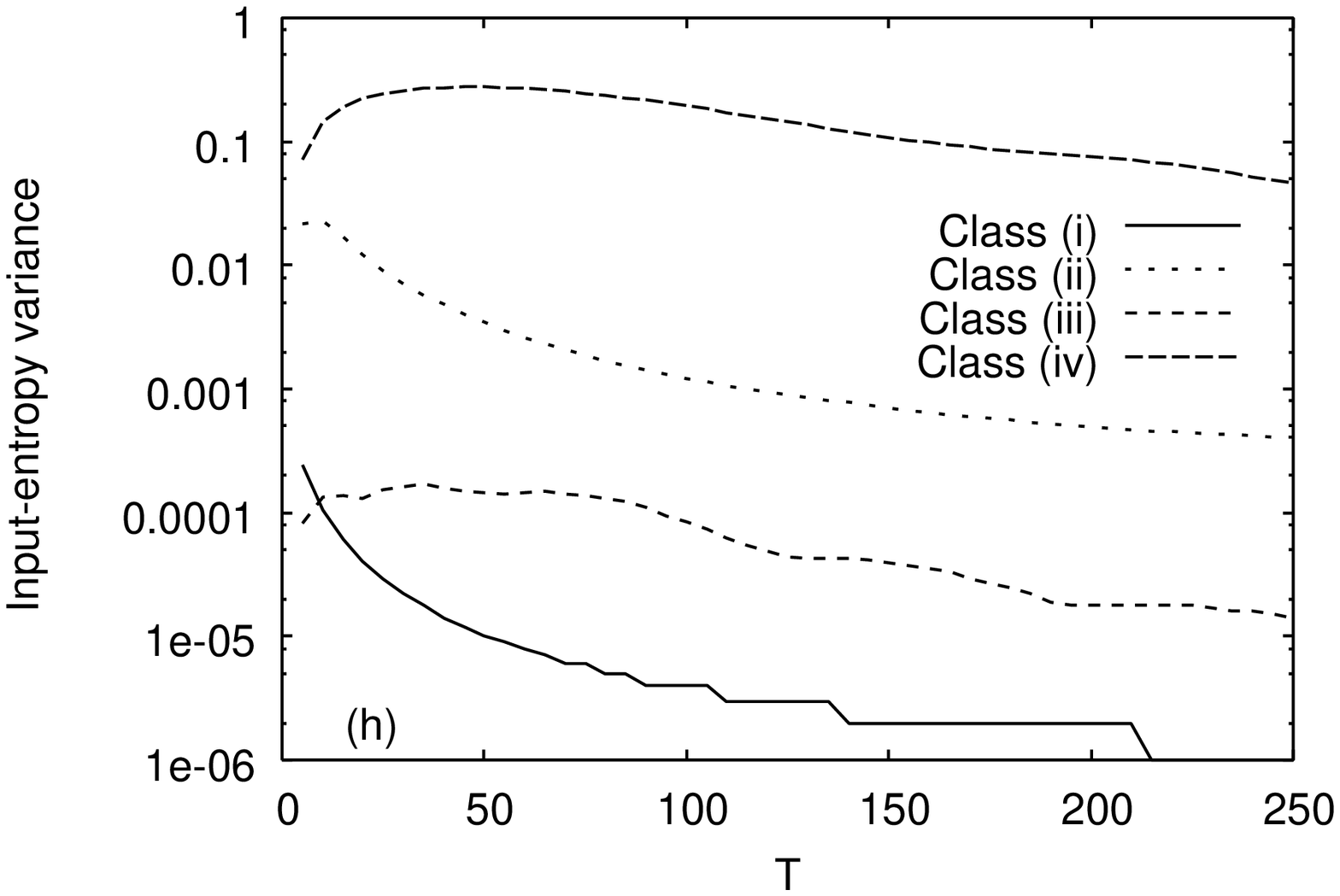}}
\end{tabular}
\caption{Mean ($\overline{C_f}$) and variance ($\sigma^2(C_f)$) of the
cell-centric input entropy as a function of $T$ under four different update
rules, one from each of classes (i) through (iv), for $d=1$. Data are given for
the $150$-cell case with $r_1=2$ (a and b), the $300$-cell case with $r_1=2$
(c and d), the $150$-cell case with $r_1=3$ (e and f), and the $300$-cell case
with $r_1=3$ (g and h).}
\label{ie-initiald1}
\end{figure}

\begin{figure}
\centering
\begin{tabular}{c@{\hspace{0.00in}}c}
\scalebox{0.300}{\includegraphics{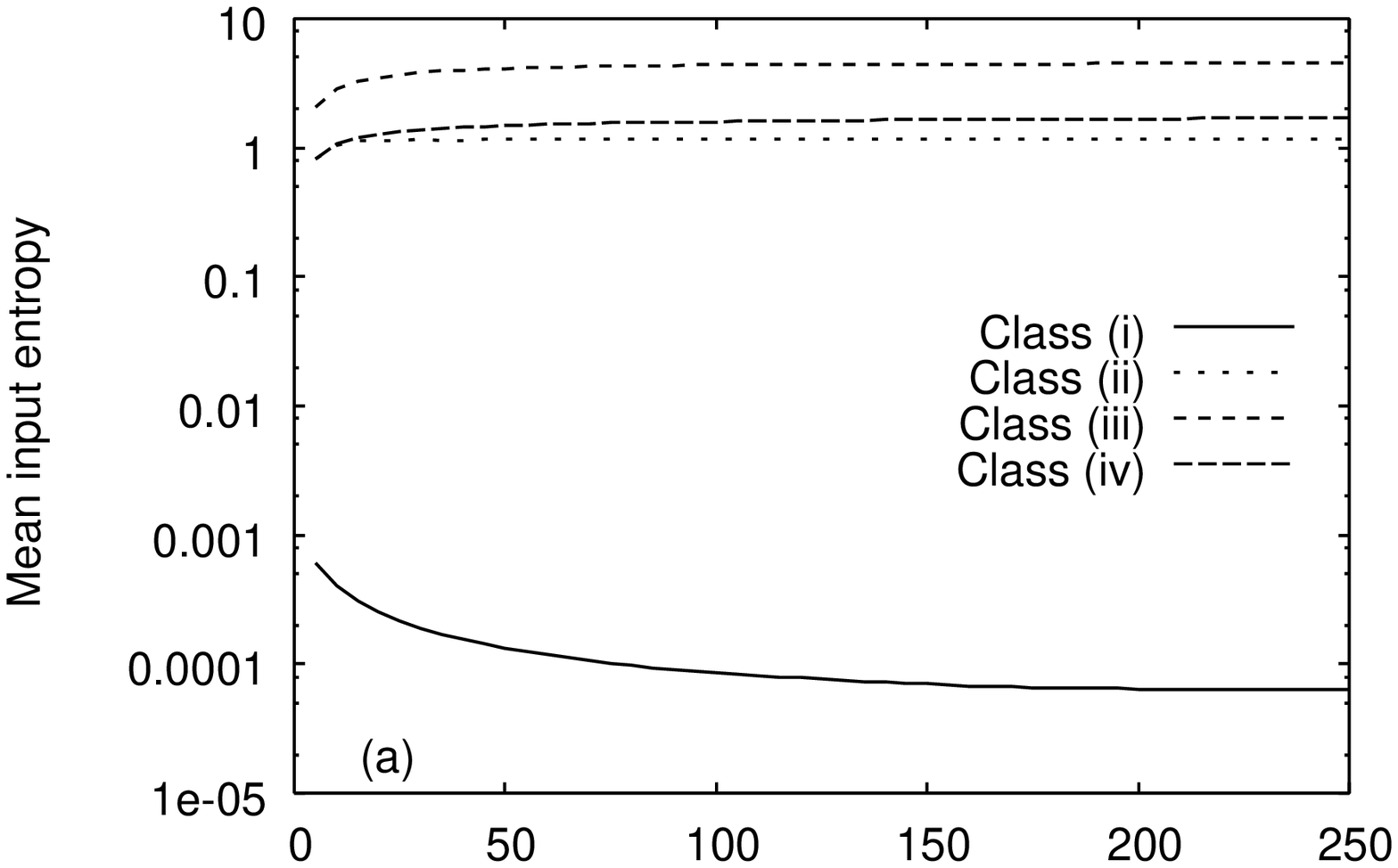}}&
\scalebox{0.300}{\includegraphics{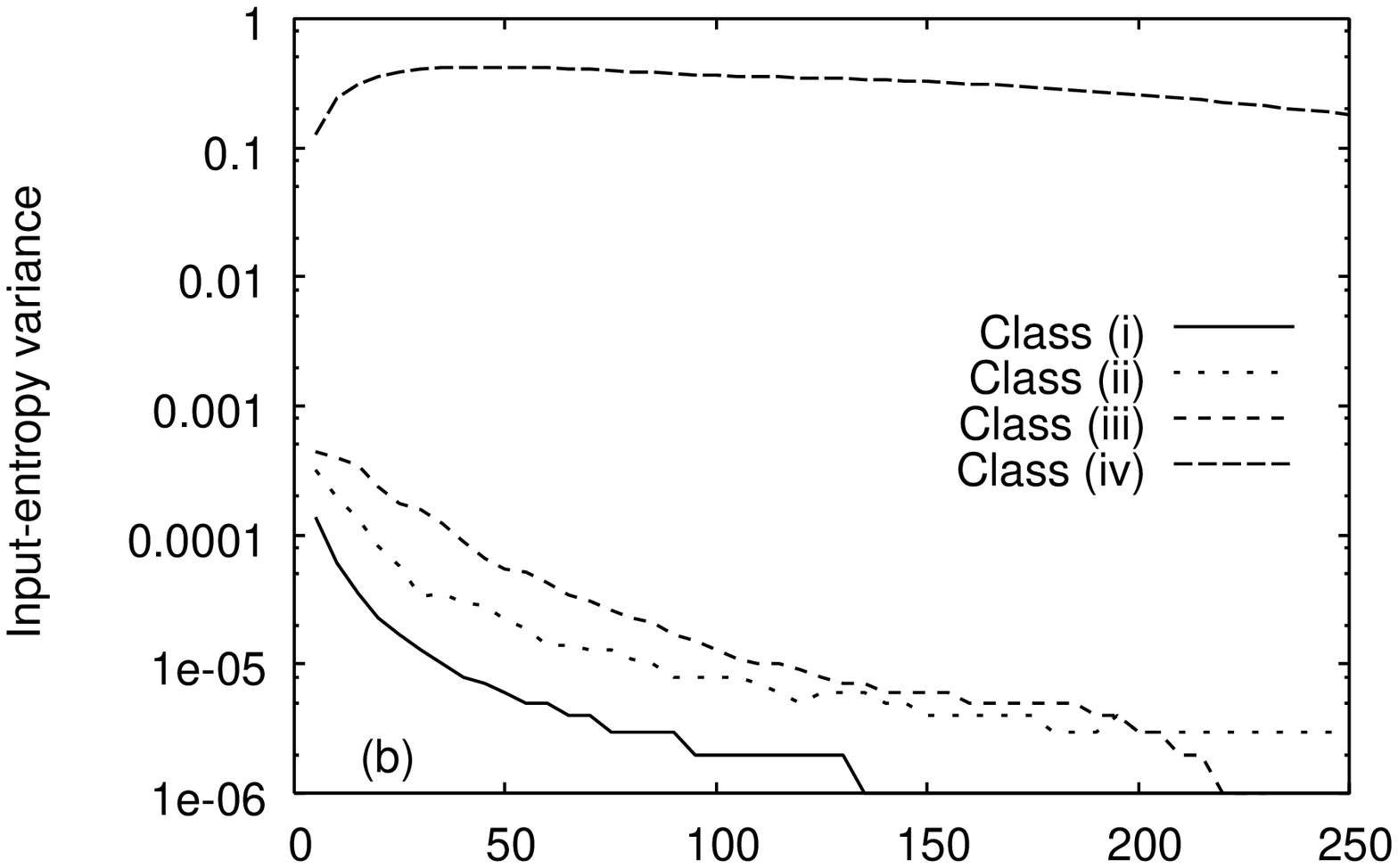}}\\
\scalebox{0.300}{\includegraphics{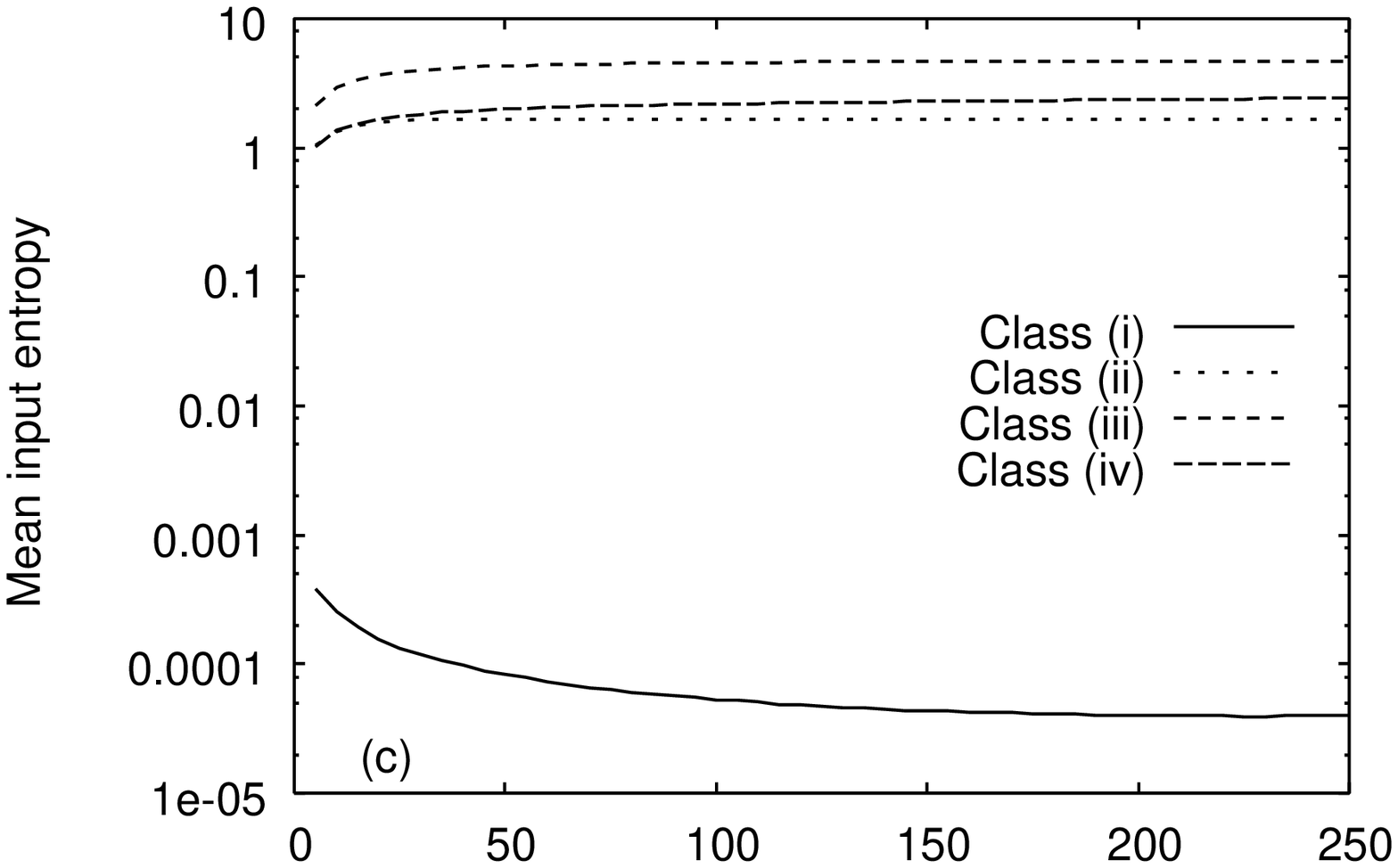}}&
\scalebox{0.300}{\includegraphics{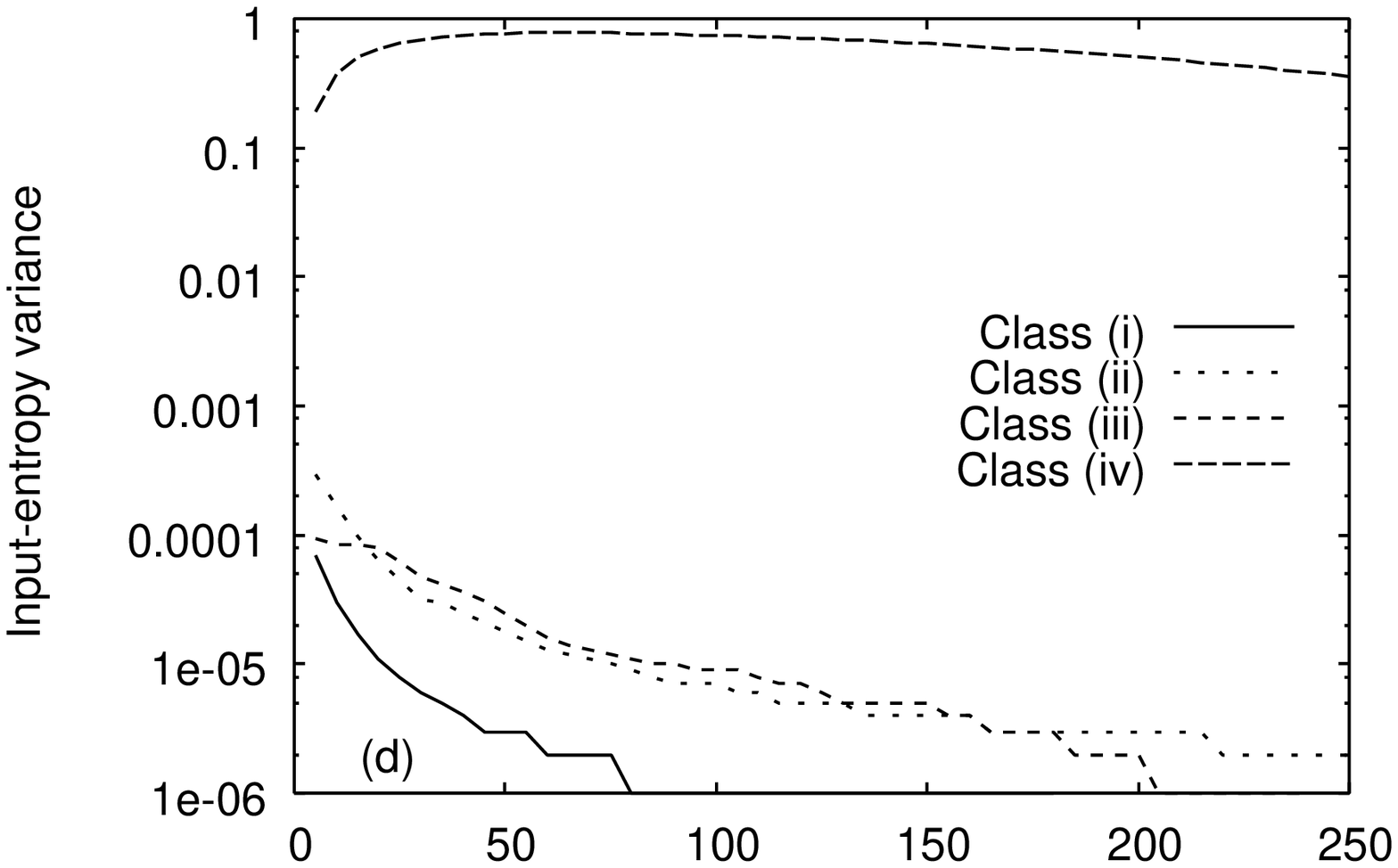}}\\
\scalebox{0.300}{\includegraphics{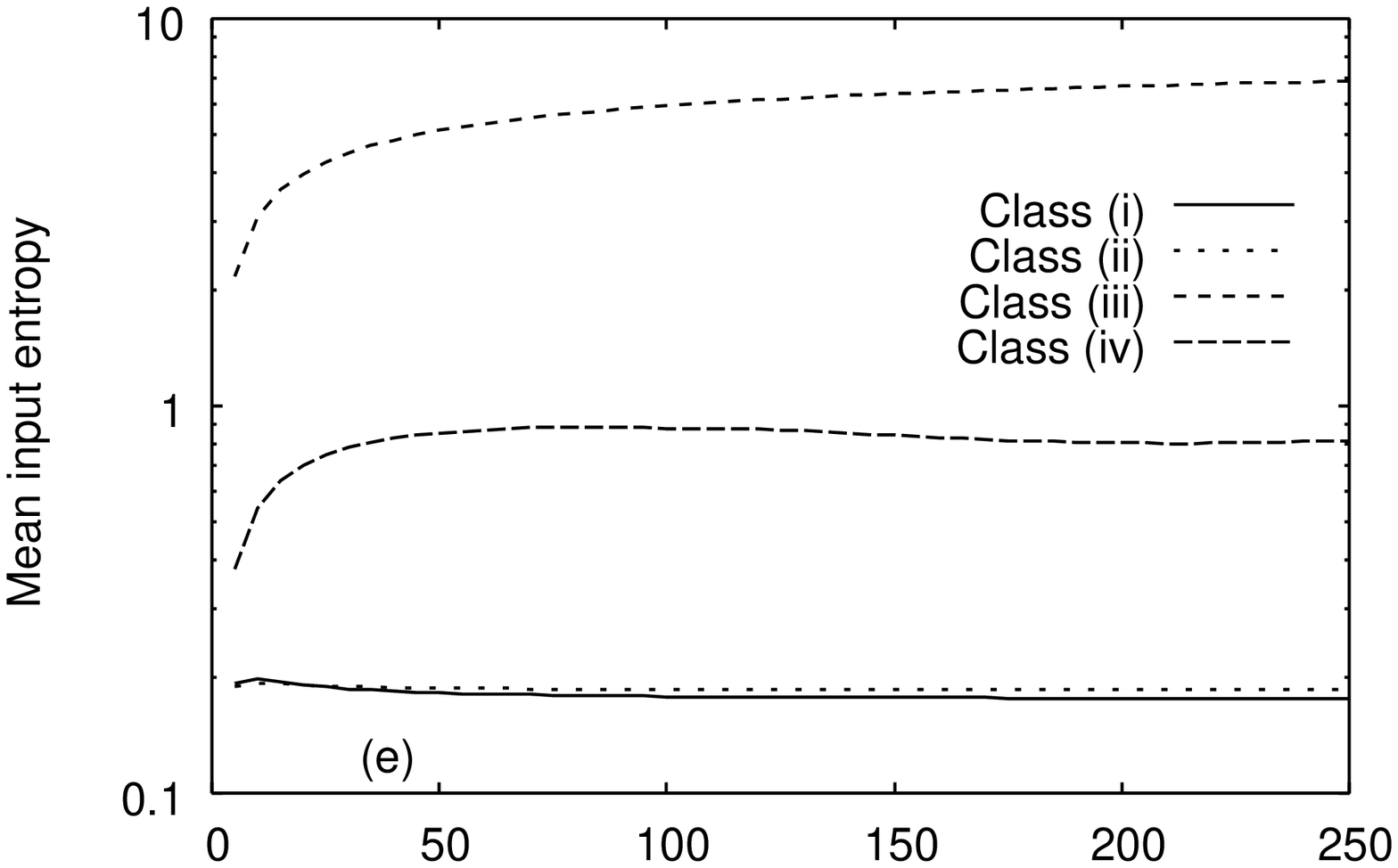}}&
\scalebox{0.300}{\includegraphics{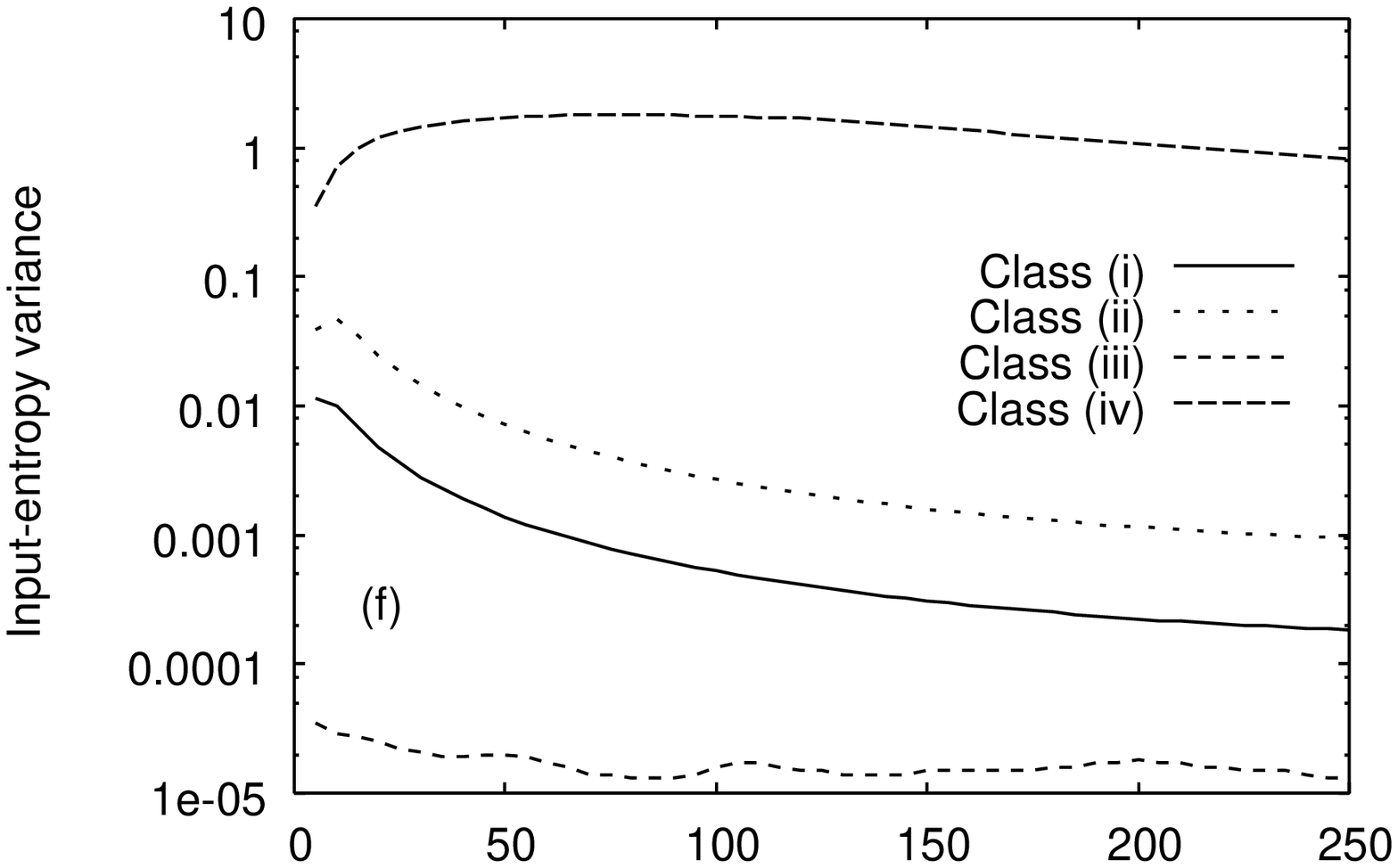}}\\
\scalebox{0.300}{\includegraphics{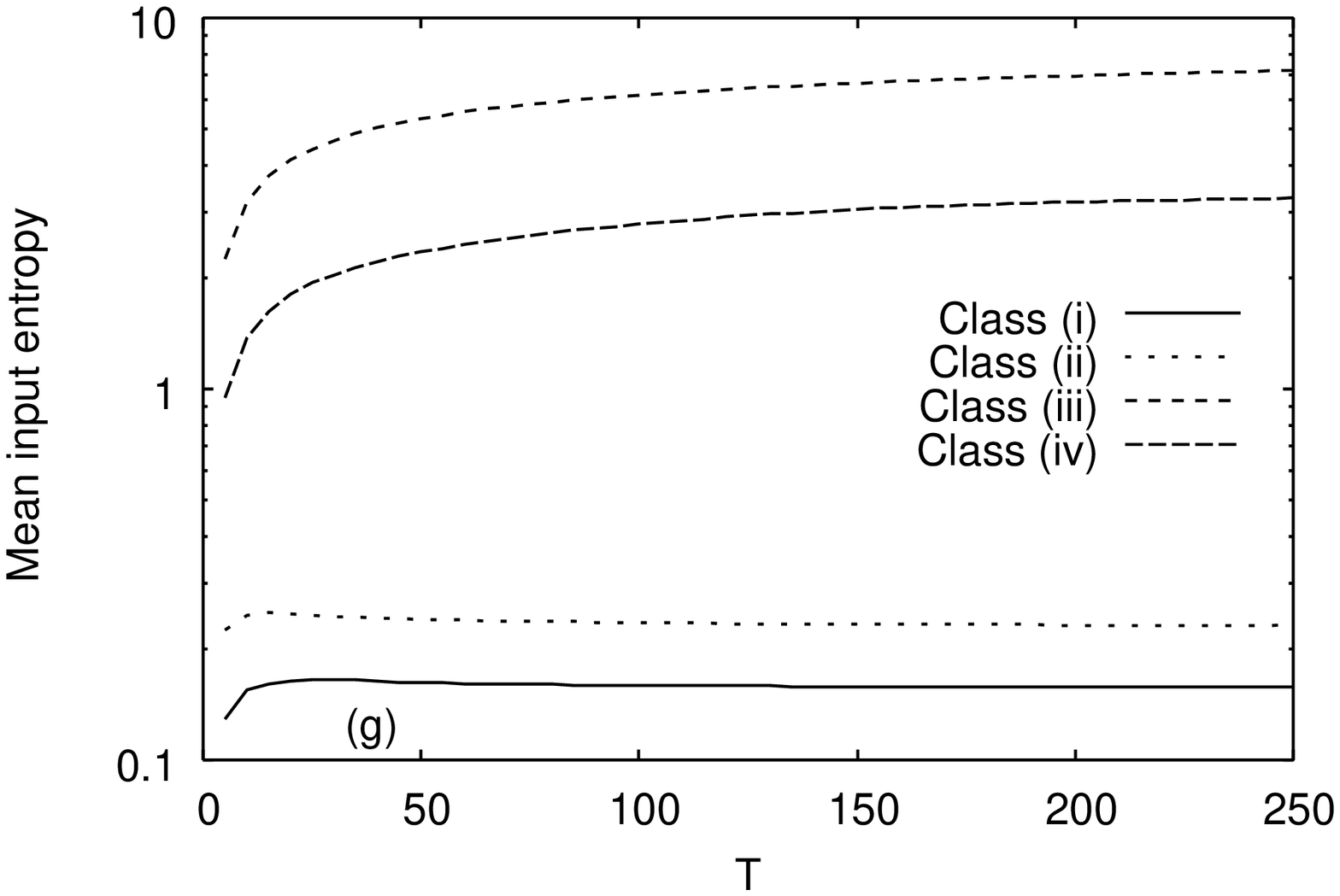}}&
\scalebox{0.300}{\includegraphics{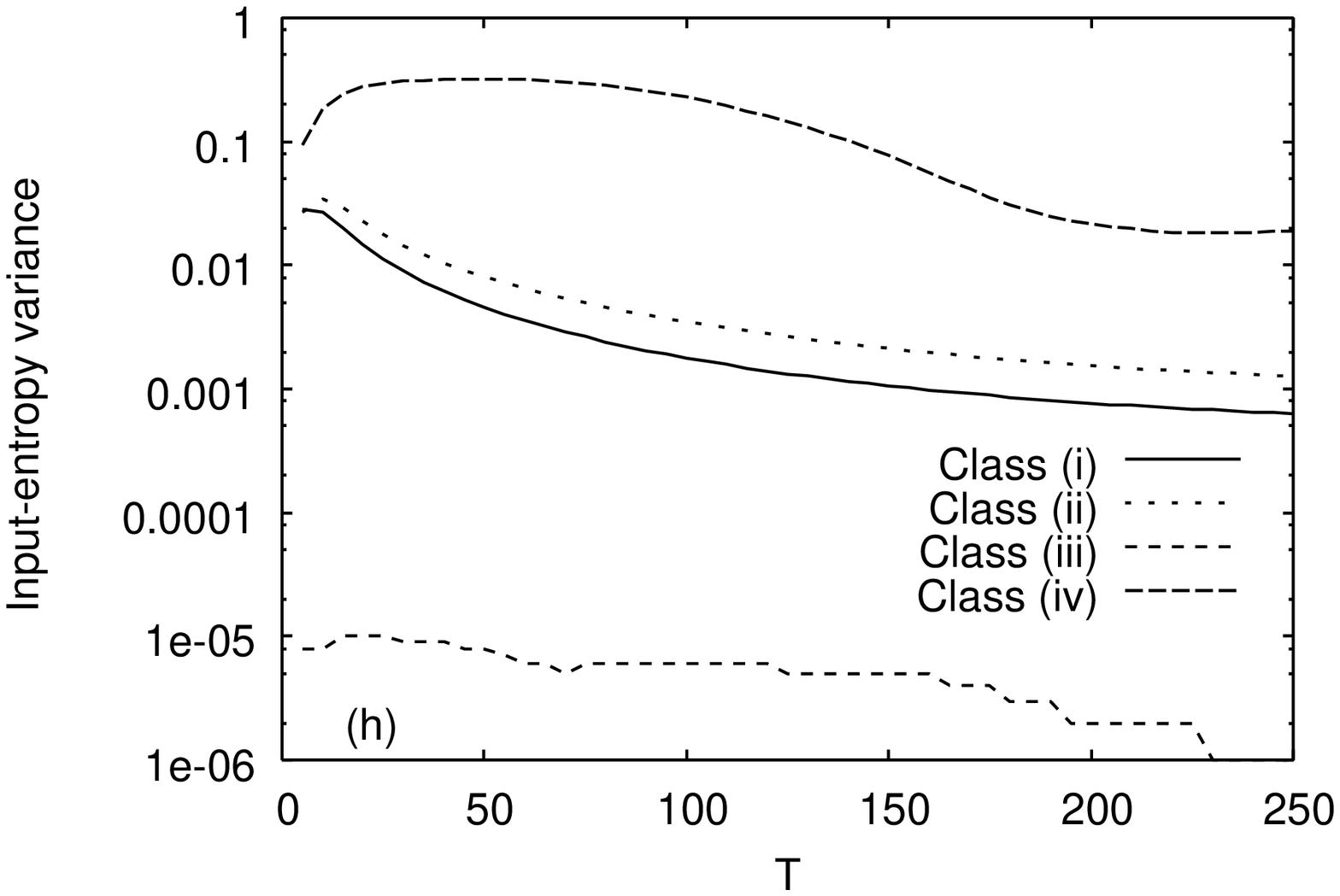}}
\end{tabular}
\caption{Mean ($\overline{C_f}$) and variance ($\sigma^2(C_f)$) of the
cell-centric input entropy as a function of $T$ under four different update
rules, one from each of classes (i) through (iv), for $d=2$. Data are given for
the $(15\times 15)$-cell von Neumann case (a and b), the $(30\times 30)$-cell
von Neumann case (c and d), the $(15\times 15)$-cell Moore case (e and f), and
the $(30\times 30)$-cell Moore case (g and h). In all cases, $r_1=r_2=1$.}
\label{ie-initiald2}
\end{figure}

\begin{figure}
\centering
\begin{tabular}{c@{\hspace{0.00in}}c}
\scalebox{0.300}{\includegraphics{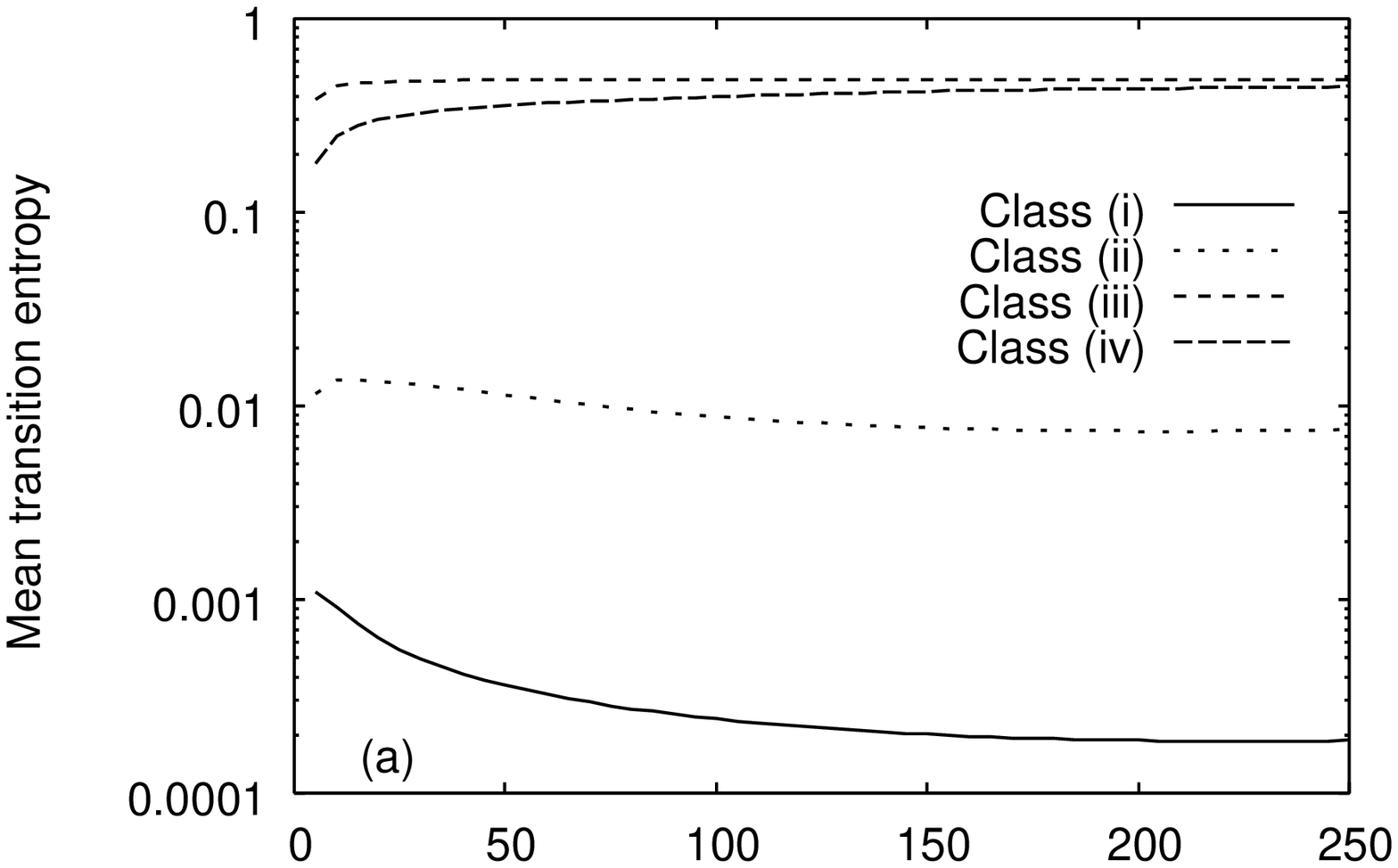}}&
\scalebox{0.300}{\includegraphics{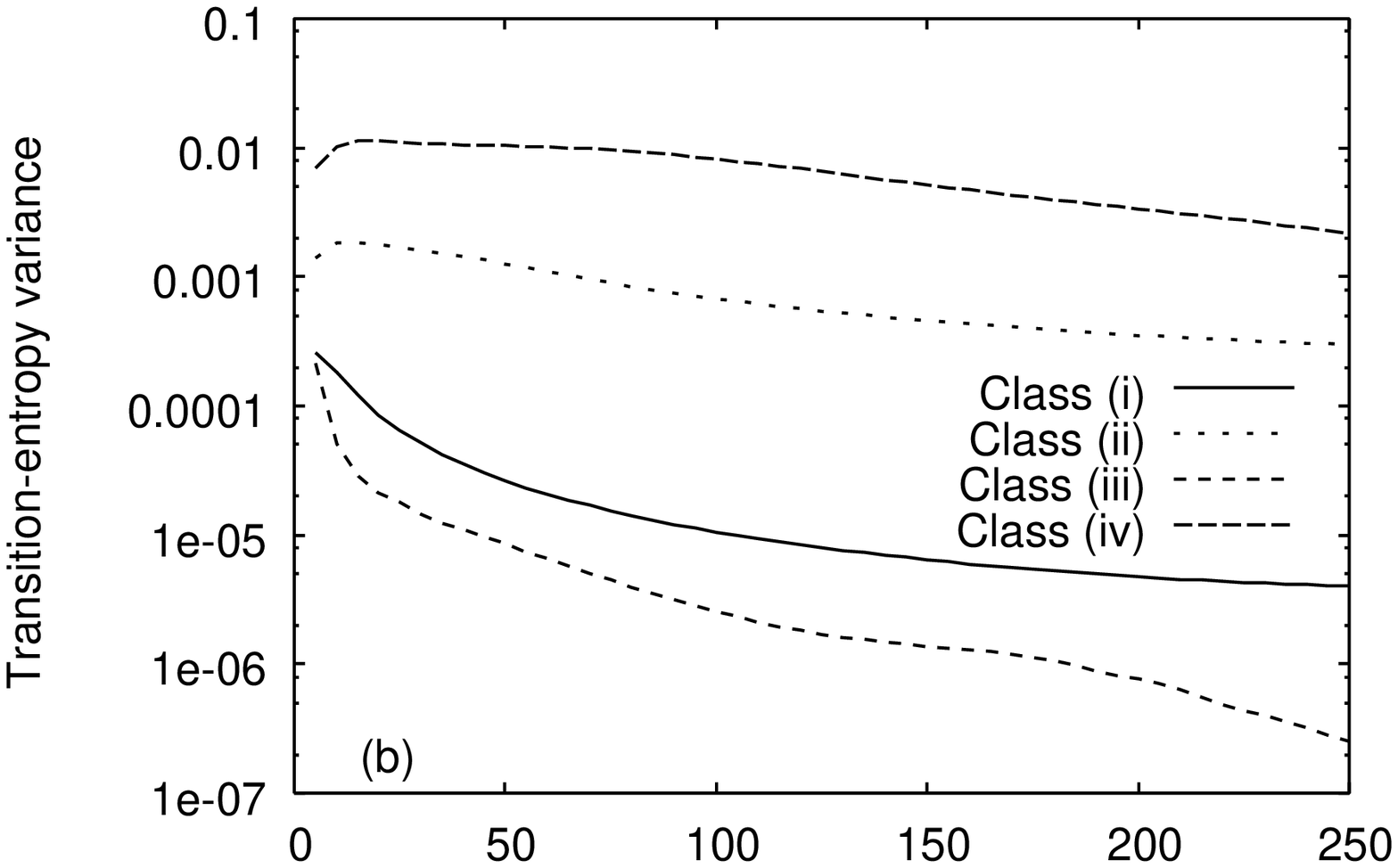}}\\
\scalebox{0.300}{\includegraphics{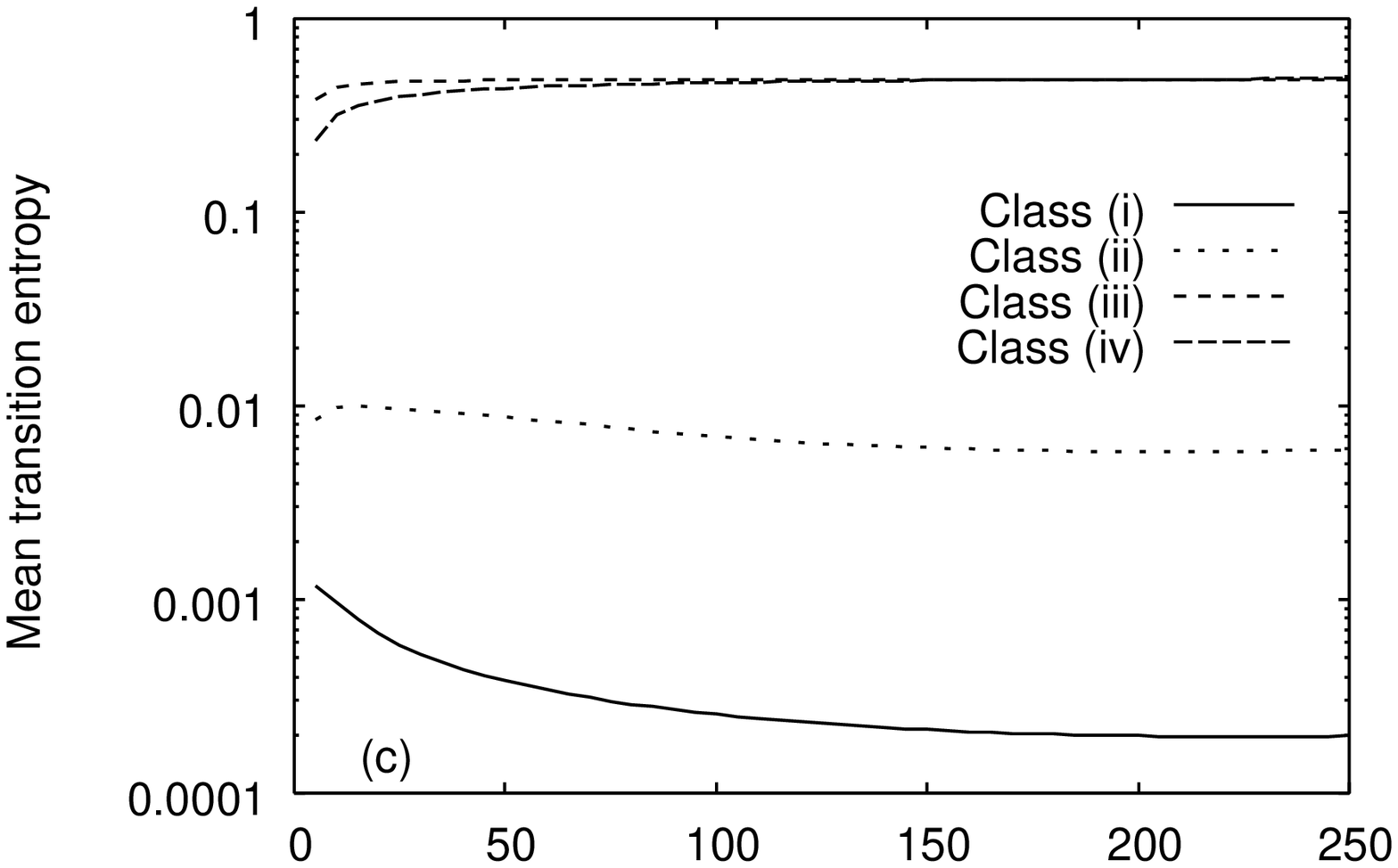}}&
\scalebox{0.300}{\includegraphics{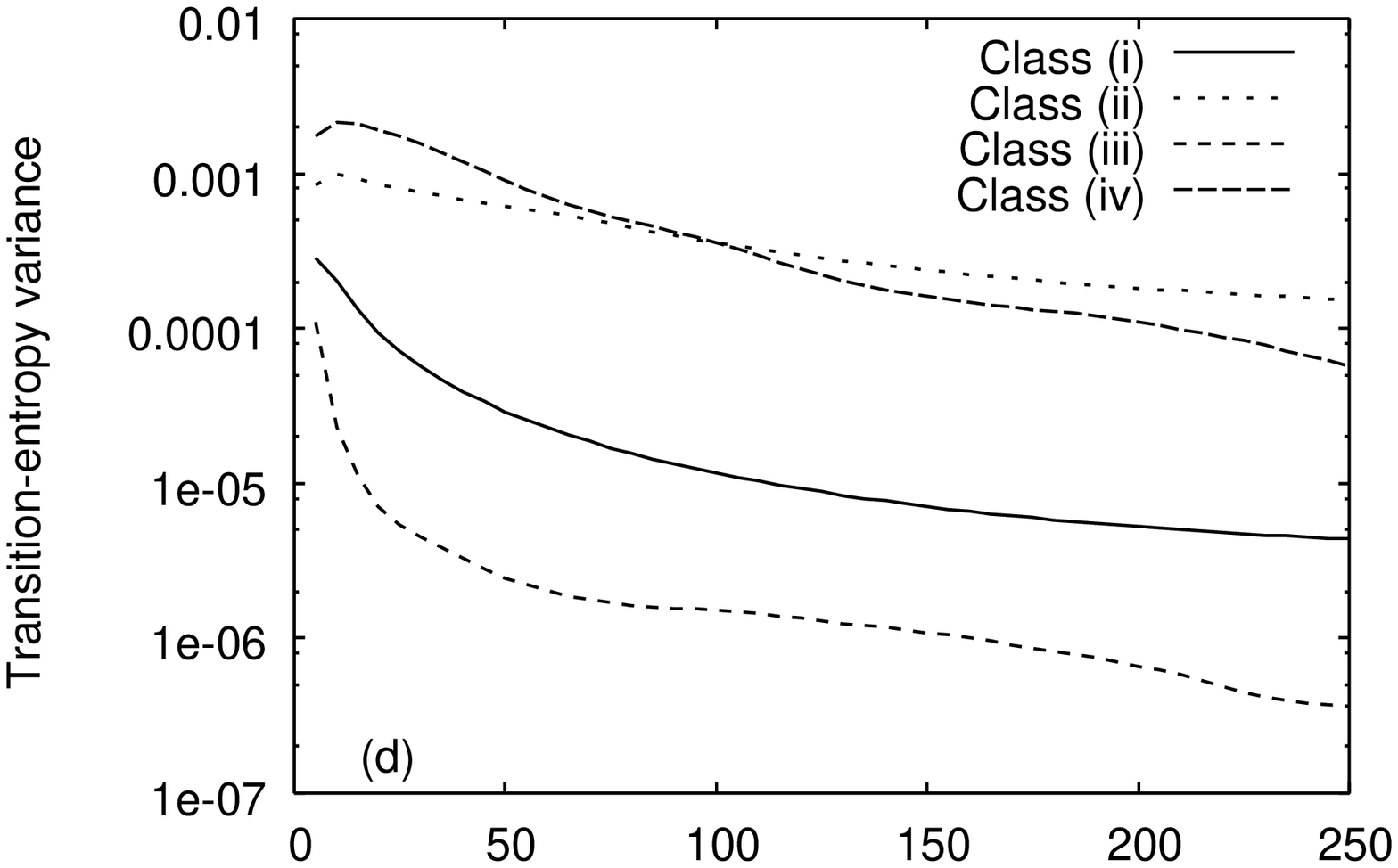}}\\
\scalebox{0.300}{\includegraphics{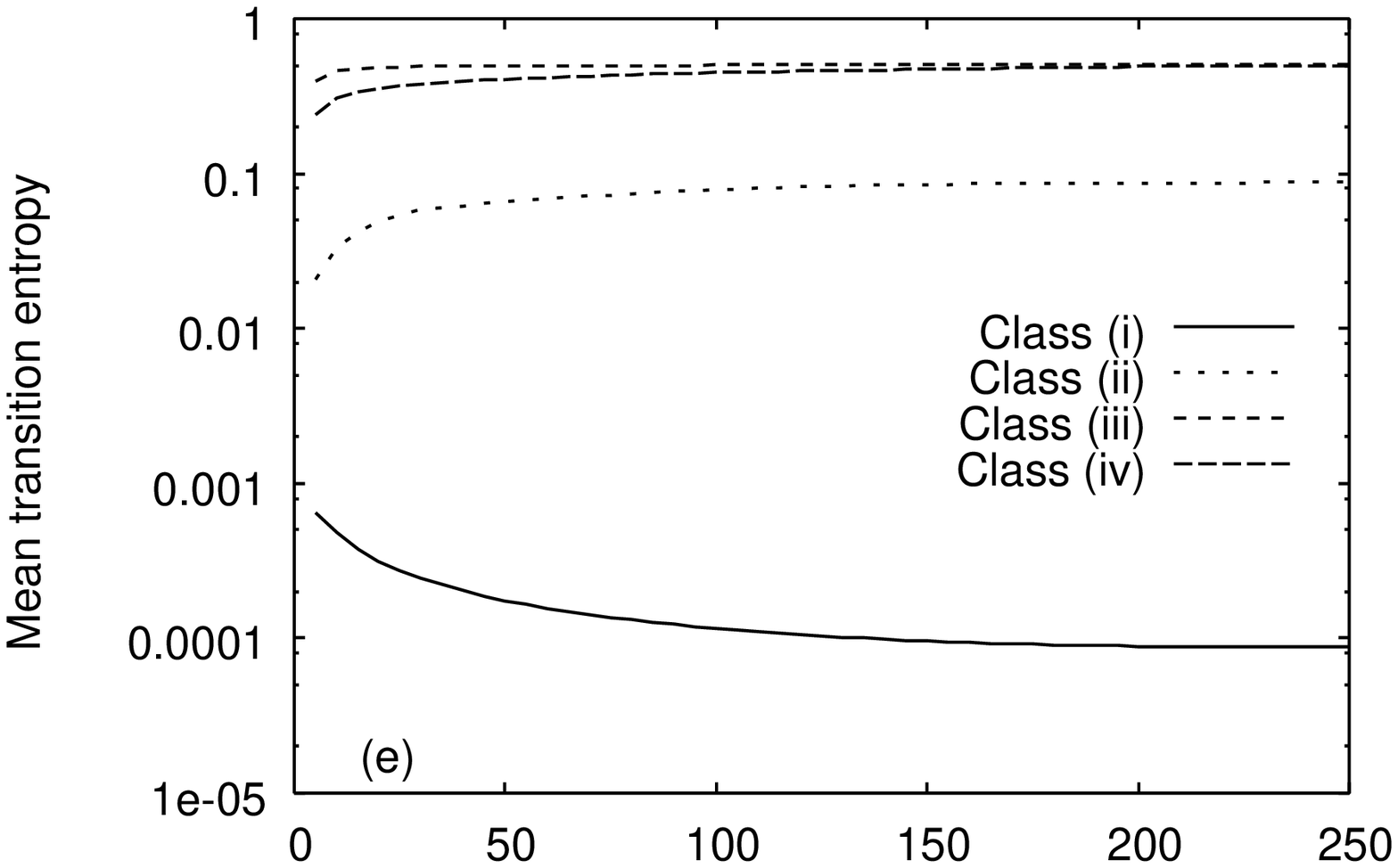}}&
\scalebox{0.300}{\includegraphics{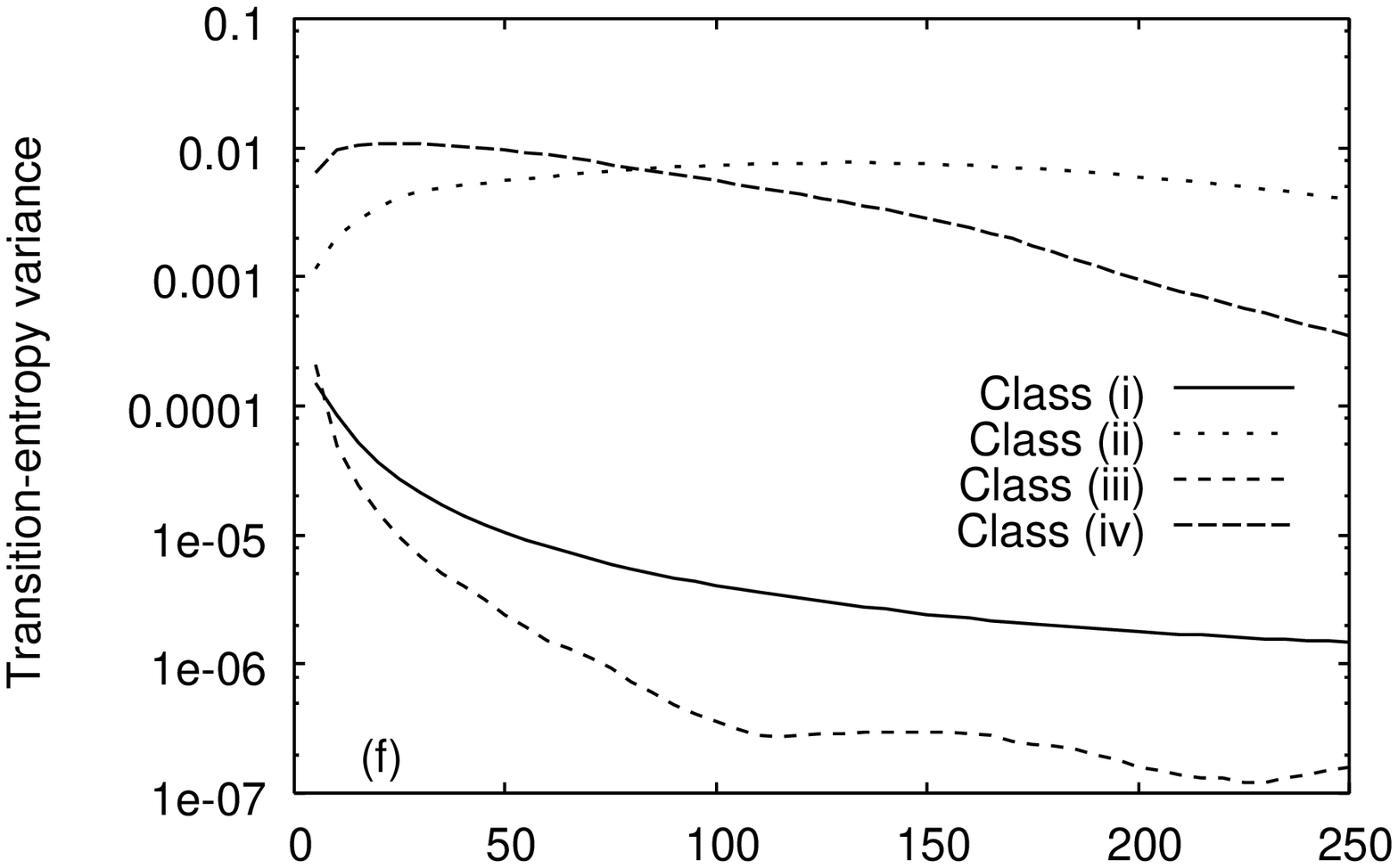}}\\
\scalebox{0.300}{\includegraphics{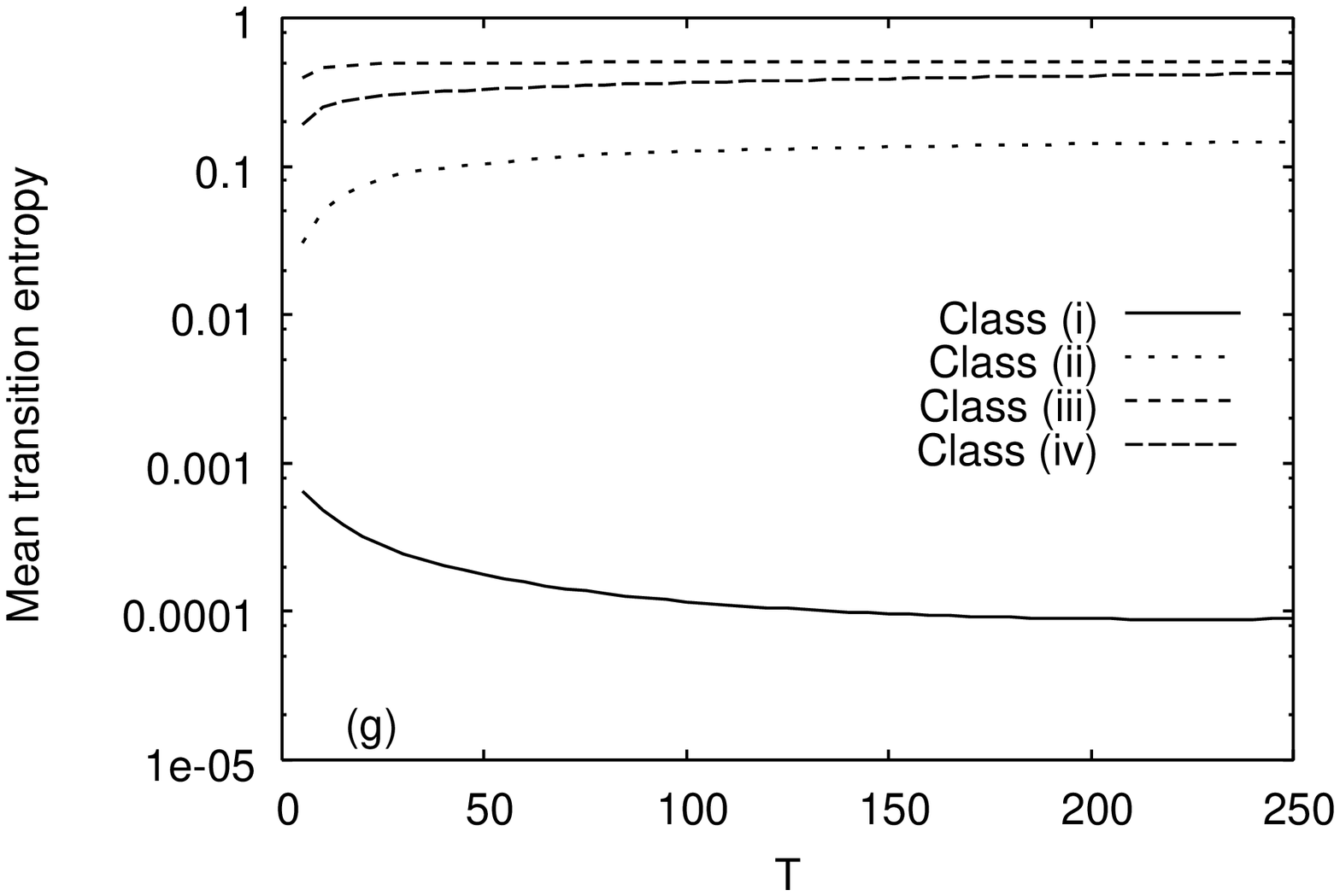}}&
\scalebox{0.300}{\includegraphics{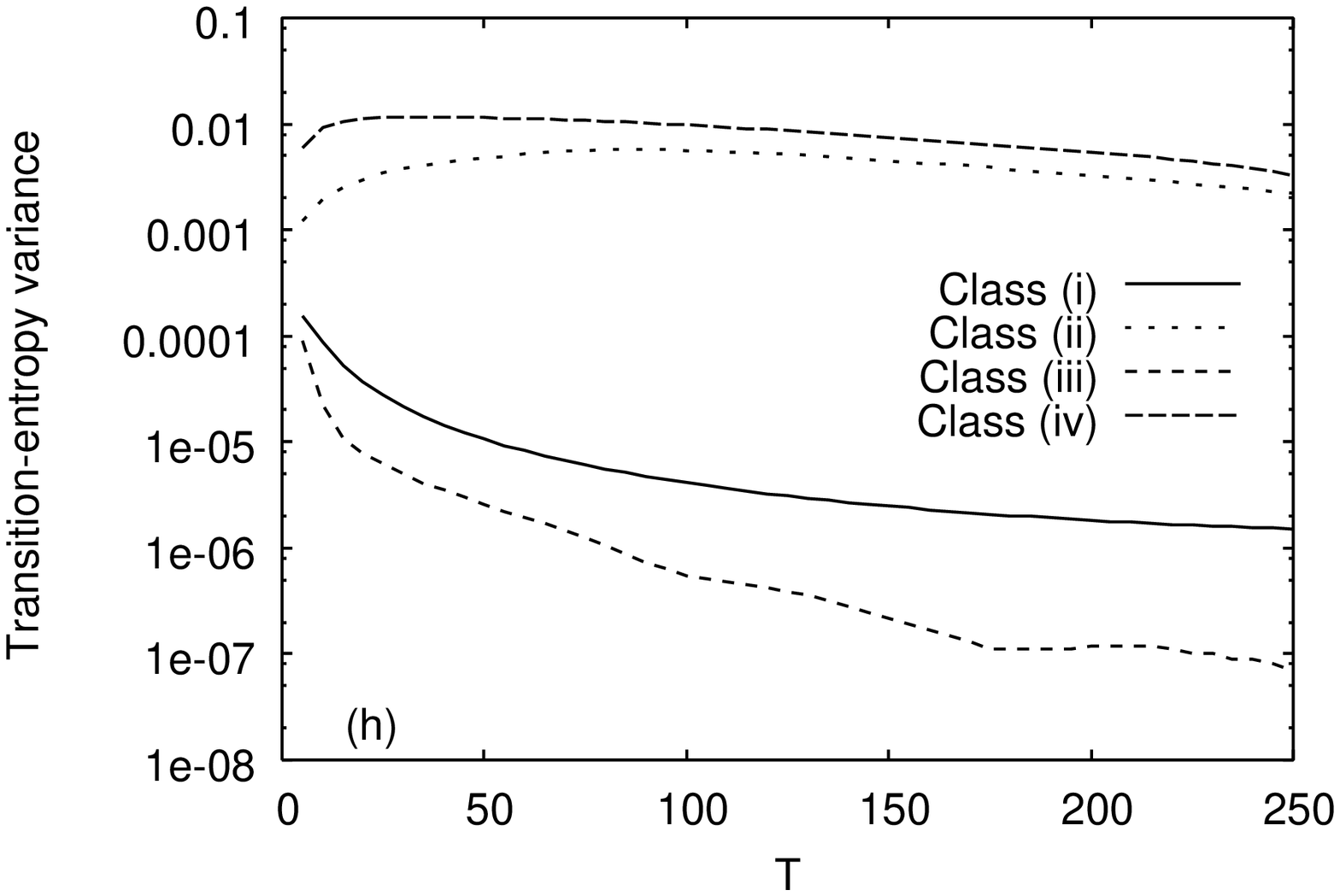}}
\end{tabular}
\caption{Mean ($\overline{T_f}$) and variance ($\sigma^2(T_f)$) of the
cell-centric transition entropy as a function of $T$ under four different update
rules, one from each of classes (i) through (iv), for $d=1$. Data are given for
the $150$-cell case with $r_1=2$ (a and b), the $300$-cell case with $r_1=2$
(c and d), the $150$-cell case with $r_1=3$ (e and f), and the $300$-cell case
with $r_1=3$ (g and h).}
\label{te-initiald1}
\end{figure}

\begin{figure}
\centering
\begin{tabular}{c@{\hspace{0.00in}}c}
\scalebox{0.300}{\includegraphics{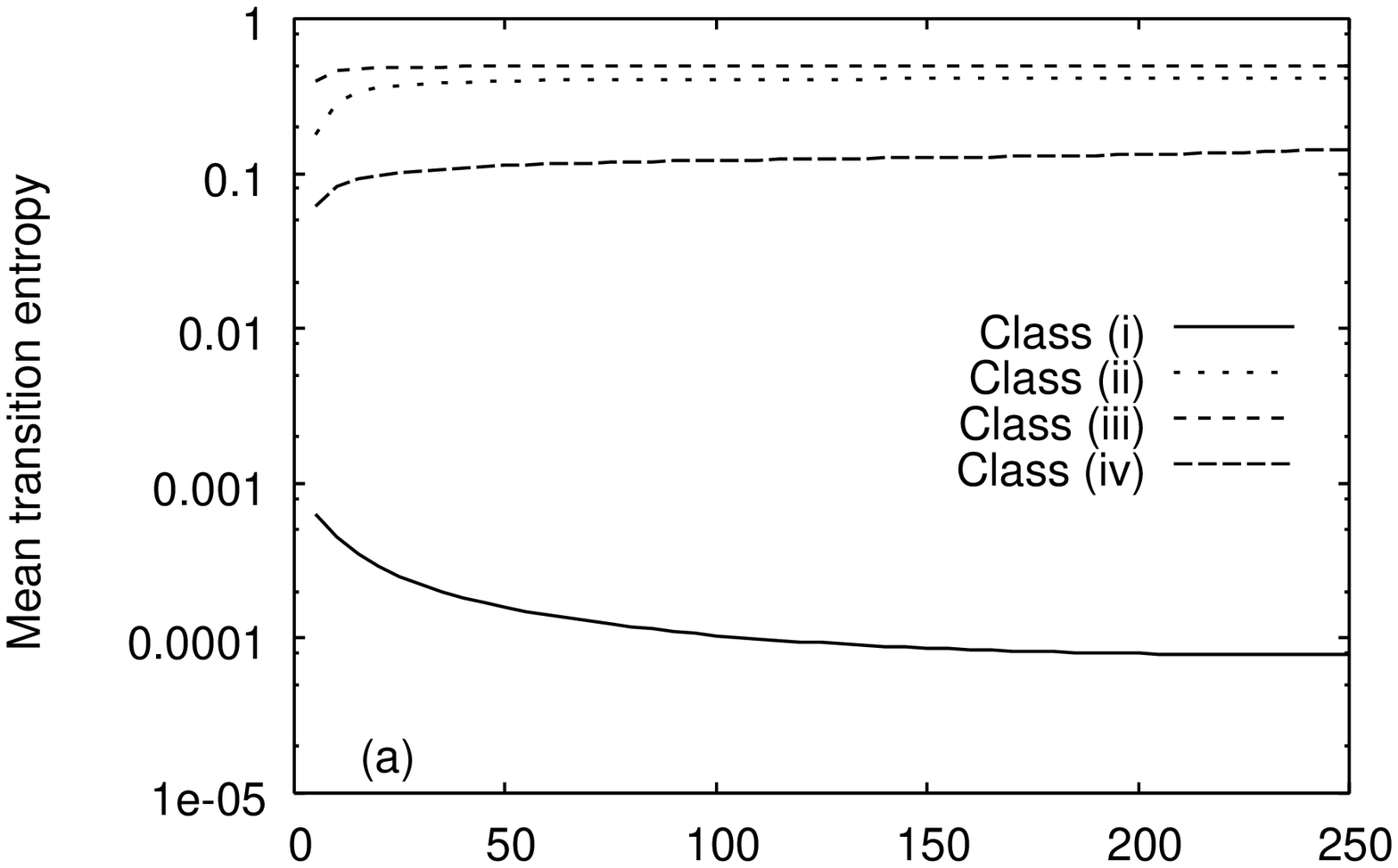}}&
\scalebox{0.300}{\includegraphics{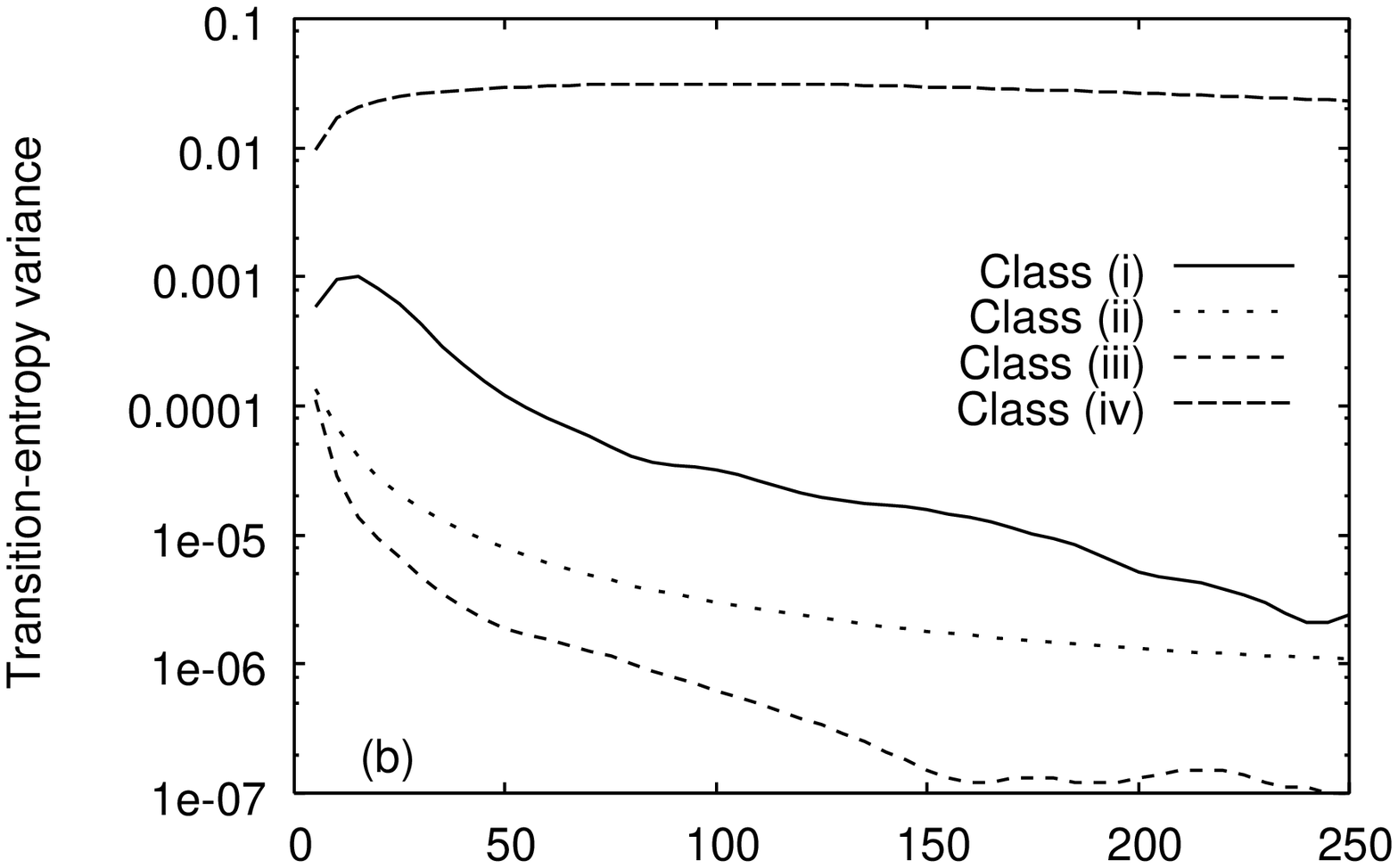}}\\
\scalebox{0.300}{\includegraphics{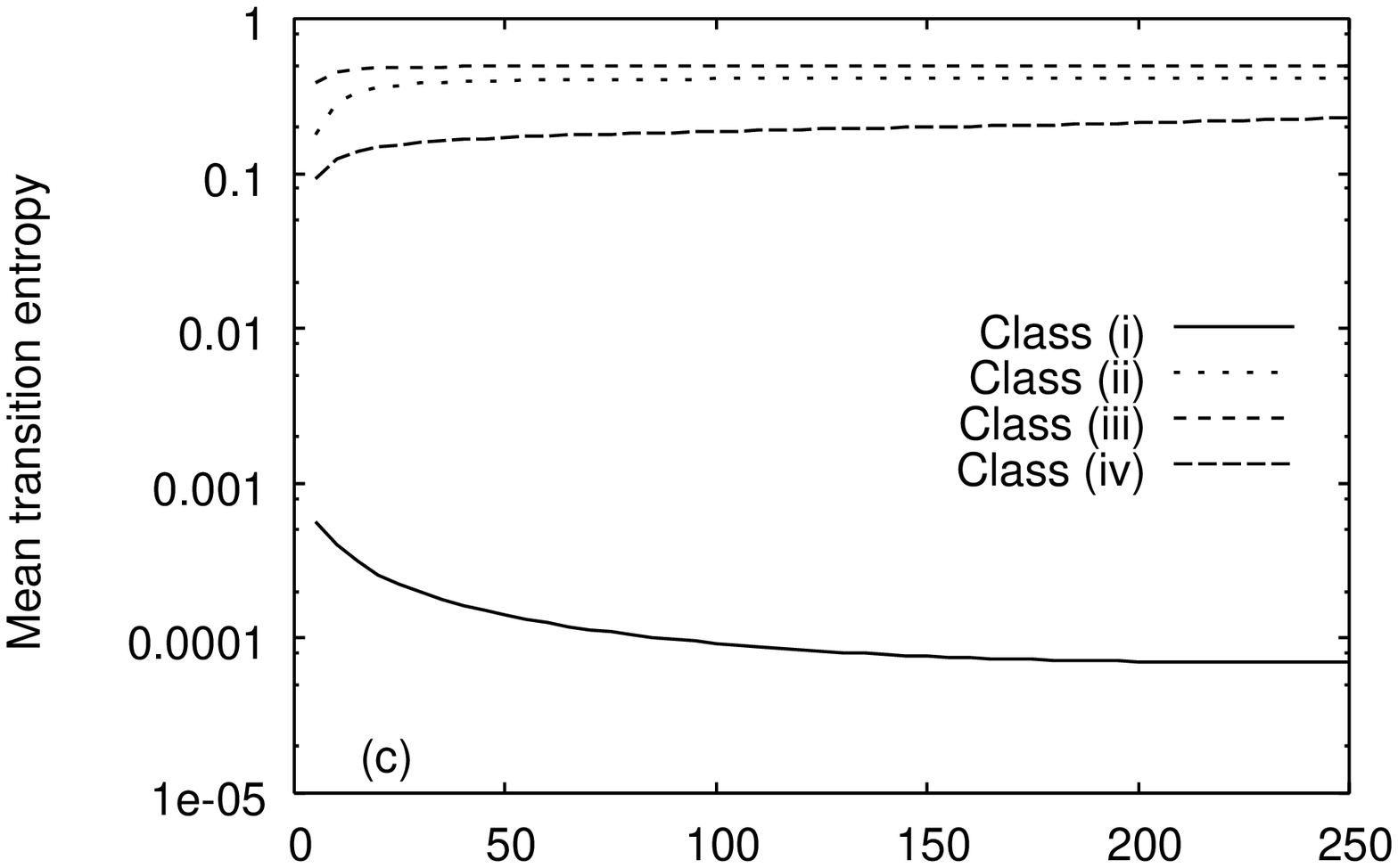}}&
\scalebox{0.300}{\includegraphics{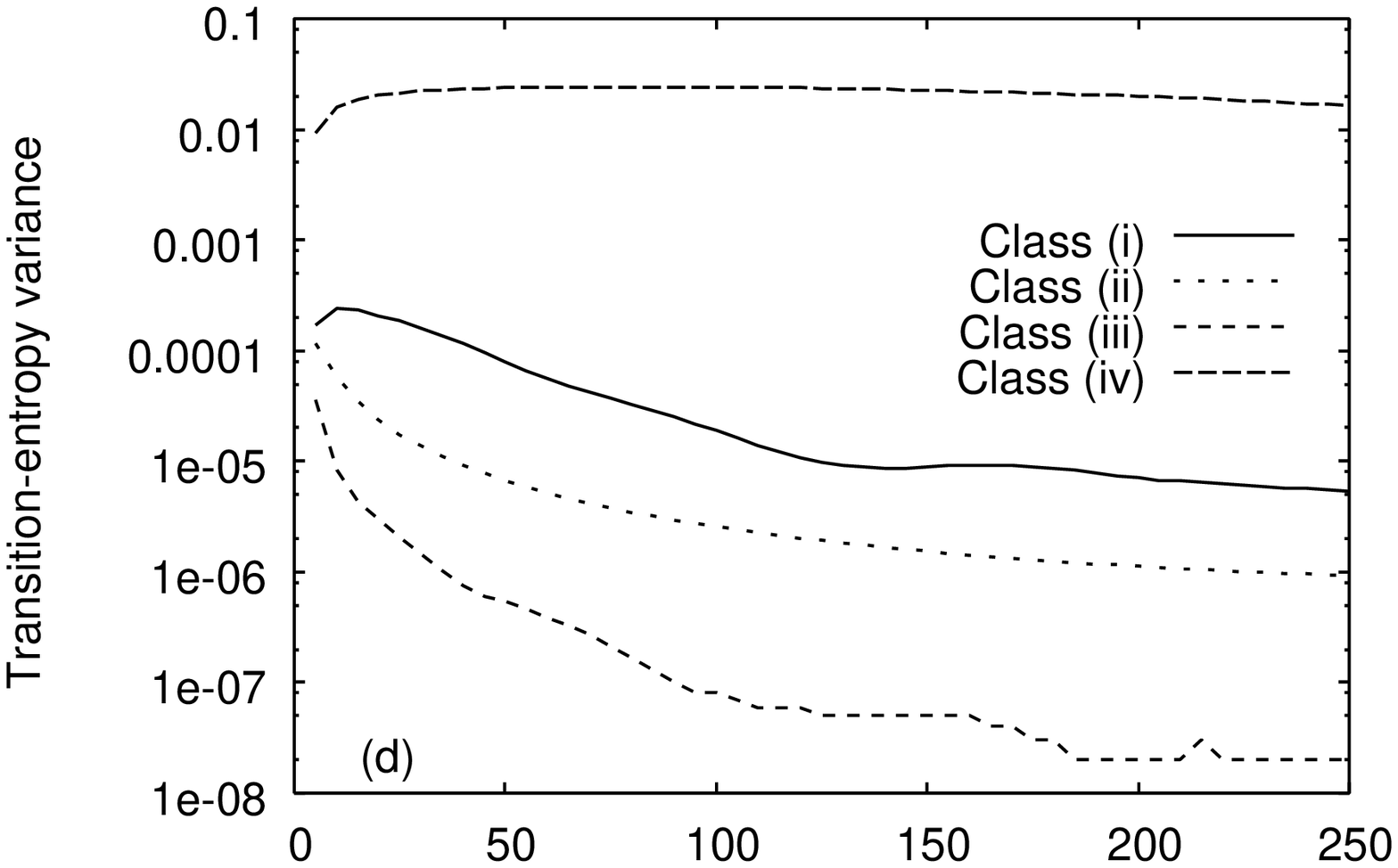}}\\
\scalebox{0.300}{\includegraphics{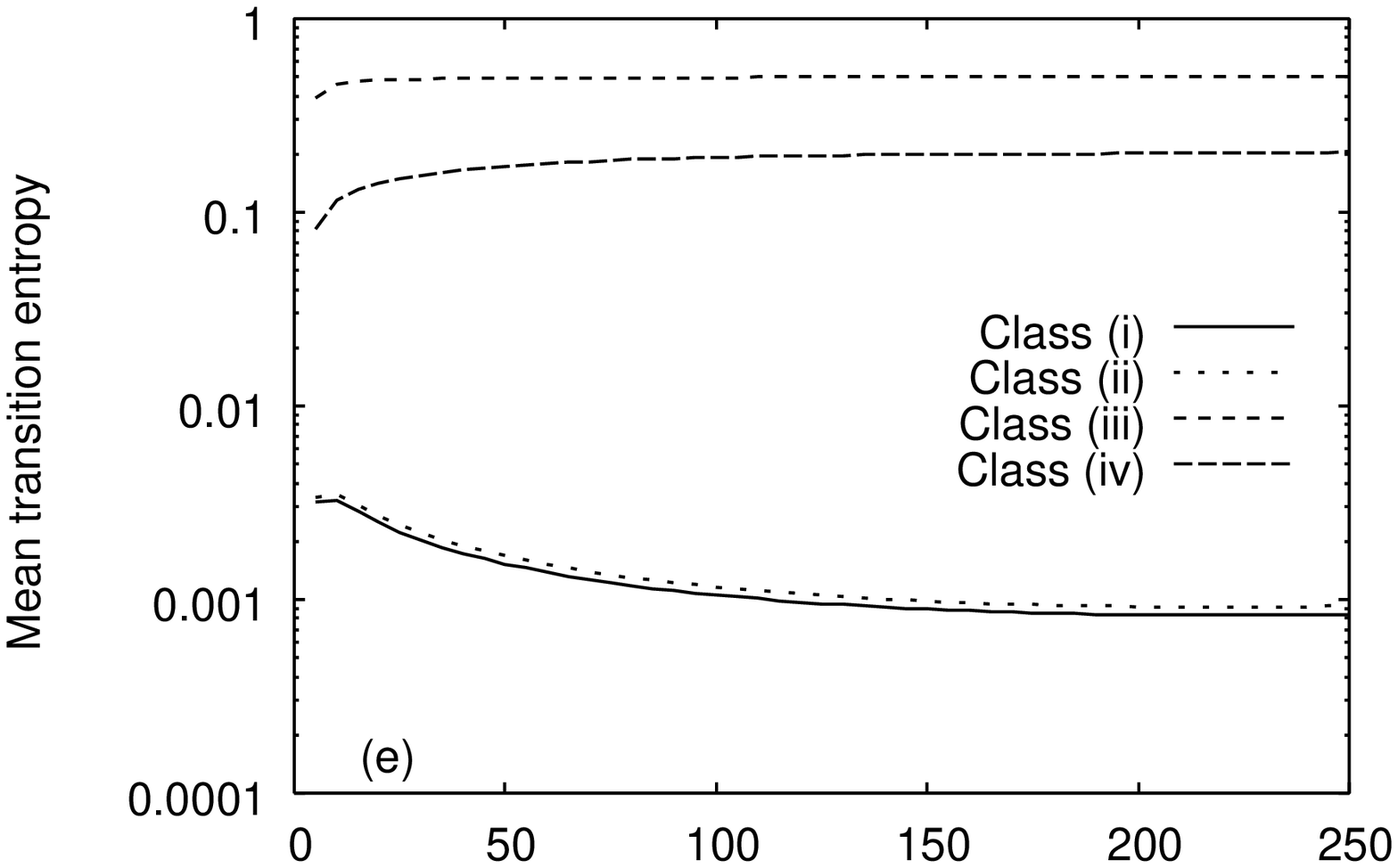}}&
\scalebox{0.300}{\includegraphics{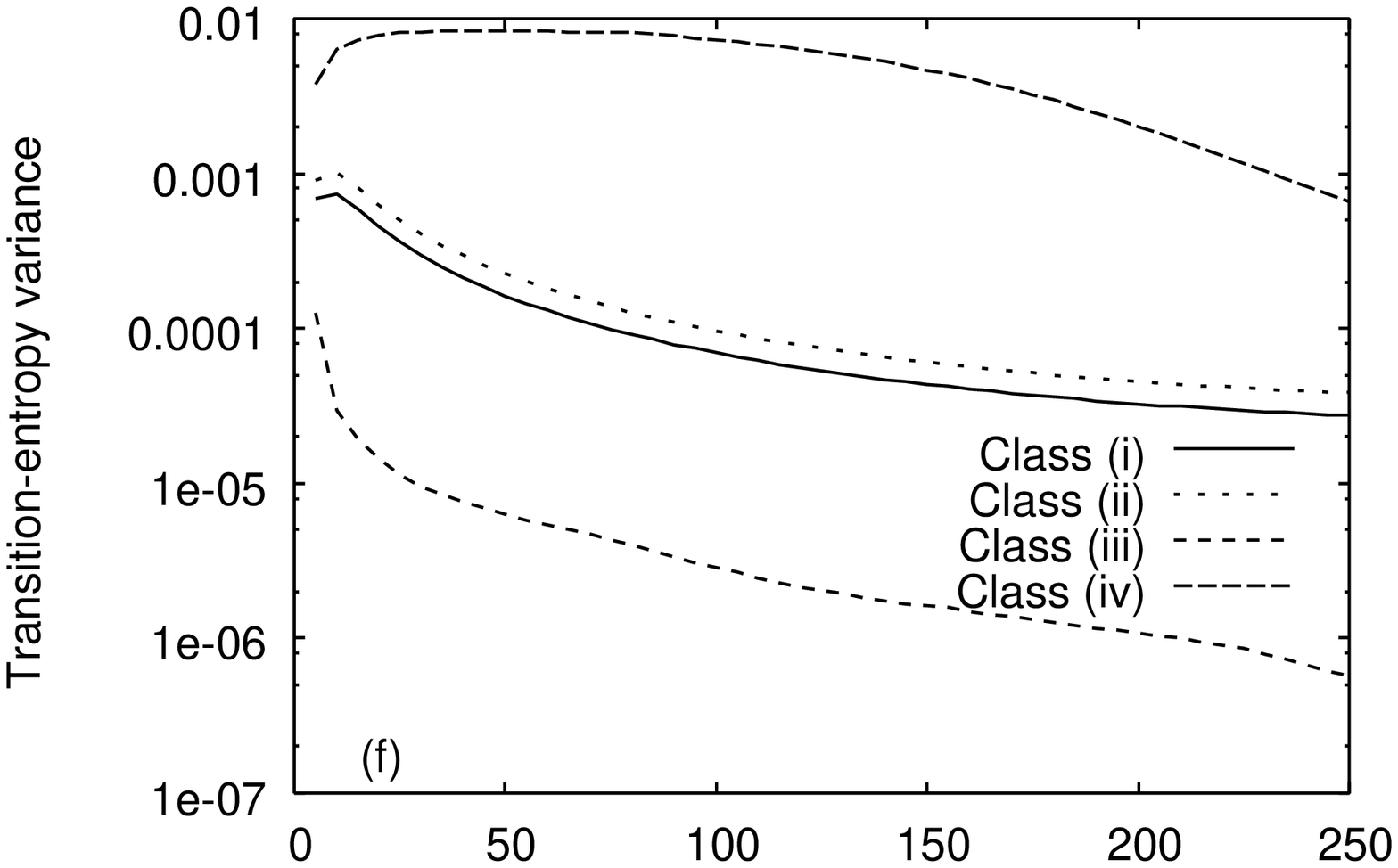}}\\
\scalebox{0.300}{\includegraphics{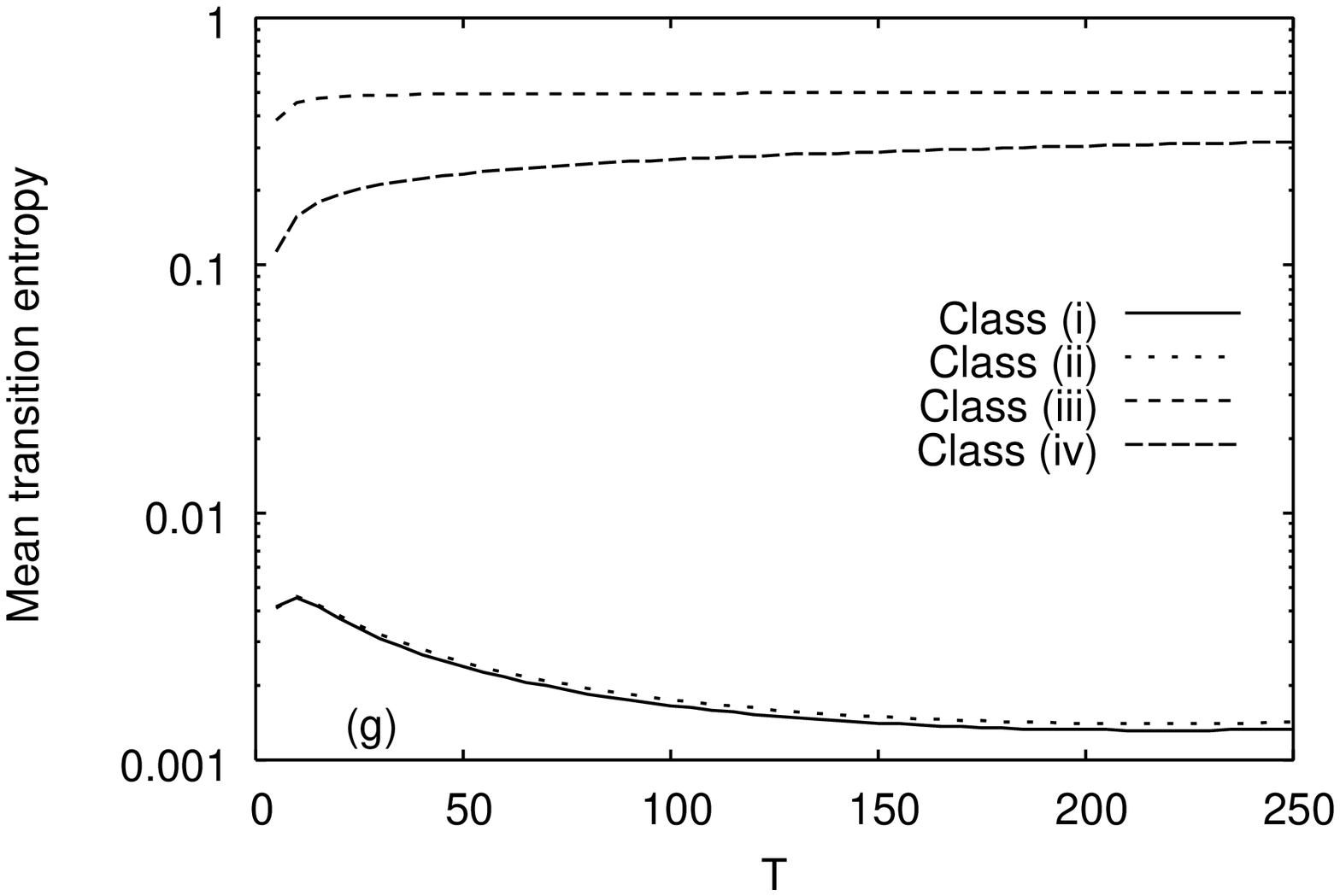}}&
\scalebox{0.300}{\includegraphics{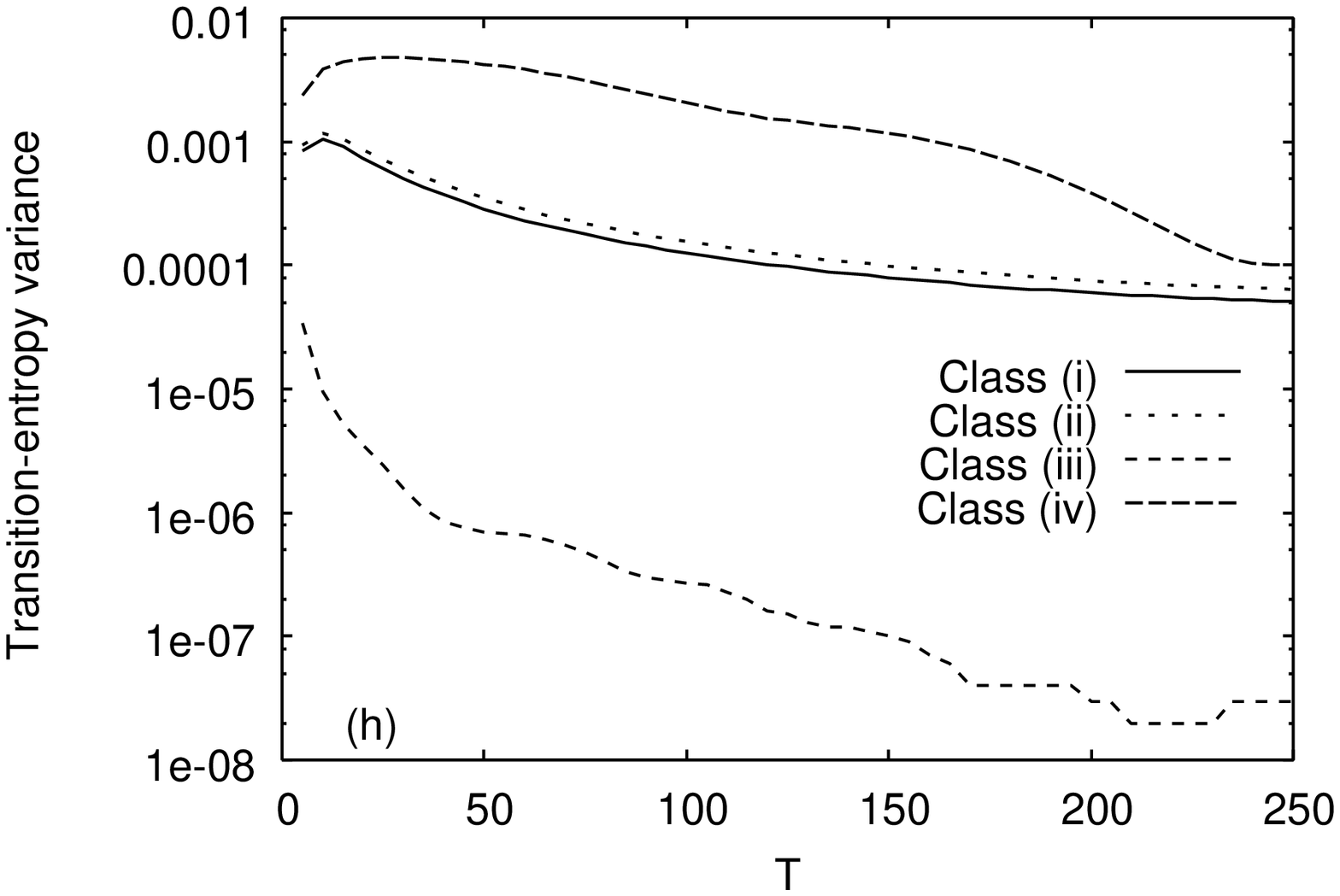}}
\end{tabular}
\caption{Mean ($\overline{T_f}$) and variance ($\sigma^2(T_f)$) of the
cell-centric transition entropy as a function of $T$ under four different update
rules, one from each of classes (i) through (iv), for $d=2$. Data are given for
the $(15\times 15)$-cell von Neumann case (a and b), the $(30\times 30)$-cell
von Neumann case (c and d), the $(15\times 15)$-cell Moore case (e and f), and
the $(30\times 30)$-cell Moore case (g and h). In all cases, $r_1=r_2=1$.}
\label{te-initiald2}
\end{figure}

Even though this first set of experiments is deprived of statistical
significance (based as it is on runs from a single initial configuration), it
provides an initial indication of the discriminatory capabilities of our
cell-centric heuristics. In fact, an examination of all mean- and variance-plot
pairs in Figures~\ref{ie-initiald1}--\ref{te-initiald2} reveals that, with a few
exceptions, classes (i)--(iv) can in the worst case be discriminated within
roughly one order of magnitude by either the mean or the variance of both the
cell-centric input and transition entropies for most values of $T$. For example,
comparing the plots in Figures~\ref{ie-initiald1}(a) and (b) indicates that the
mean cell-centric input entropy provides good discrimination among the four
classes, except between classes (iii) and (iv), which nonetheless can be told
apart easily by the variance of that entropy. The exceptions are the
two-dimensional cases with Moore neighborhoods, in which neither of the
cell-centric heuristics seems to be able to capture the distinction between
classes (i) and (ii).

We return to this discussion of our heuristics' discriminatory capabilities
shortly, after we have provided more significant data. But returning to the
original goal of this initial set of experiments, we see from the plots in
Figures~\ref{ie-initiald1}--\ref{te-initiald2} that several possibilities exist
for choosing a value for $T$. We note that, naturally, choosing as small a value
as possible has the advantage of alleviating the processing demands for
computing the entropy figures. With these observations in mind, our choice
hereafter is to use $T=25$.

\subsection{Experimental results}

Simulating a cellular automaton in parallel is essentially an exercise in
designing a simple synchronous distributed algorithm, in the sense described in
\cite{b96}, employing for synchronization the technique of
$\alpha$-synchronization of \cite{a85}. Within this general framework, several
proposals have been put forward (cf., e.g., \cite{b93,bh93,t00}).

Our parallel simulator is no exception and has been designed and implemented
within this same framework for one- and two-dimensional cellular automata. Each
simulation is initiated by partitioning the automaton into the $N$ available
processors. The hardware we have used in all the experiments described
henceforth has $N=8$. All processors have the capability of communicating
directly with all others. For $d=1$, the cells are partitioned equitably among
the processors in such a way that each processor receives a contiguous set of
cells to simulate; for $d=2$, the automaton is subdivided into rectangles of
contiguous cells by slicing it equitably along the dimension that has the least
number of cells.

Notice that the neighborhood relation among cells as given by the lattice that
underlies the cellular automaton automatically implies a neighborhood relation
among processors, too. Specifically, two processors are neighbors whenever at
least one cell that one of them lodges is a neighbor of a cell lodged by the
other. Obviously, some of the cells that a processor lodges are distinguished
in that their states are needed by the processor's neighbors; we refer to such
cells as frontier cells.

The simulation proper starts at each processor with the assignment of a
randomly chosen initial state to each of the cells it lodges and the sending of
the initial states of all frontier cells to the neighbor processors at which
they are needed. The processor then iterates as $t$ is incremented from $0$
through $t_+$: for each $t$, new states are computed for all the cells that the
processor lodges and so are the portions of (\ref{ccentropy}) and
(\ref{cctentropy}) corresponding to those of its cells that are observed,
provided $t\ge T-1$; then the new states of the processor's frontier cells are
sent where they are needed. At the end, each processor that lodges at least one
observed cell forwards its $2(t_+-T+2)$ entropy results to a previously
designated processor for computation of the two means and variances (viz.\ the
means $\overline{C_f}$ and $\overline{T_f}$ and the variances $\sigma^2(C_f)$
and $\sigma^2(T_f)$).

\subsubsection*{One-dimensional cellular automata}

Our setup for the one-dimensional experiments is based on either $r_1=2$ or
$r_1=3$. For $r_1=2$, the setup has $X_1=2000$, $t_+=500$, and $T=25$. The
overall number of cells to simulate is then $X_1+2r_1t_+=4000$, so the number of
observed cells constitutes half of the total. For $r_1=3$, our setup has
$X_1=2400$, $t_+=400$, and $T=25$. In this case, the total number of cells in
the simulation is $4800$, and once again the observed cells account for half the
total number of cells.

\begin{figure}
\centering
\begin{tabular}{c@{\hspace{0.00in}}c}
\resizebox{!}{1.75in}{\includegraphics{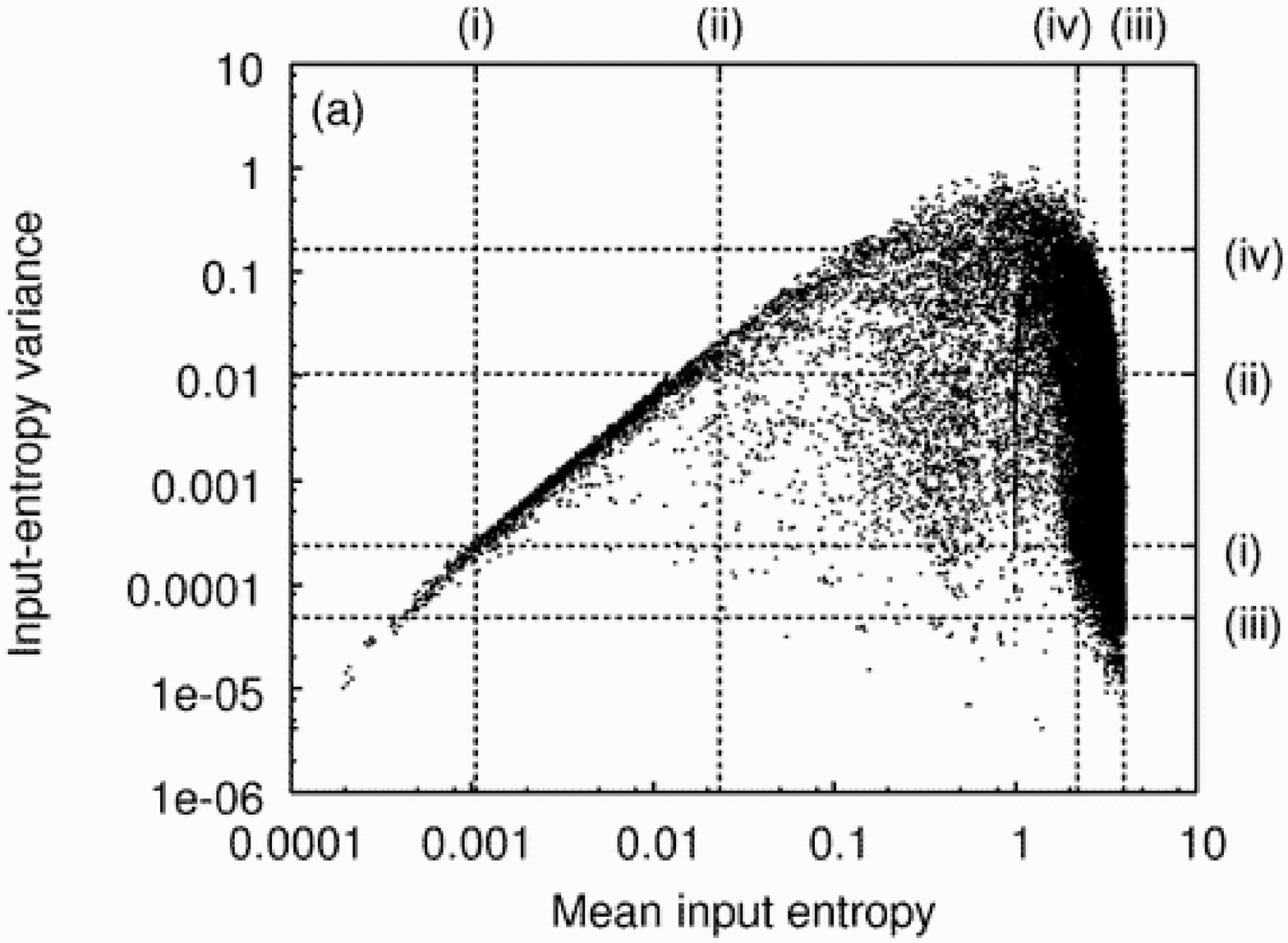}}&
\resizebox{!}{1.75in}{\includegraphics{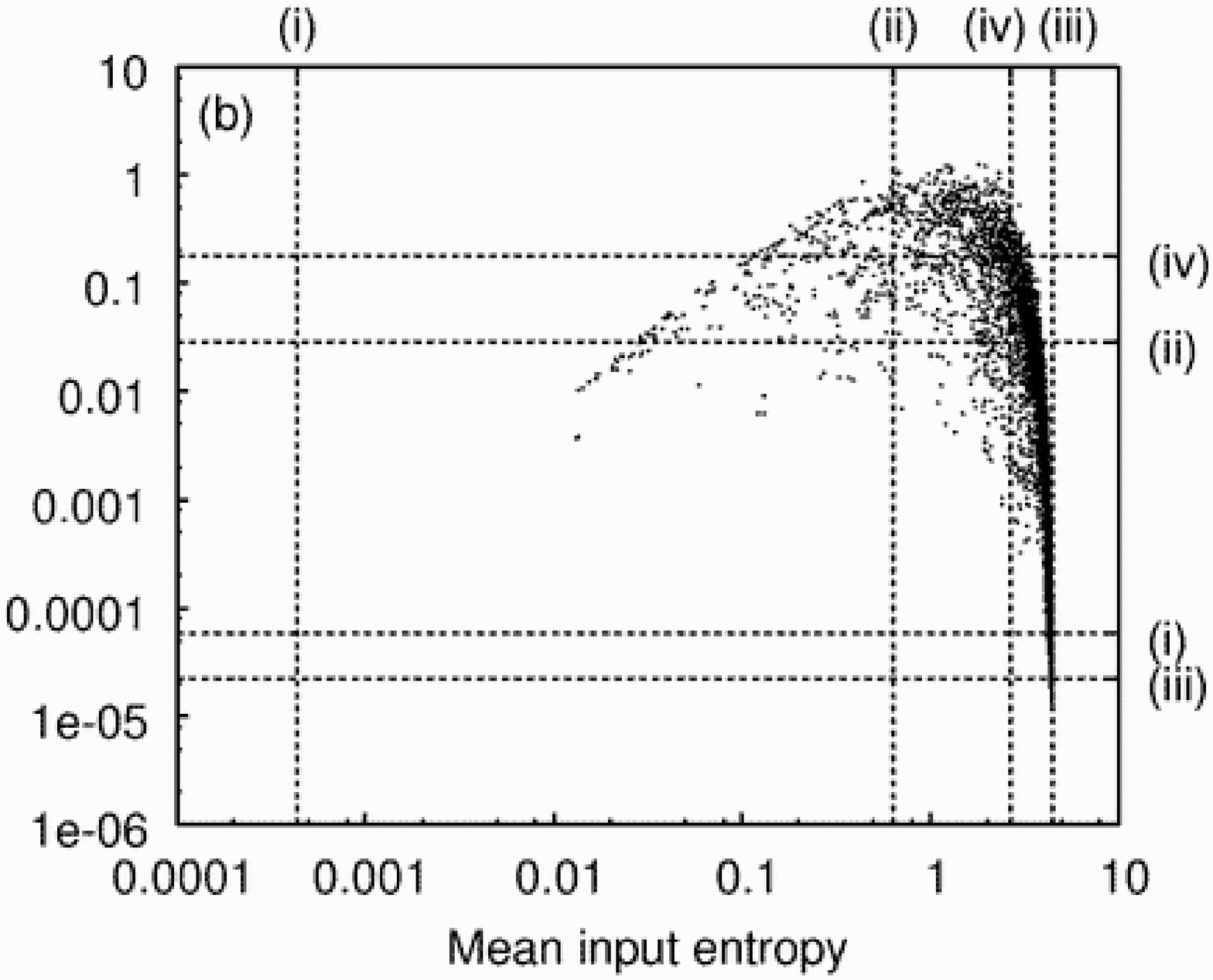}}\\
\resizebox{!}{1.75in}{\includegraphics{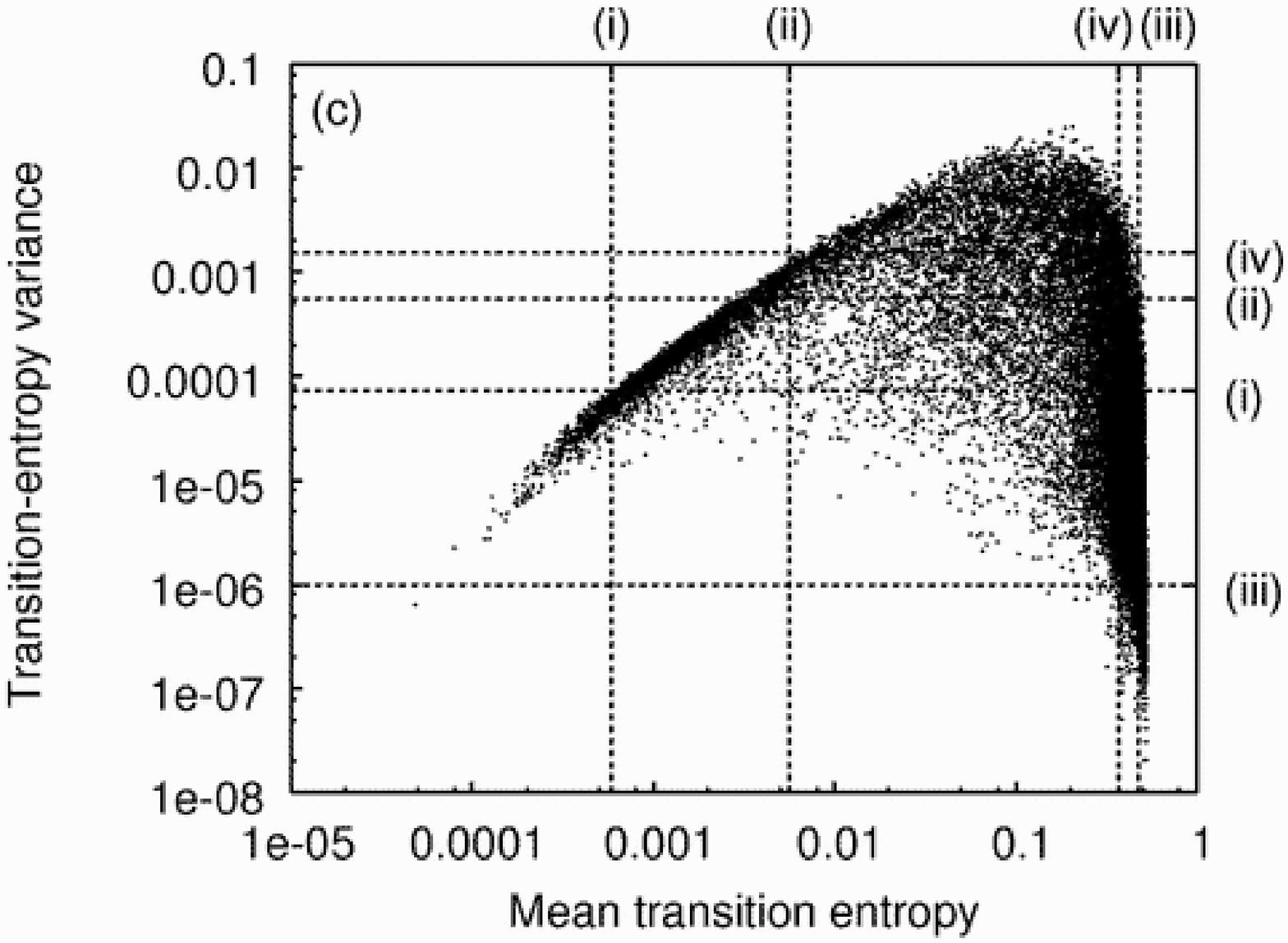}}&
\resizebox{!}{1.75in}{\includegraphics{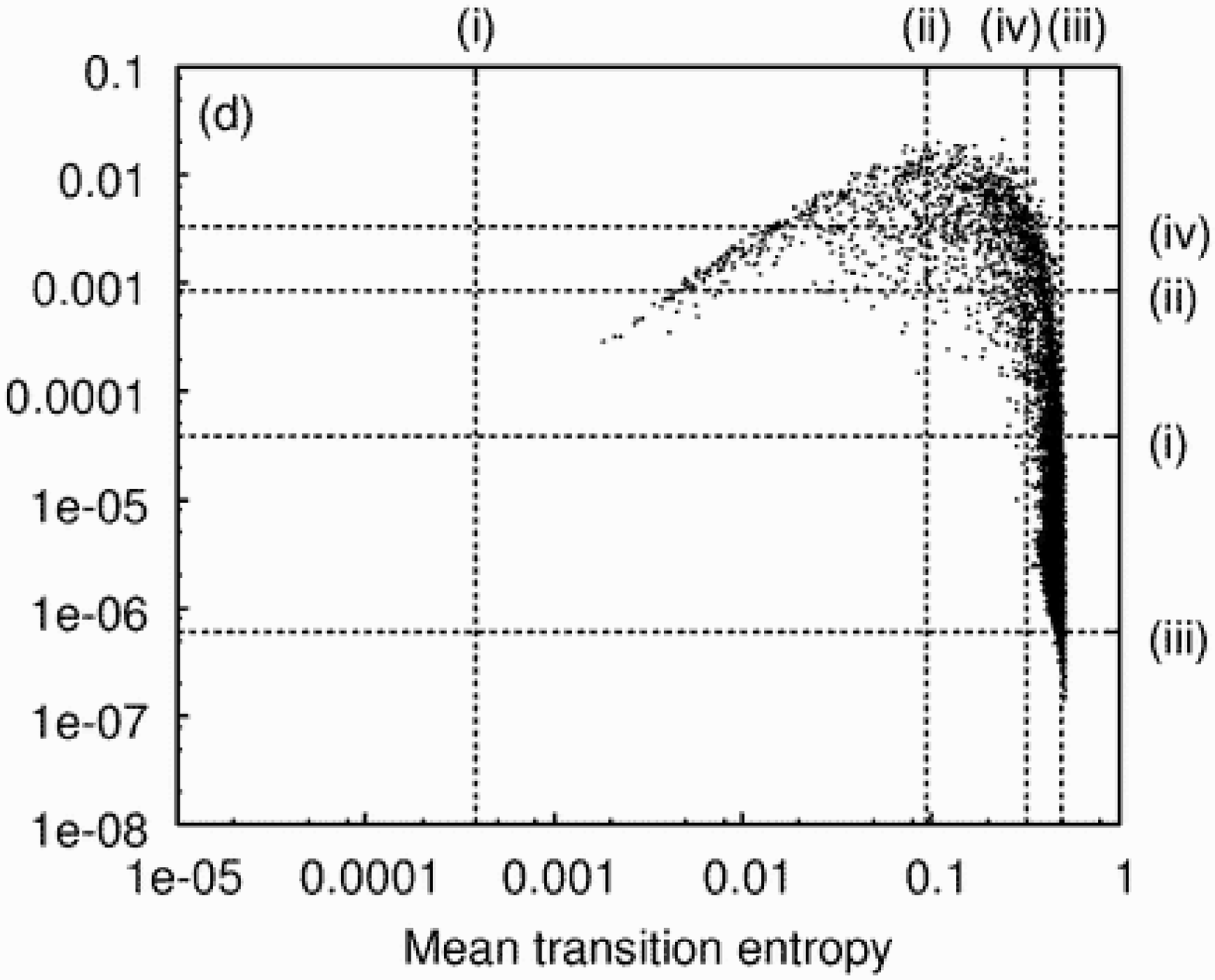}}
\end{tabular}
\caption{Occurrence of mean-variance pairs within the general experimental setup
for one-dimensional cellular automata. Data are given for the cell-centric input
entropy with $r_1=2$ (a) and $r_1=3$ (b) as plots of $\sigma^2(C_f)$ against
$\overline{C_f}$, and also for the cell-centric transition entropy with $r_1=2$
(c) and $r_1=3$ (d) as plots of $\sigma^2(T_f)$ against $\overline{T_f}$. Each
plot contains $50000$ points, each point corresponding to a randomly chosen
update rule and to an average over $5$ randomly chosen initial configurations.
The one-dimensional update rules of Table~\ref{updtrules} are also shown within
the same experimental setup, but not as points: instead, they are singled out
with an indication at their coordinates of which of classes (i)--(iv) they
belong to.}
\label{map-1d}
\vspace{0.45in}
\end{figure}

In the one-dimensional case, the number of distinct update rules is given by
$2^{2^{1+2r_1}}$, that is, $2^{32}$ for $r_1=2$ and $2^{128}$ for $r_1=3$. Our
results are based on $50000$ update rules randomly chosen out of those and are
shown in Figure~\ref{map-1d} as plots of the variances $\sigma^2(C_f)$ and
$\sigma^2(T_f)$ against the means $\overline{C_f}$ and $\overline{T_f}$,
respectively. The points that correspond to the one-dimensional update rules of
Table~\ref{updtrules} are not shown explicitly but are singled out by
indications, at their coordinates, of the classes to which the update rules
belong.

One crucial information that has been left out of the plots in
Figure~\ref{map-1d} in order to avoid any further cluttering is the density of
points at any particular mean-variance region. We provide some of this
information next. First we choose, for each of the plots, a value for the mean
entropy that separates the update rules labeled (i) or (ii) from those labeled
(iii) or (iv). In Figures~\ref{map-1d}(a) and (b), which refer to the
cell-centric input entropy, this mean entropy can be taken to be $1$; in
Figures~\ref{map-1d}(c) and (d), it can be taken to be $0.1$. Selecting this
value partitions the plot into two regions and for each one we now select an
entropy variance that can be used to separate the update rules labeled (i) and
(ii) on the left, and another that can likewise be used for those labeled (iii)
and (iv). In parts (a) and (b) of the figure, our choices are $0.001$ and $0.1$,
respectively on the left and right sides; in parts (c) and (d) the corresponding
values are $0.0001$ and $0.001$. At the end, in each plot we are left with a
partition into four regions, each containing exactly one of the update rules
labeled (i)--(iv).

We may then provide the missing information. In part (a), $2.30\%$ of the update
rules are inside the (i) region, $8.10\%$ in the (ii) region, $86,96\%$ in the
(iii) region, and $2.64\%$ in the (iv) region. Part (b) contains no update rules
inside the (i) region, $0.80\%$ of the update rules in region (ii), $97.85\%$ in
(iii), and $1.35\%$ in (iv). In part (c) we have the figures $2.54\%$, $9.30\%$,
$83.47\%$, and $4.69\%$. In part (d) we once again have no update rules inside
the (i) region and the remaining figures are $0.80\%$, $97.59\%$, and $1.61\%$.
The well-known preponderance of class-(iii) update rules, as well as the
relative rarity of class-(iv) update rules, particularly as $r_1$ is increased
from parts (a) and (c) to parts (b) and (d), are then confirmed.

\subsubsection*{Two-dimensional cellular automata}

For the two-dimensional experiments we use $r_1=r_2=1$ throughout. Regardless of
the neighborhood type (von Neumann or Moore), our experiments' setup has
$X_1=X_2=100$, $t_+=50$, and $T=25$. The total number of cells to be simulated
is therefore $(X_1+2r_1t_+)(X_2+2r_2t_+)=40000$, so the number of observed cells
is one quarter of the total number.

\begin{figure}
\centering
\begin{tabular}{c@{\hspace{0.00in}}c}
\resizebox{!}{1.75in}{\includegraphics{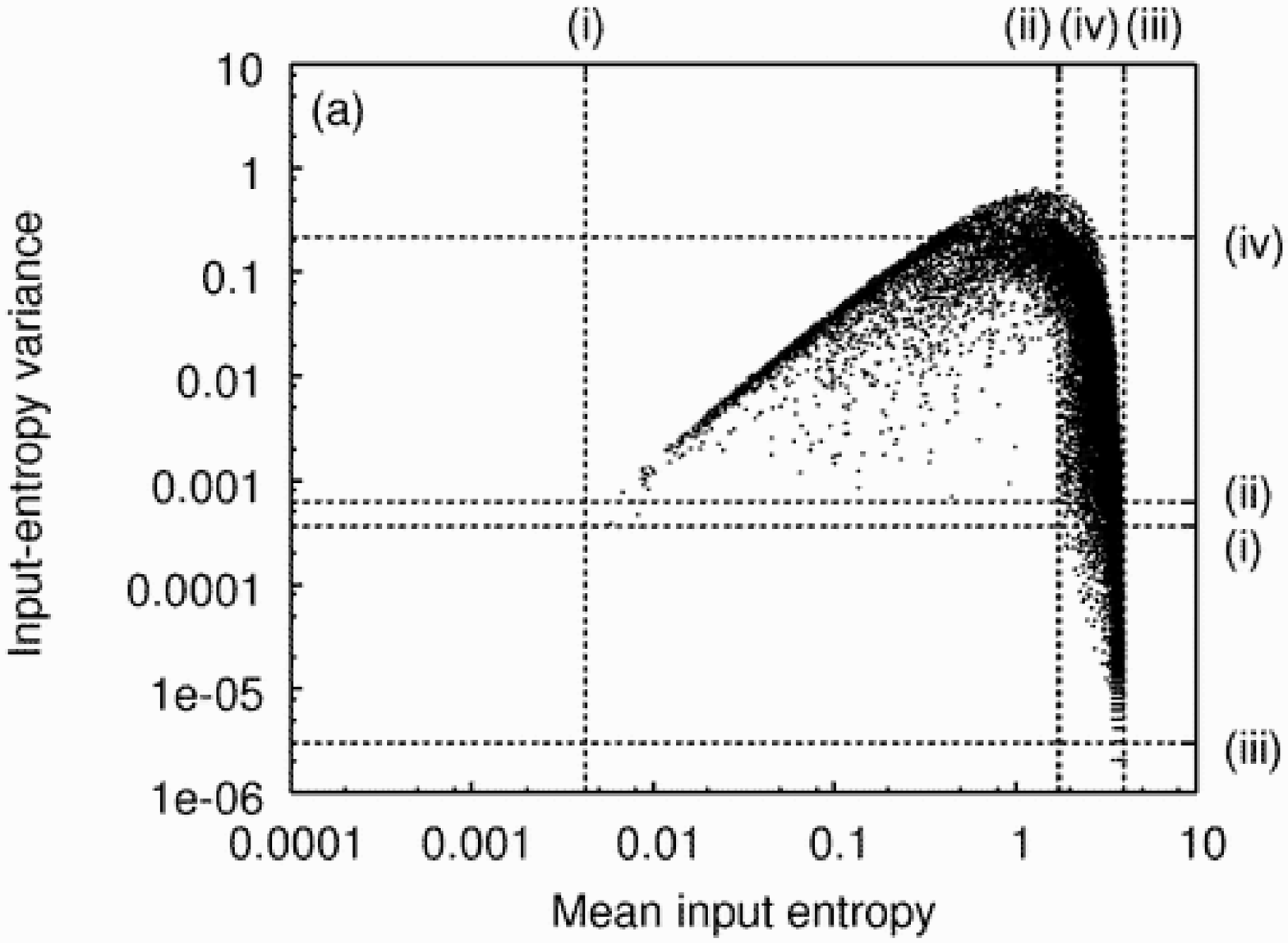}}&
\resizebox{!}{1.75in}{\includegraphics{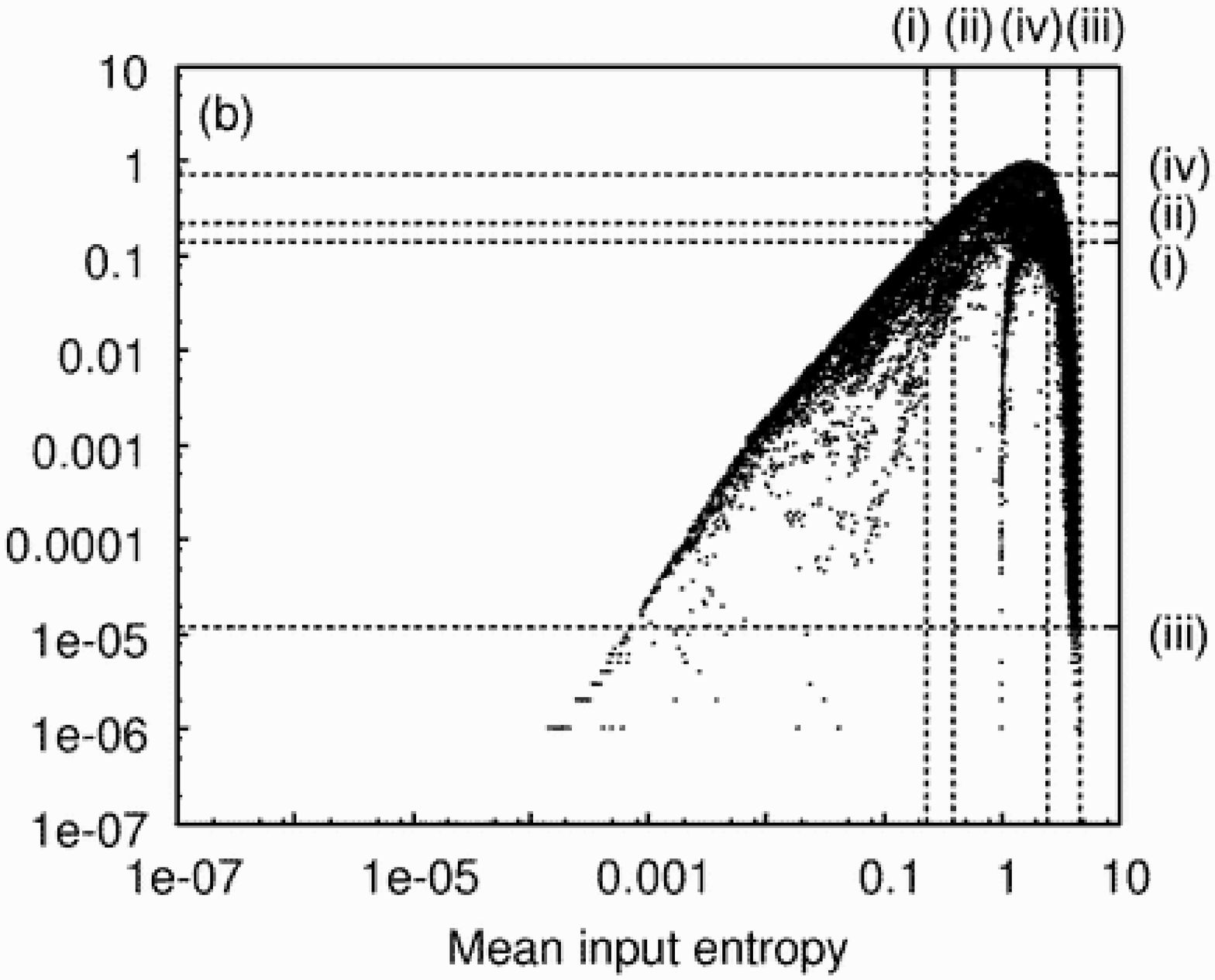}}\\
\resizebox{!}{1.75in}{\includegraphics{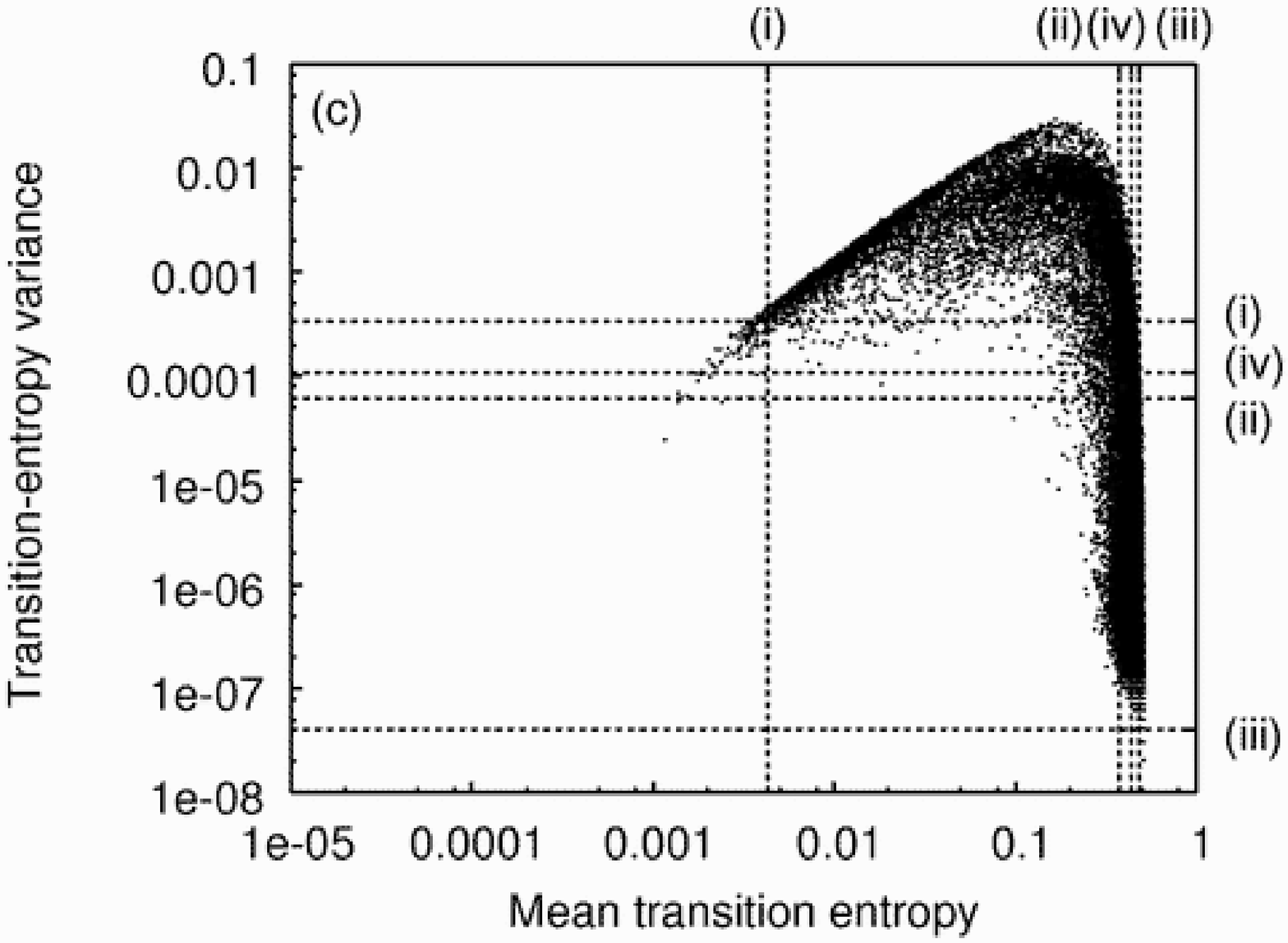}}&
\resizebox{!}{1.75in}{\includegraphics{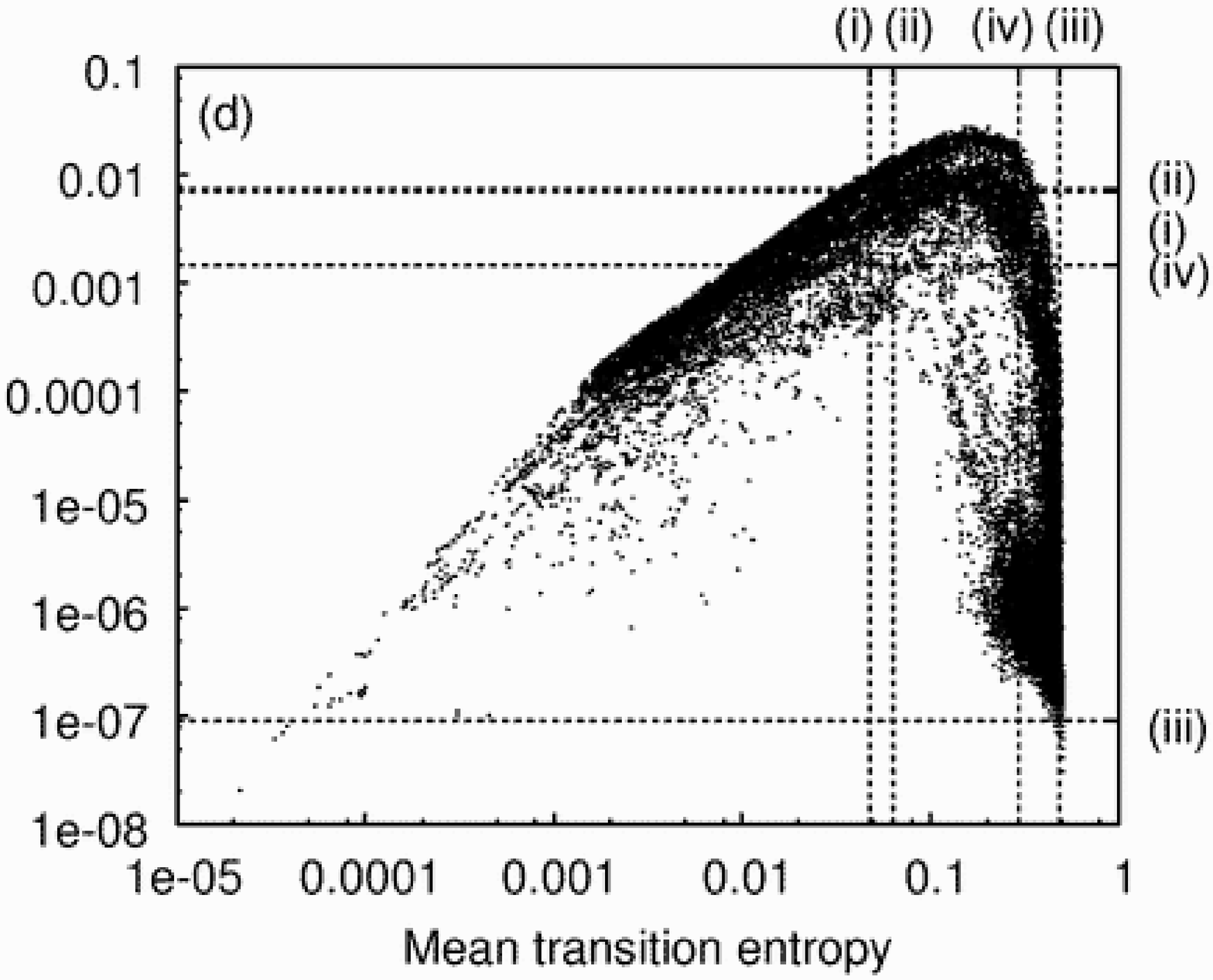}}
\end{tabular}
\caption{Occurrence of mean-variance pairs within the general experimental setup
for two-dimensional cellular automata. Data are given for the cell-centric input
entropy with the von Neumann (a) and the Moore (b) neighborhoods as plots of
$\sigma^2(C_f)$ against $\overline{C_f}$, and also for the cell-centric
transition entropy with the von Neumann (c) and the Moore (d) neighborhoods as
plots of $\sigma^2(T_f)$ against $\overline{T_f}$. Each plot contains $50000$
points, each point corresponding to a randomly chosen update rule and to an
average over $5$ randomly chosen initial configurations. The two-dimensional
update rules of Table~\ref{updtrules} are also shown within the same
experimental setup, but not as points: instead, they are singled out with an
indication at their coordinates of which of classes (i)--(iv) they belong to.}
\label{map-2d}
\vspace{0.45in}
\end{figure}

Similarly to the one-dimensional case, in the two-dimensional case with von
Neumann neighborhoods there are $2^{2^{1+2(r_1+r_2)}}=2^{32}$ distinct update
rules. With Moore neighborhoods, and considering only outer-totalistic update
rules, the number of distinct update rules is $2^{2(1+2r_1)(1+2r_2)}=2^{18}$.
Once again, in both cases our results are based on $50000$ update rules randomly
chosen out of the corresponding sets.\footnote{Note that restricting
Moore-neighborhood update rules to lie within the set of outer-totalistic update
rules is a means to ensure that these $50000$ samples have some statistical
representativeness. In the absence of this restriction, the number of possible
update rules becomes $2^{2^{(1+2r_1)(1+2r_2)}}$. This number, with values for
$r_1$ and $r_2$ as we have adopted, is $2^{512}$, which is larger by more than a
hundred orders of magnitude than the number of distinct update rules in any of
our other experiments.} They are shown in the plots of Figure~\ref{map-2d} in
the same style as Figure~\ref{map-1d}. As in the case of that figure, the
marginal indications (i)--(iv) give the coordinates at which the points
corresponding to the two-dimensional update rules of Table~\ref{updtrules} would
be found, had they been plotted explicitly. Parts (a) and (c) of the figure
refer to a von Neumann neighborhood, parts (b) and (d) to a Moore neighborhood.

Once again additional information regarding the density of points in the four
plots must be given on the side. Following the same methodological steps as for
the one-dimensional cases, first we select a mean-entropy value for each plot to
separate the update rules labeled (i) or (ii) from those labeled (iii) or (iv),
and then we select an entropy-variance value to separate each pair of labeled
update rules. As the four plots in Figure~\ref{map-2d} indicate, this may prove
a harder task than in the one-dimensional cases, since it is now common to find
two or more labels clustered together along one of the axes.

Let us begin with part (a), which refers to the cell-centric input entropy under
a von Neumann neighborhood. If we select $1.73$ as the first separator, and then
select $0.0005$ on the left and $0.1$ on the right, then we are left with no
update rules lying within class-(i) region, while $7.92\%$ of the update rules
are in region (ii), $90.47\%$ in region (iii), and $1.61\%$ in region (iv).
Moving to part (b) we select the separators $1$, $0.007$, and $0.1$, which
yields the percentages $4.42\%$, $12.08\%$, $77.19\%$, and $6.31\%$,
respectively for regions (i)--(iv), relative to the cell-centric input entropy
under a Moore neighborhood. The remaining two plots, in parts (c) and (d), are
both relative to the cell-centric transition entropy, respectively under a von
Neumann and a Moore neighborhood. Selecting the separators $0.4$, $0.0002$, and
$0.0001$ in the former case yields $6.67\%$, $16.07\%$, $73.27\%$, and $3.98\%$.
In the latter case, we select $0.1$, $0.0072$, and $0.0001$, obtaining
$19.02\%$, $1.82\%$, $70.74\%$, and $8.42\%$. As in the one-dimensional case,
indications are once again clear concerning the relative predominance and rarity
of classes (iii) and (iv), respectively.

\section{Discussion}\label{disc}

The data shown in Figures~\ref{map-1d} and \ref{map-2d}, respectively for one-
and two-dimensional cellular automata, tend all to exhibit the following
behavior regarding classes (i)--(iv). When plotted on doubly-logarithmic scales,
they appear clustered roughly as a boomerang whose traversal from the low-mean,
low-variance tip leads us through class-(i) update rules, then class (ii), then
class (iv) near the middle bend, and finally class (iii) past the bend. Entropy
means increase at varying rates along the traversal, while variances increase at
first but fall down again after the middle bend in the cluster's shape.

The one-dimensional cases, depicted in Figure~\ref{map-1d}, indicate
unequivocally that the eight one-dimensional update rules of
Table~\ref{updtrules} can be told apart by at least one full order of magnitude
of the mean entropy or the entropy variance, often both, regardless of which
cell-centric entropy is being used. While the same holds unchanged for the case
shown in Figure~\ref{map-2d}(a), which refers to the cell-centric input entropy
under a von Neumann neighborhood, the remaining three parts, (b)--(d), must be
examined in more detail. The case of part (b), in which the input entropy is
still the one in use but now under a Moore neighborhood, allows proper
separation between classes (iii) and (iv), but apparently leave classes (i) and
(ii) mixed up together. The picture as we move to part (c), which corresponds to
the cell-centric transition entropy under a von Neumann neighborhood, is once
again subject to mix-ups, this time between classes (ii) and (iv). The final
case is that of part (d), corresponding to the transition entropy and to a Moore
neighborhood. In this case, as in the case of part (b), classes (i) and (ii) are
hard to tell apart.

We envisage two major trends underlying these class mixtures in the
mean-variance plots for the two-dimensional cases. The first one has to do with
part (c) of Figure~\ref{map-2d}, where a mix-up of classes (ii) and (iv) under
a von Neumann neighborhood turns up when the transition entropy is used. What
may be happening is that the relatively low value of $50$ chosen for $t_+$ (for
the strictly practical reason of keeping our run times within reasonable bounds
while the plots of the figure were produced) is insufficient for the automaton
to settle into a more typical class-(ii) behavior (and hence mean and variance
values of the transition entropy that are commensurate with that class). This is
somewhat supported by an examination of Figure~\ref{patterns-inftyd2}, but
clearly calls for additional investigation (more on this in
Section~\ref{concl}).

The second trend concerns parts (b) and (d) of Figure~\ref{map-2d} mainly, and
thus has to do with the mix-up between classes (i) and (ii) when a Moore
neighborhood is in use, but may also be related to the case of part (c) we just
discussed. Our expectation that the Wolfram classification carries on naturally
to the two-dimensional case comes from Wolfram's own investigations on
two-dimensional cellular automata \cite{pw85}, but this has been challenged on
the grounds that such a classification scheme fails to recognize the real sign
of complex behavior in outer-totalistic, two-dimensional update rules, which is
the presence of the so-called gliders, that is, the structures that are seen to
``glide'' across the two-dimensional lattice as time elapses
\cite{e02,eppstein-url}.\footnote{More generally, there have been arguments
calling for classification schemes that take the particular application area
under study into consideration more seriously (cf., e.g., \cite{mhk97}).} If
this is the case, then what is out of place is not the mix-up of classes (i) and
(ii), but rather the separation between the two classes, since none of the two
exhibits gliders and should therefore be coalesced together into one single
class.

But beyond these slight conflicts, and whichever of the competing trends may win
at the end, we perceive our experiments' outcomes as expressed in
Figures~\ref{map-1d} and \ref{map-2d} as laying out an overall methodology for
the classification of cellular-automaton update rules, one that in many senses
confirms the initial conclusions of \cite{w99}. First and foremost is an
examination of the mean entropies vis-\`{a}-vis the bounds made available in
(\ref{ccebound}) through (\ref{ccebound2}). For the one-dimensional cases,
(\ref{ccebound}) predicts that no mean cell-centric input entropy goes beyond
$1+2r_1$, while predicting $1+2(r_1+r_2)$ as the maximum for the von Neumann
two-dimensional case. Thus, the upper bound turns out to be $5$ in the case of
part (a) of Figure~\ref{map-1d}, $7$ for part (b), and $5$ for
Figure~\ref{map-2d}(a). The tighter number given by (\ref{ccebound2}) for
outer-totalistic, Moore-neighborhood update rules yields approximately $6.87$
for Figure~\ref{map-2d}(b). While the four plots respect the corresponding
upper bounds (this may not be immediate from the figures, owing to the
logarithmic scale, but we know it from our files and refrain from presenting
further plots), none of the $50000$ randomly chosen update rules comes very near
its bound. Perhaps this is due to the difficulty of sampling an update rule
whose mean input entropy comes sufficiently near the bound, but the fact remains
that comparing an update-rule's mean input entropy to its known upper bound may
be of little help towards classifying the update rule.

The case of parts (c) and (d) of both Figures~\ref{map-1d} and \ref{map-2d},
being as they are based on the cell-centric transition entropy, is different. In
this case, meeting the upper bound of approximately $0.53$ given by
(\ref{cctebound}) does not seem to depend on serendipitously finding any
particular update rule. In fact, update rules whose mean transition entropy
approach the bound closely occur frequently, as once again can be seen in the
figures. When using the transition entropy, then, a useful first step is to
compare the update rule's mean entropy with this bound: if close enough, almost
surely the update rule is a class-(iii) one.

Beyond this initial test against known upper bounds, what remains of the
aforementioned overall methodology is essentially a
cladistics-like\footnote{Here we allude to the method known as cladistics for
hypothesizing relationships among (extant or extinct) organisms. Beyond its core
assumptions, the method in essence relies on examining several characters of the
organisms and employing them for grouping the organisms into the desired taxa
\cite{cladistics-url}.} buildup of relationships among update rules given their
cell-centric (input and/or transition) entropy means and variances. The crux
here is that classification is the product of comparison, thence the fundamental
importance of update rules such as the ones in Table~\ref{updtrules}, for which
we are capable of providing a desired classification a priori so they can
function as seeds in the larger classification process.

\section{Conclusions}\label{concl}

We have in this paper addressed the automatic classification of the update rules
of cellular automata. Our departing point has been the notion of input entropy,
on which we built by the introduction of two novel entropy measures, both
inspired by, and targeted at, the simulation of cellular automata by
message-passing parallel machines. Our two new measures are the cell-centric
input entropy and the cell-centric transition entropy. For both of them we
provided extensive experimental results on both one- and two-dimensional
cellular automata. Within our assumed classification context, that of Wolfram's
four-class scheme, these results demonstrated that the two new measures provide
satisfactory discriminatory capabilities in the one-dimensional case, while in
the two-dimensional case it is also a good discriminator but in addition
helps support other authors' suggestions that a better classification scheme may
be needed.

Our experimental results were the product of a parallel implementation of a
simulator coupled with a module for calculating the two cell-centric entropies.
We finalize by commenting on some performance-related aspects of this simulator.
The results presented in Section~\ref{exp} were obtained on an eight-computer
cluster, each based on an Intel Pentium 4 processor running at 1.8 GHz and
having $1$ gigabyte of memory. The eight computers are fully interconnected by a
gigabit-ethernet switch. On this cluster, each of the eight test suites of
Section~\ref{exp}, comprising $5$ independent runs for each of $50000$ update
rules, requires somewhere from three to six days to complete, depending on which
of the four update-rule categories (one-dimensional with two possible radii, von
Neumann two-dimensional, Moore two-dimensional) and which of the cell-centric
measures (input or transition entropy) are being used.

The fact that we are simulating infinite cellular-automata, as explained right
at the beginning of Section~\ref{exp}, is naturally the source of considerable
load imbalance among the processors. We have paid no heed to this issue, but
clearly it has to be reckoned with by anyone undertaking the parallel
simulation of large-scale cellular automata if the effects of infinite
boundaries are to be taken into account. There are two kinds of load imbalance
to be considered. First is the fact that only those processors that lodge some
of the observed cells do actually perform entropy-related calculations; among
these, those that lodge more of those cells are more loaded by that kind of
computation. Secondly, cells that are not observed but do nonetheless
participate in the simulation for the sole sake of providing the illusion of an
infinite cellular automaton do not have to be simulated for all the $t_+$ steps;
instead, as time elapses progressively less of such cells need to be simulated.
Once these two types of load imbalance are taken into account, there are all
sorts of policies that can be adopted to re-balance the computational load
among the processors. We dwell on the issue no further in this paper, but it is
clearly important and should be considered upon embarking in a more
performance-aware implementation.

Another important aspect that ultimately is closely related to these performance
issues is whether the need really exists to undertake the simulation of cellular
automata with all the extra load for providing the illusion of infinity. While
unquestionably this seems the right way to approach the simulation when a new
classification scheme is first being tested, perhaps once it is established the
infinity requirement may be dropped and cylindrical boundary conditions adopted
instead. We have performed a few experiments with this trade-off in mind; their
outcomes are shown in Table~\ref{cylindrical}. Examining the table carefully
reveals clearly that both of our cell-centric heuristics retain the same
discriminatory capabilities we found them to possess in Section~\ref{exp}, even
though occasionally the relative positioning of the classes with respect to the
mean or variance of some entropy may not be the same, possibly due to the
different numbers of cells in the two sets of experiments. In fact, a quick
examination of Figures~\ref{patterns-cyld1} and \ref{patterns-cyld2}, which
depict the spatiotemporal patterns of some cellular automata with cylindrical
boundaries, reveals the same features we have come to associate with classes
(i)--(iv), despite the artificial periodicity that appears in some cases as a
result of assuming finite boundaries. But, as demonstrated by the results in
Table~\ref{cylindrical}, such periodicity appears to have no noticeable effect
on our cell-centric entropies. One immediate consequence of this is that
considerably larger cellular automata can now be simulated with the same overall
processing effort, and also that the sources of load imbalance we discussed
earlier become moot. Likewise, the simulation of two-dimensional cellular
automata for significantly larger values of $t_+$ becomes more viable, which
perhaps may lead to a clarification of the mix-up between classes (ii) and (iv)
under a von Neumann neighborhood alluded to in Section~\ref{disc}.

\begin{table}
\centering
\caption{Means and variances averaged over $5$ runs from randomly chosen initial
configurations with cylindrical boundaries for $t_+=500$ and $T=25$. Experiment
codes are as follows.
I: $d=1$, $150$ cells, $r_1=2$;
II: $d=1$, $300$ cells, $r_1=2$;
III: $d=1$, $150$ cells, $r_1=3$;
IV: $d=1$, $300$ cells, $r_1=3$;
V: $d=2$, von Neumann neighborhood, $15\times 15$ cells, $r_1=r_2=1$;
VI: $d=2$, von Neumann neighborhood, $30\times 30$ cells, $r_1=r_2=1$;
VII: $d=2$, Moore neighborhood, $15\times 15$ cells, $r_1=r_2=1$;
VIII: $d=2$, Moore neighborhood, $30\times 30$ cells, $r_1=r_2=1$.
Update rules are as given in Table~\ref{updtrules} for each of classes
(i)--(iv). Numbers are truncated to six decimal places.}
\begin{tabular}{lcccc}
\hline
Experiment&\multicolumn{4}{c}{$\overline{C_f}$}\\
\cline{2-5}
&(i)&(ii)&(iii)&(iv)\\
\cline{2-5}
I&0.000419&0.004730&4.671842&2.726404\\
II&0.001030&0.023110&3.997883&2.396267\\
III&0.000170&0.685005&5.757584&3.195973\\
IV&0.000146&0.723408&5.755847&3.527819\\
V&0.000214&1.434876&3.692984&1.854994\\
VI&0.000245&1.543048&3.843658&1.506388\\
VII&0.066639&0.110987&4.268379&1.206355\\
VIII&0.134897&0.148938&4.431762&1.452313\\
\cline{2-5}
&\multicolumn{4}{c}{$\sigma^2(C_f)$} \\
\cline{2-5}
&(i)&(ii)&(iii)&(iv)\\
\cline{2-5}
I&0.000039&0.000917&0.000102&0.442224\\
II&0.000251&0.009366&0.000273&0.250179\\
III&0.000010&0.003644&0.000166&0.749215\\
IV&0.000008&0.004534&0.000076&0.686746\\
V&0.000019&0.000068&0.000195&0.444950\\
VI&0.000023&0.000045&0.000058&0.721707\\
VII&0.010898&0.032257&0.000031&0.805210\\
VIII&0.009618&0.021399&0.000009&0.352702\\
\hline
\end{tabular}

\label{cylindrical}
\vspace{0.45in}
\end{table}

\addtocounter{table}{-1}
\begin{table}
\centering
\caption{(Continued).}
\begin{tabular}{lcccc}
\hline
Experiment&\multicolumn{4}{c}{$\overline{T_f}$}\\
\cline{2-5}
&(i)&(ii)&(iii)&(iv)\\
\cline{2-5}
I&0.000642&0.006438&0.474852&0.340475\\
II&0.000643&0.003924&0.474737&0.389267\\
III&0.000307&0.137617&0.489661&0.355899\\
IV&0.000310&0.102894&0.490509&0.290770\\
V&0.000237&0.335283&0.469148&0.231000\\
VI&0.000237&0.366104&0.484582&0.185903\\
VII&0.002447&0.002894&0.484342&0.042994\\
VIII
&0.002582
&0.003393
&0.484713&0.094744\\
\cline{2-5}
&\multicolumn{4}{c}{$\sigma^2(T_f)$} \\
\cline{2-5}
&(i)&(ii)&(iii)&(iv)\\
\cline{2-5}
I&0.000080&0.000779&0.000010&0.005762\\
II&0.000079&0.000433&0.000007&0.002972\\
III&0.000031&0.000544&0.000009&0.006885\\
IV&0.000032&0.000585&0.000004&0.005747\\
V&0.000019&0.000005&0.000001&0.012932\\
VI&0.000019&0.000003&0.000000&0.016717\\
VII&0.000473&0.000532&0.000010&0.010095\\
VIII&0.000497&0.000609&0.000002&0.009627\\
\hline
\end{tabular}

\end{table}

\clearpage
\appendix

\section{Selected spatiotemporal patterns}\label{patterns}

In this appendix, we provide illustrations of the spatiotemporal patterns
resulting from some of the evolutions based on the update rules of
Table~\ref{updtrules}. In all illustrations, the color white is associated with
the $0$ state, the color black with the $1$ state. All spatiotemporal plots are
framed for increased ease of reference.

\begin{figure}
\fboxsep=0pt
\centering
\begin{tabular}{c@{\hspace{0.200in}}c@{\hspace{0.200in}}c@{\hspace{0.200in}}c}
\fbox{\scalebox{0.400}{\includegraphics{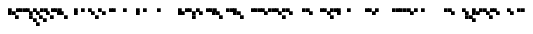}}}&
\fbox{\scalebox{0.400}{\includegraphics{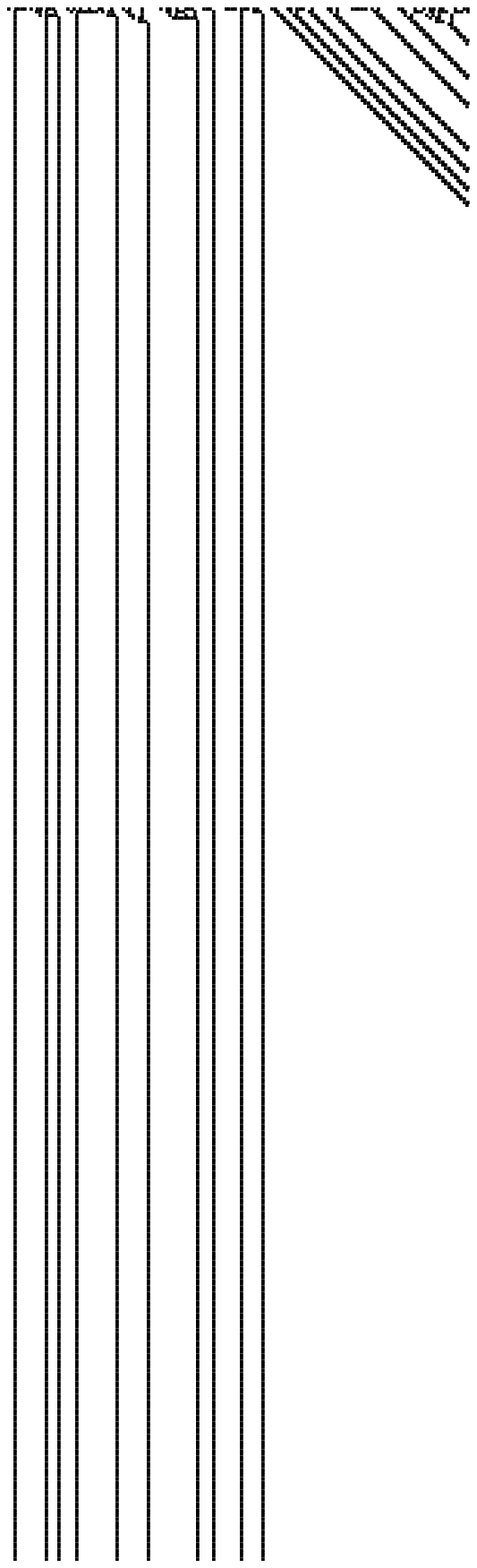}}}&
\fbox{\scalebox{0.400}{\includegraphics{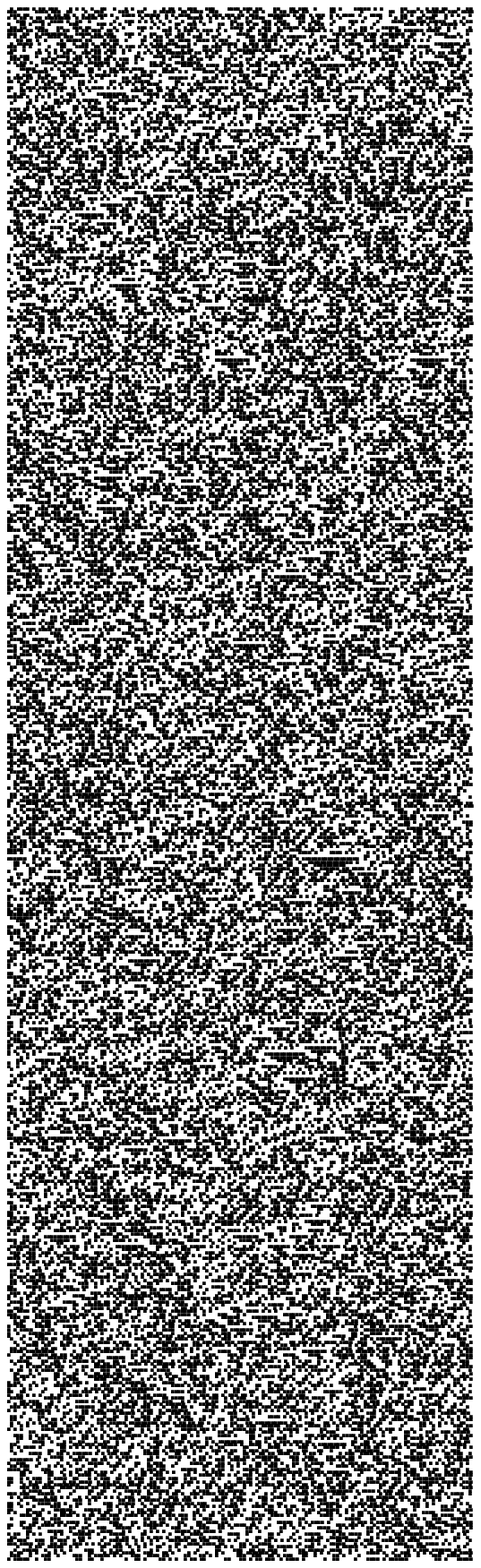}}}&
\fbox{\scalebox{0.400}{\includegraphics{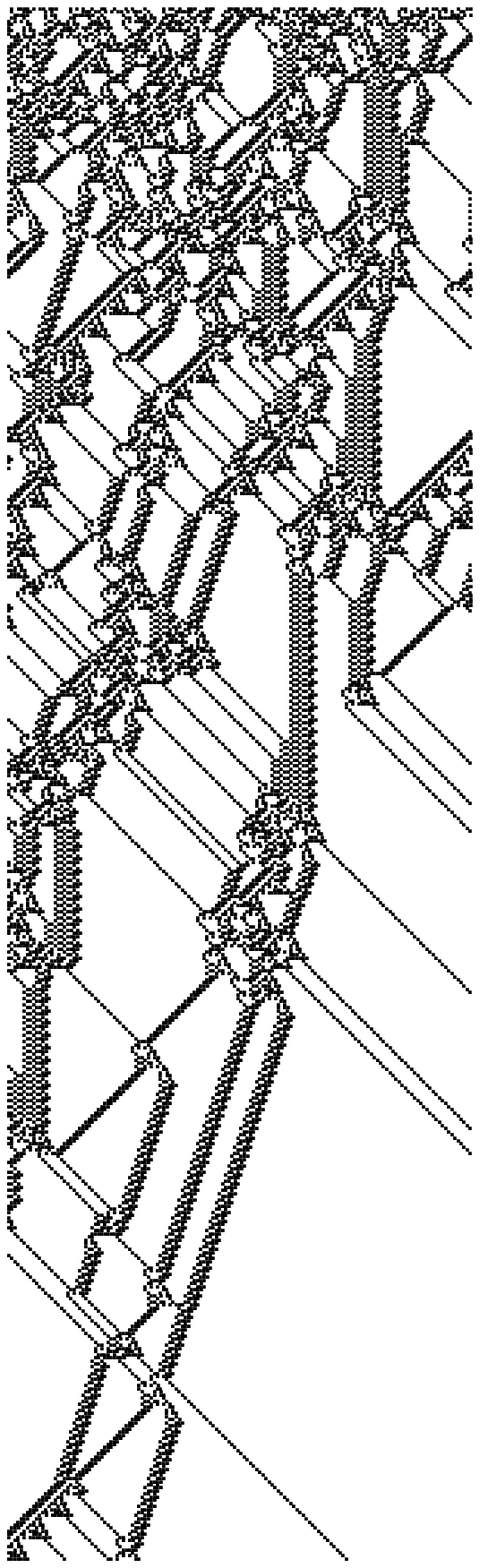}}}\\\\
\fbox{\scalebox{0.400}{\includegraphics{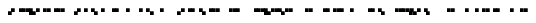}}}&
\fbox{\scalebox{0.400}{\includegraphics{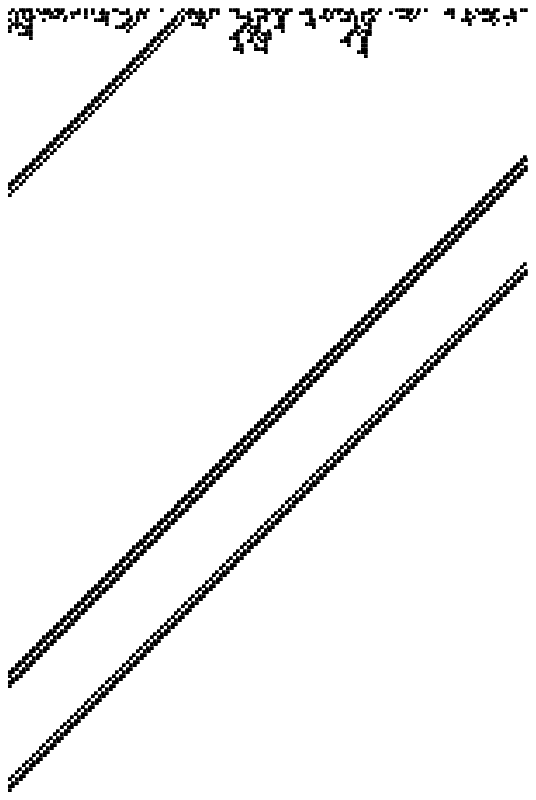}}}&
\fbox{\scalebox{0.400}{\includegraphics{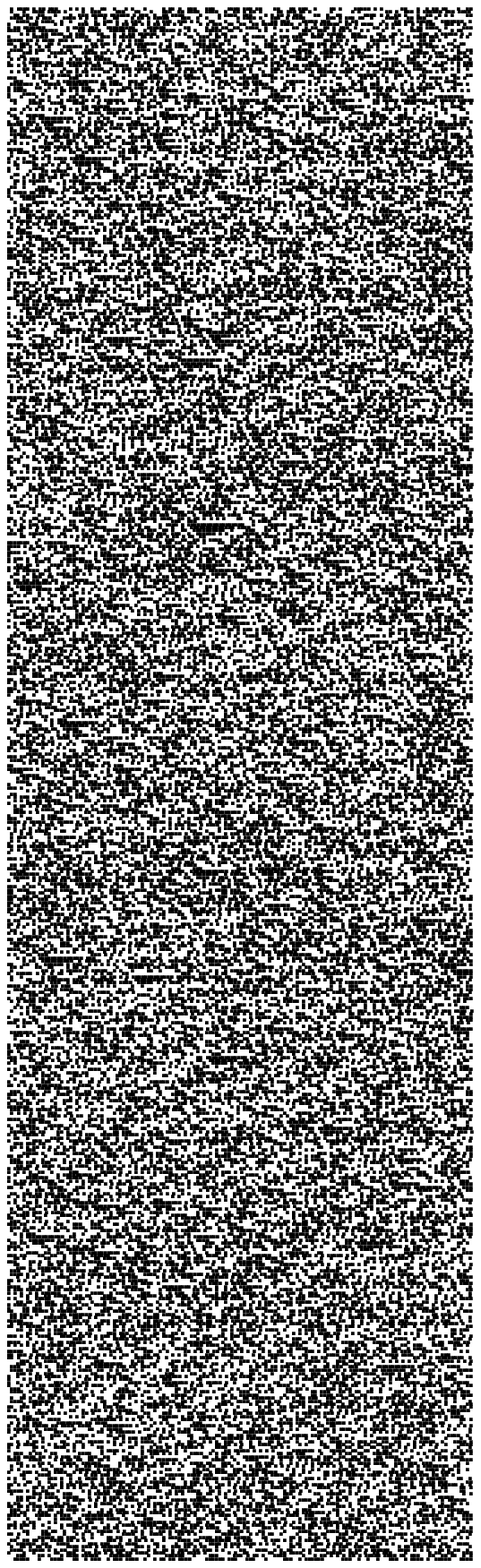}}}&
\fbox{\scalebox{0.400}{\includegraphics{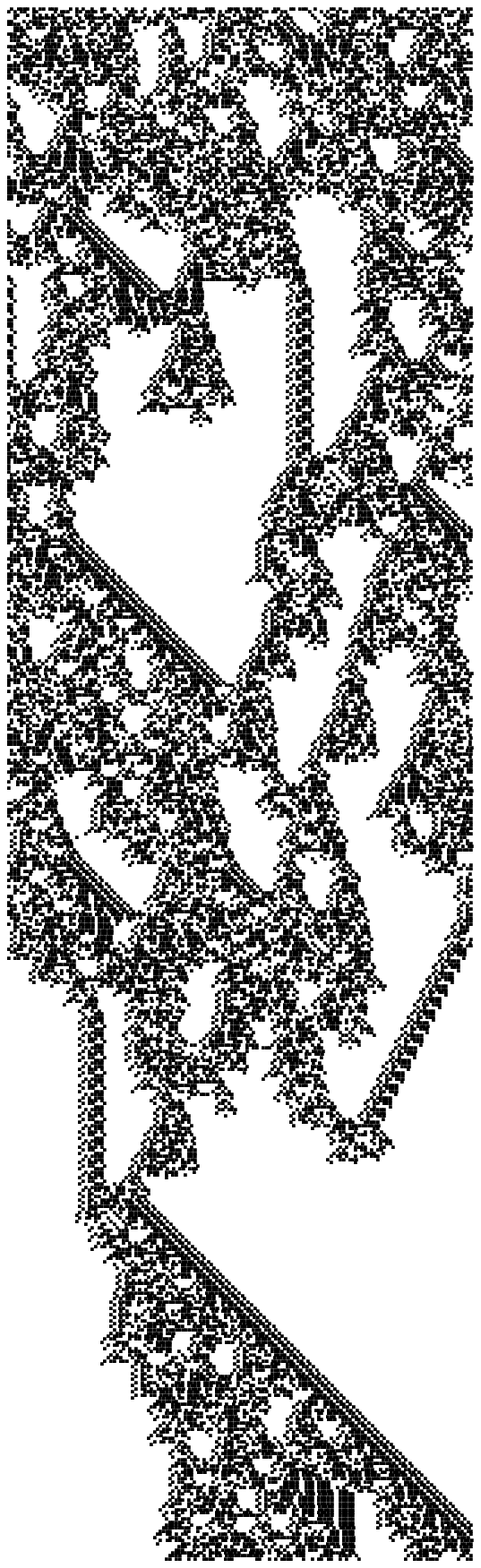}}}
\end{tabular}
\caption{Sample spatiotemporal patterns for the update rules of
Table~\ref{updtrules} in the infinite, $d=1$ cases with $150$ cells observed for
$500$ time steps. Each plot displays cell states horizontally for each time
step; time grows from top to bottom. The topmost row of plots corresponds to
$r_1=2$, the bottommost to $r_1=3$. Within each row, from left to right, the
update rules belong each to classes (i)--(iv).}
\label{patterns-inftyd1}
\end{figure}

\begin{figure}
\fboxsep=0pt
\centering
\begin{tabular}{c@{\hspace{0.200in}}c@{\hspace{0.200in}}c@{\hspace{0.200in}}c}
\fbox{\scalebox{0.400}{\includegraphics{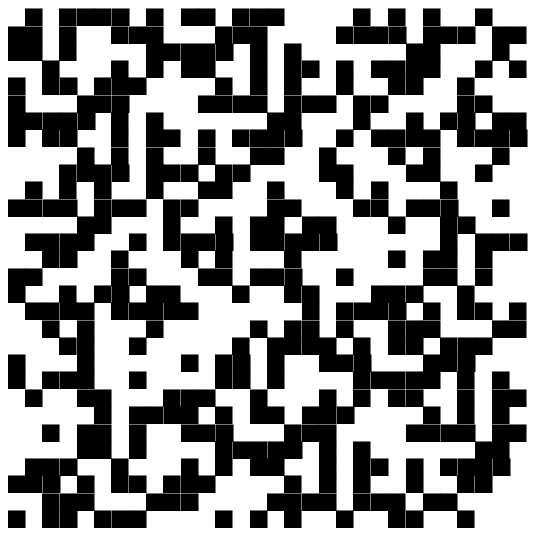}}}&
\fbox{\scalebox{0.400}{\includegraphics{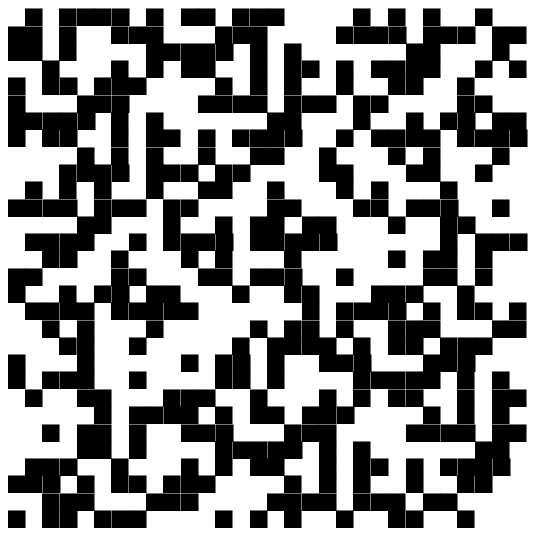}}}&
\fbox{\scalebox{0.400}{\includegraphics{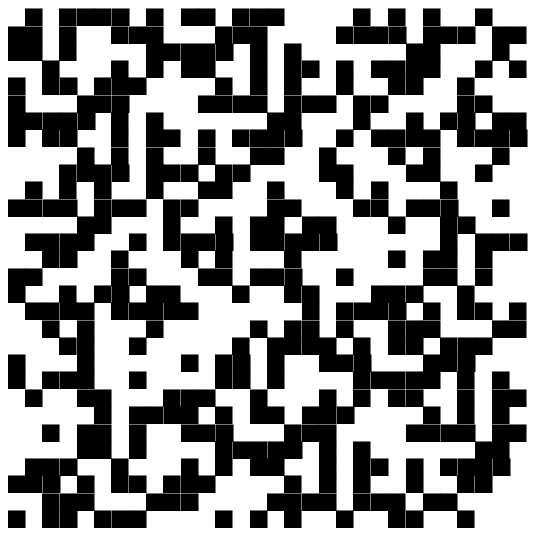}}}&
\fbox{\scalebox{0.400}{\includegraphics{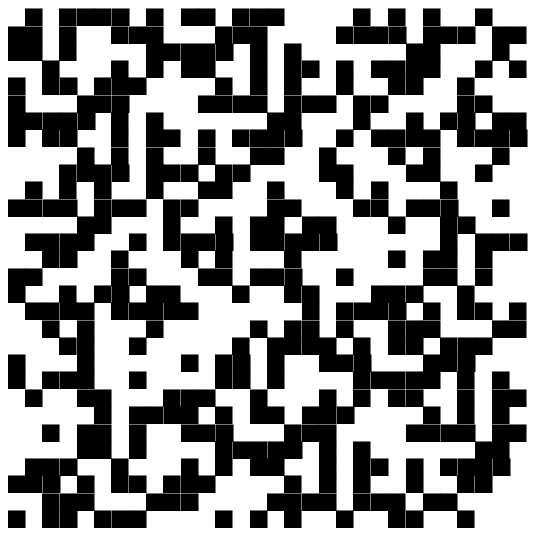}}}\\
\fbox{\scalebox{0.400}{\includegraphics{pat-30-v-ib.ps}}}&
\fbox{\scalebox{0.400}{\includegraphics{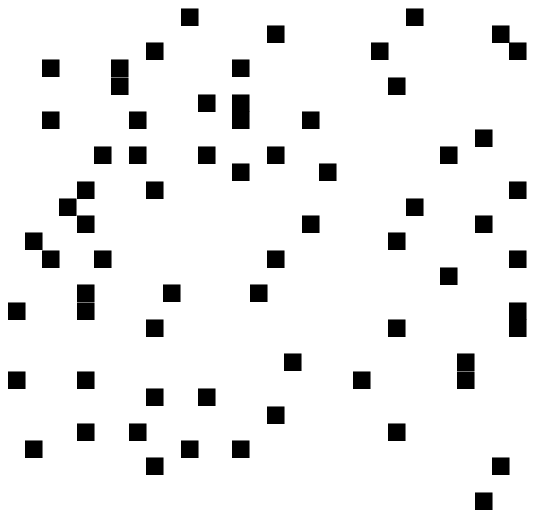}}}&
\fbox{\scalebox{0.400}{\includegraphics{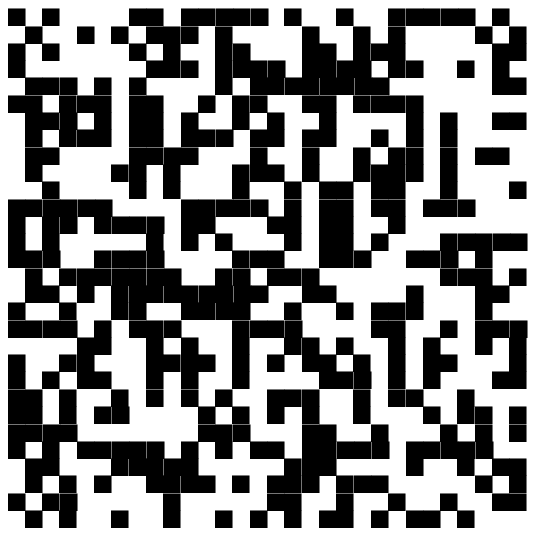}}}&
\fbox{\scalebox{0.400}{\includegraphics{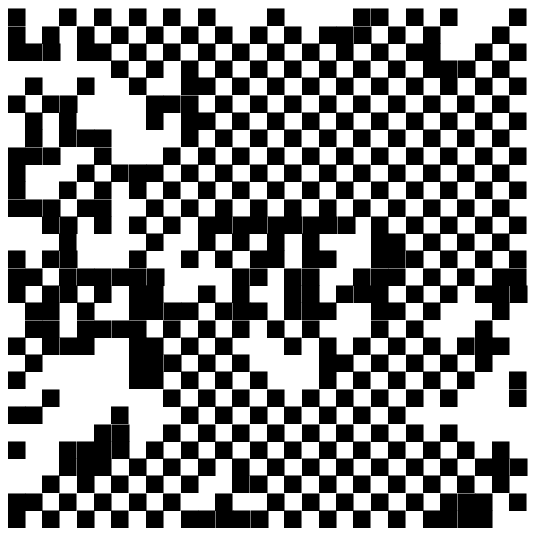}}}\\
\fbox{\scalebox{0.400}{\includegraphics{pat-30-v-ic.ps}}}&
\fbox{\scalebox{0.400}{\includegraphics{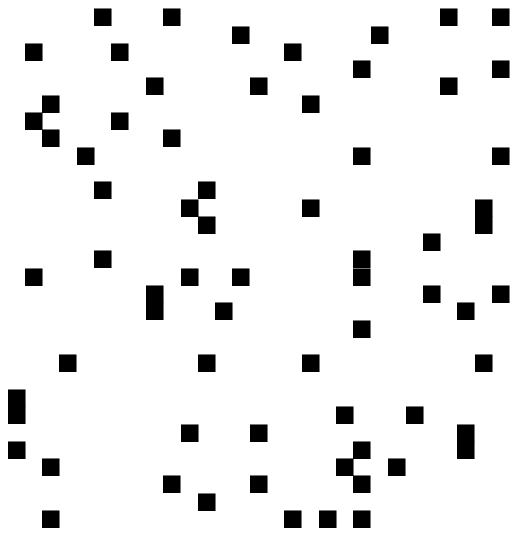}}}&
\fbox{\scalebox{0.400}{\includegraphics{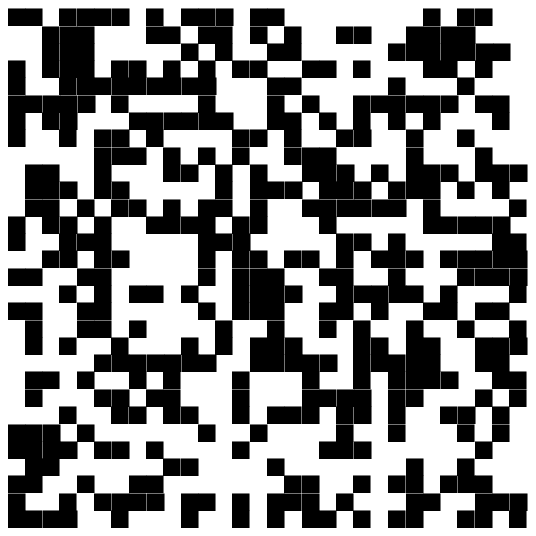}}}&
\fbox{\scalebox{0.400}{\includegraphics{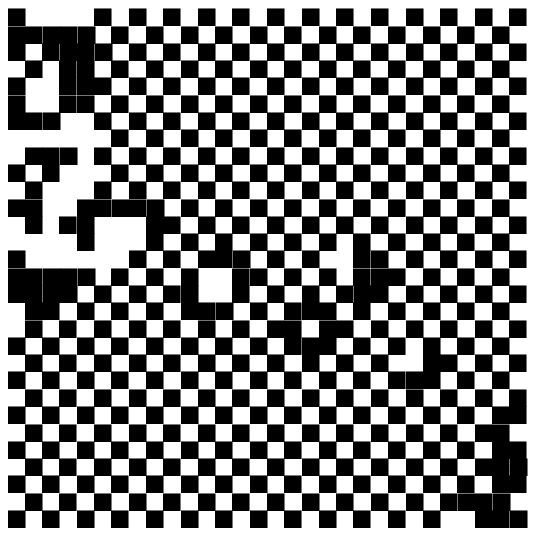}}}\\\\
\fbox{\scalebox{0.400}{\includegraphics{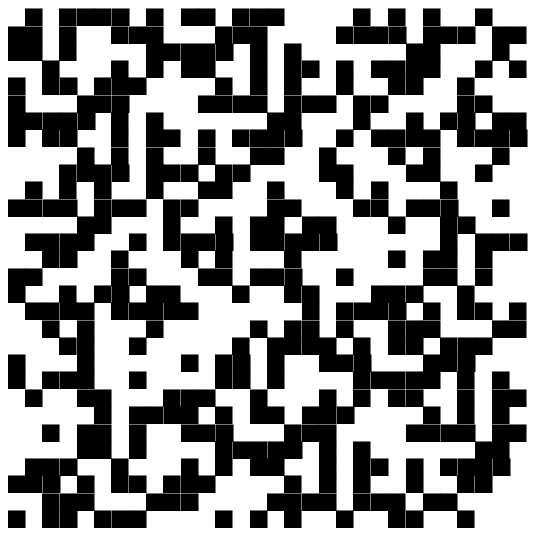}}}&
\fbox{\scalebox{0.400}{\includegraphics{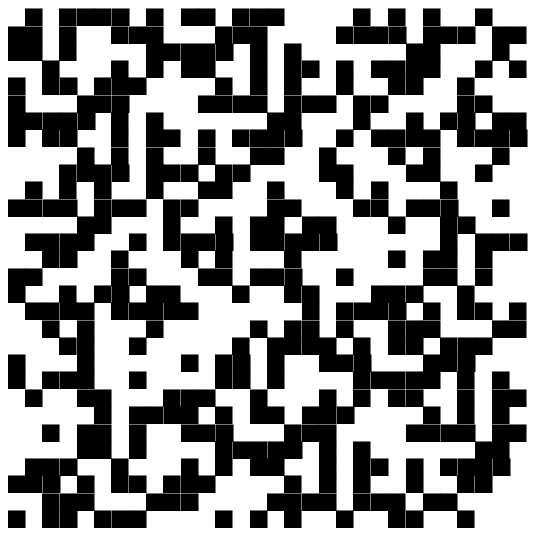}}}&
\fbox{\scalebox{0.400}{\includegraphics{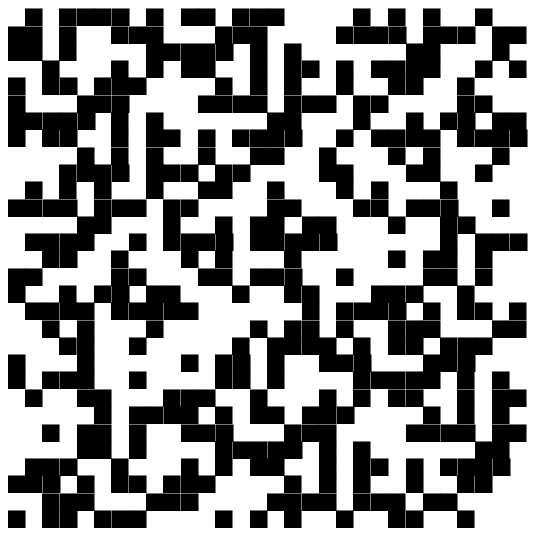}}}&
\fbox{\scalebox{0.400}{\includegraphics{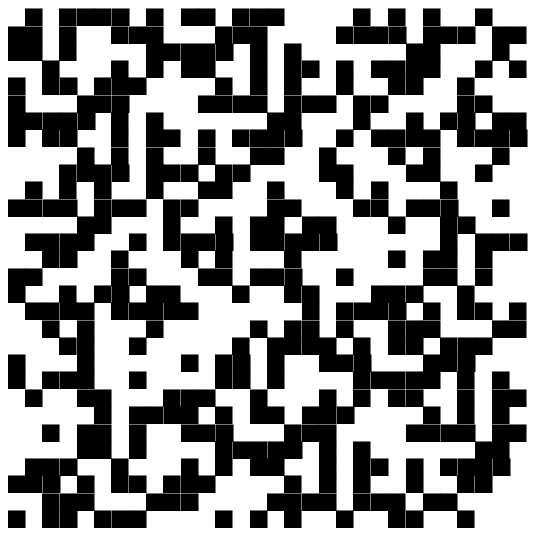}}}\\
\fbox{\scalebox{0.400}{\includegraphics{pat-30-m-ib.ps}}}&
\fbox{\scalebox{0.400}{\includegraphics{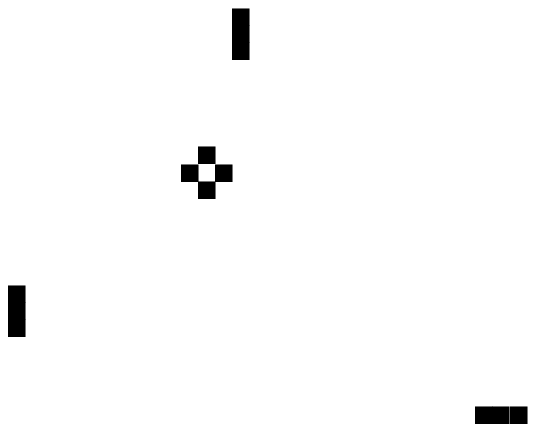}}}&
\fbox{\scalebox{0.400}{\includegraphics{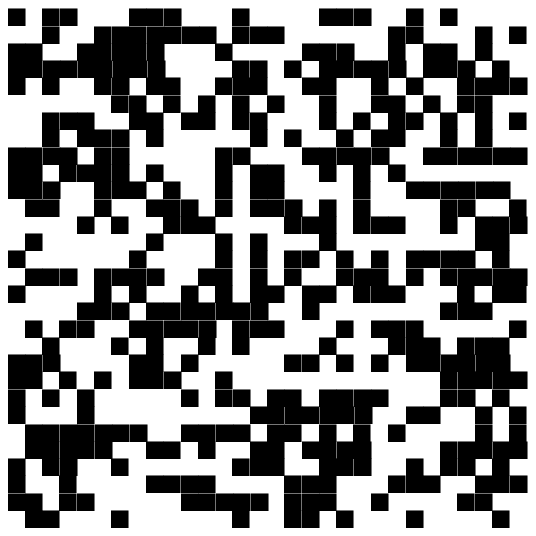}}}&
\fbox{\scalebox{0.400}{\includegraphics{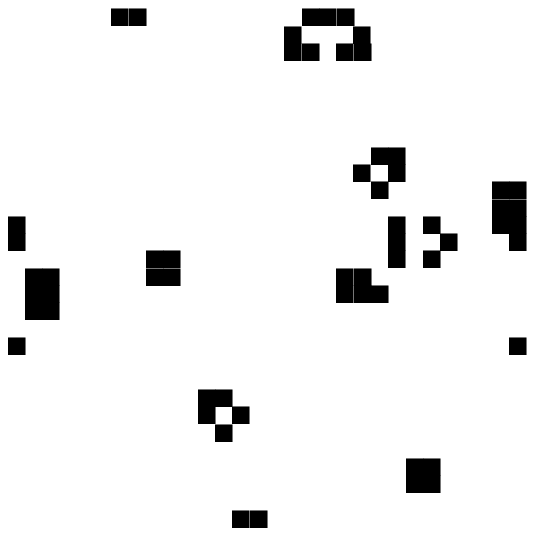}}}\\
\fbox{\scalebox{0.400}{\includegraphics{pat-30-m-ic.ps}}}&
\fbox{\scalebox{0.400}{\includegraphics{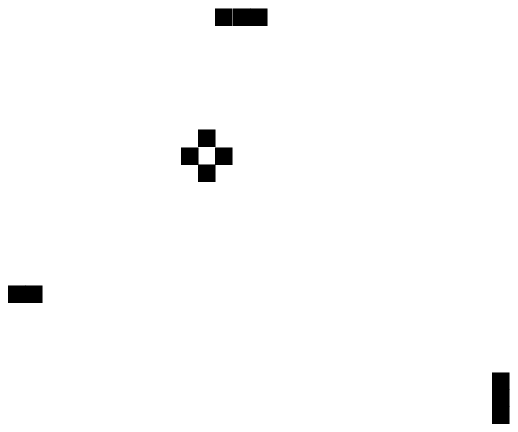}}}&
\fbox{\scalebox{0.400}{\includegraphics{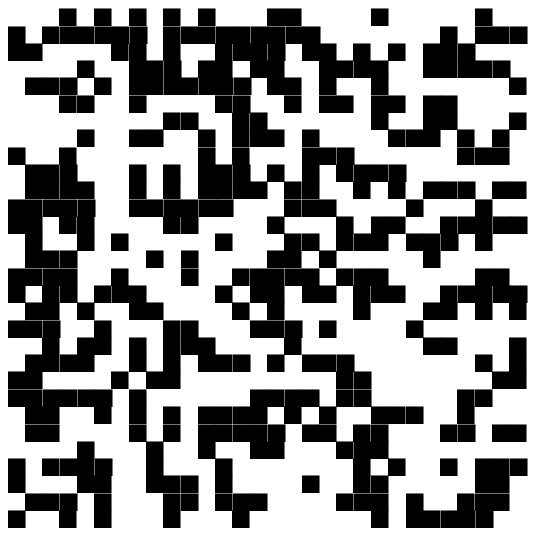}}}&
\fbox{\scalebox{0.400}{\includegraphics{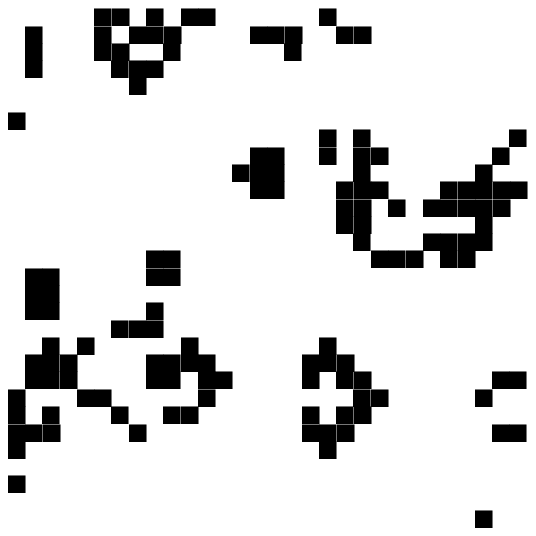}}}
\end{tabular}
\caption{Sample spatiotemporal patterns for the update rules of
Table~\ref{updtrules} in the infinite, $d=2$ cases with $30\times 30$ observed
cells. Each plot displays a configuration during the evolution of the automaton.
The topmost three rows of plots are relative to the von Neumann update rules,
the bottommost three rows to the Moore update rules. Within each triple of rows,
the topmost row corresponds to $t=0$, the middle one to $t=125$, and the
bottommost to $t=250$. Within each row, from left to right, the update rules
belong each to classes (i)--(iv).}
\label{patterns-inftyd2}
\end{figure}

The first set of illustrations corresponds to some of the evolutions to which
Figures~\ref{ie-initiald1}--\ref{te-initiald2} refer. They are therefore for
infinite cellular automata and correspond to the evolutions of the observed
cells, that is, the cells whose states contributed to the entropy calculations.
This set is shown in Figures~\ref{patterns-inftyd1} and \ref{patterns-inftyd2},
respectively for the one- and two-dimensional cases.

\begin{figure}
\fboxsep=0pt
\centering
\begin{tabular}{c@{\hspace{0.200in}}c@{\hspace{0.200in}}c@{\hspace{0.200in}}c}
\fbox{\scalebox{0.400}{\includegraphics{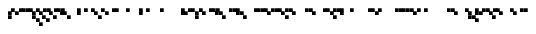}}}&
\fbox{\scalebox{0.400}{\includegraphics{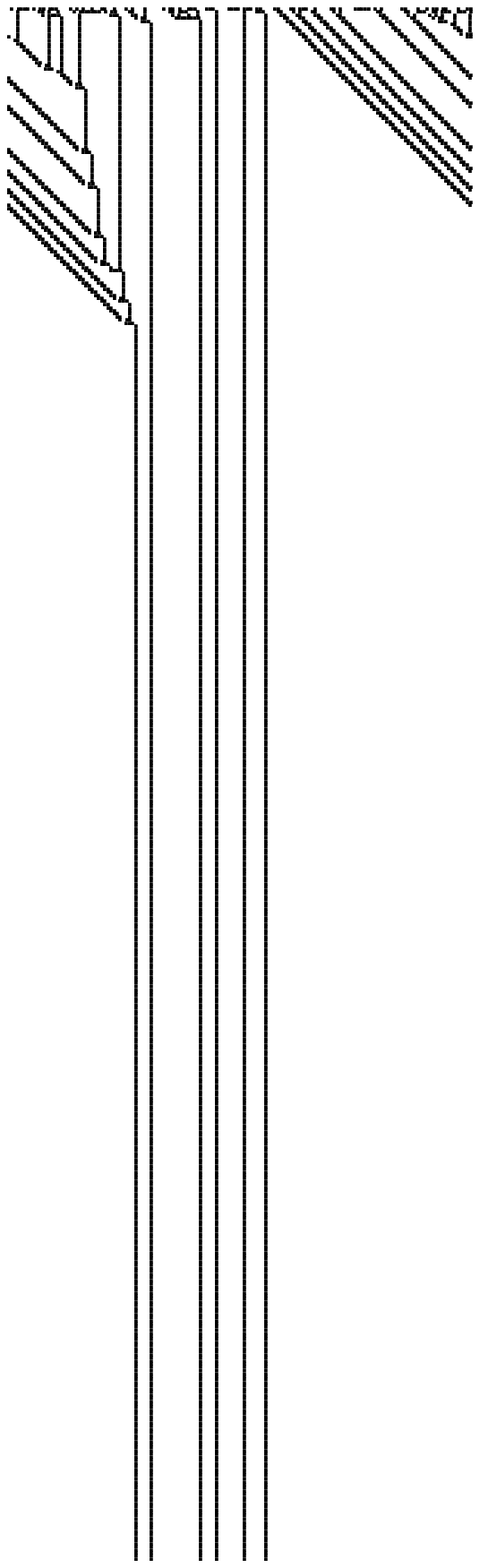}}}&
\fbox{\scalebox{0.400}{\includegraphics{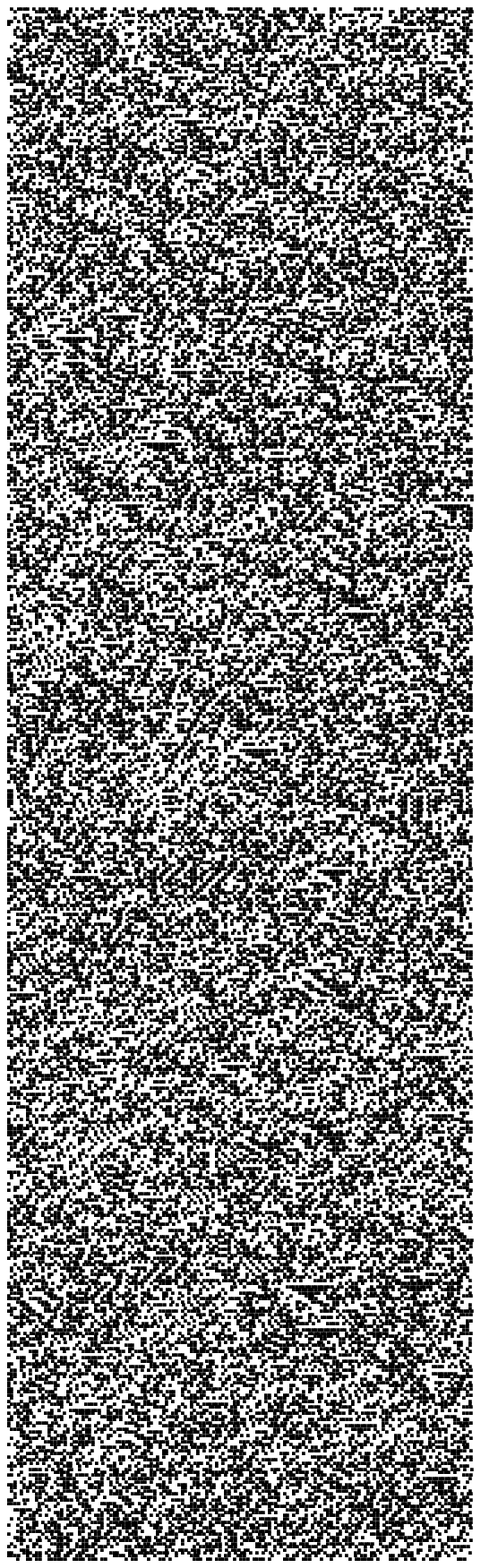}}}&
\fbox{\scalebox{0.400}{\includegraphics{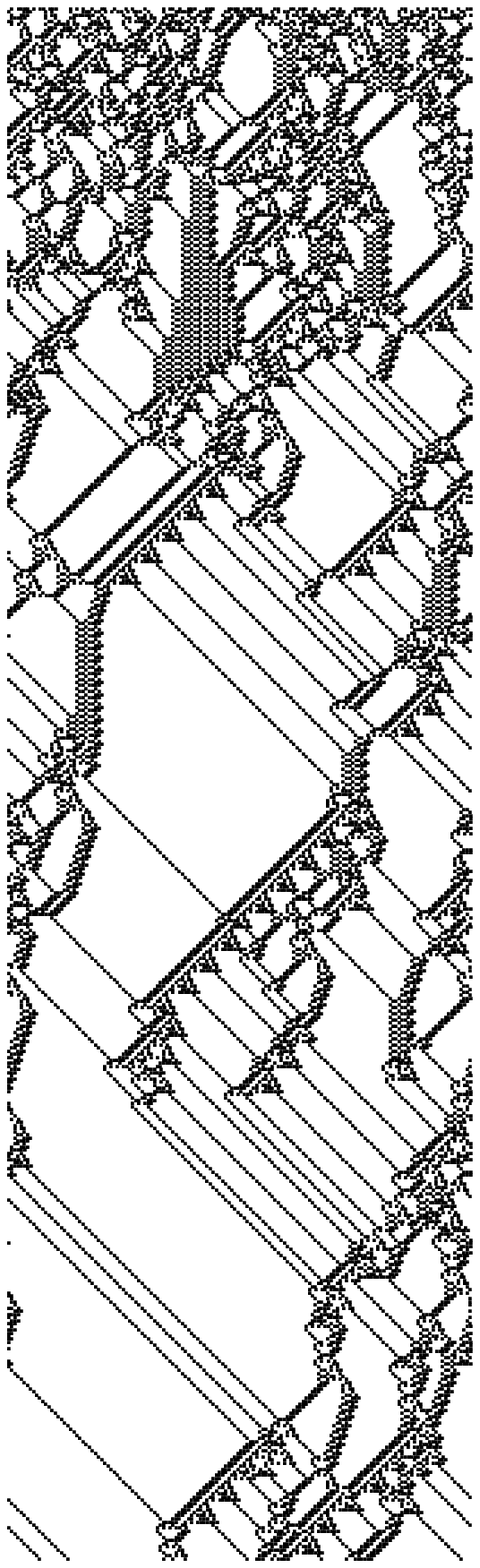}}}\\\\
\fbox{\scalebox{0.400}{\includegraphics{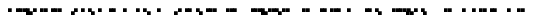}}}&
\fbox{\scalebox{0.400}{\includegraphics{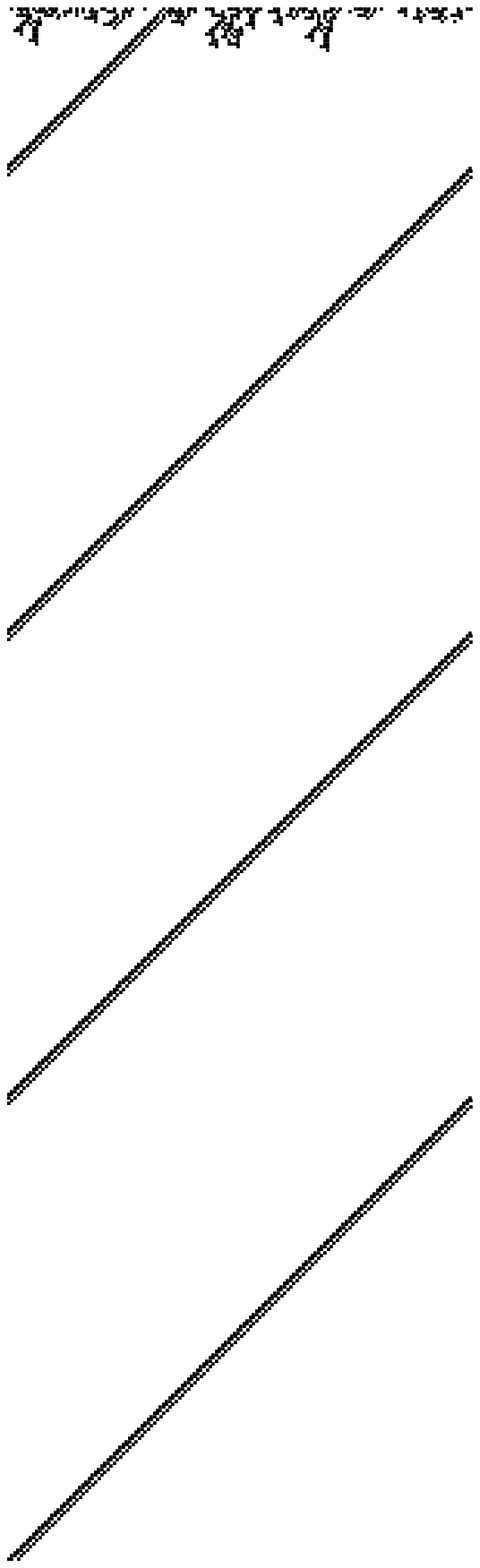}}}&
\fbox{\scalebox{0.400}{\includegraphics{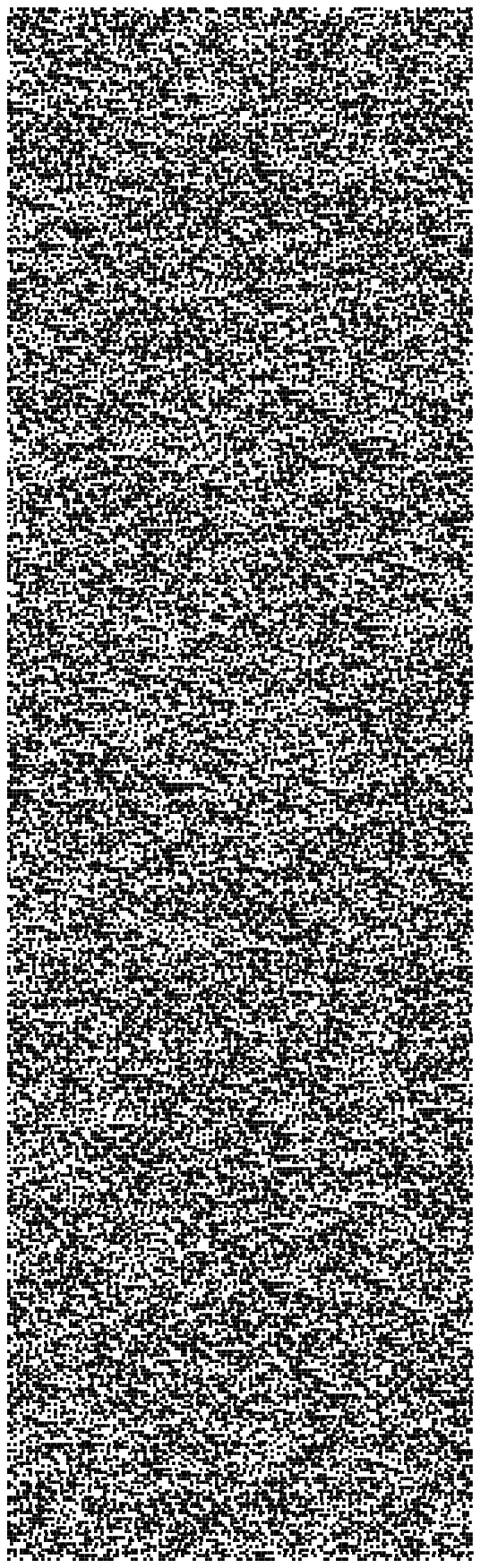}}}&
\fbox{\scalebox{0.400}{\includegraphics{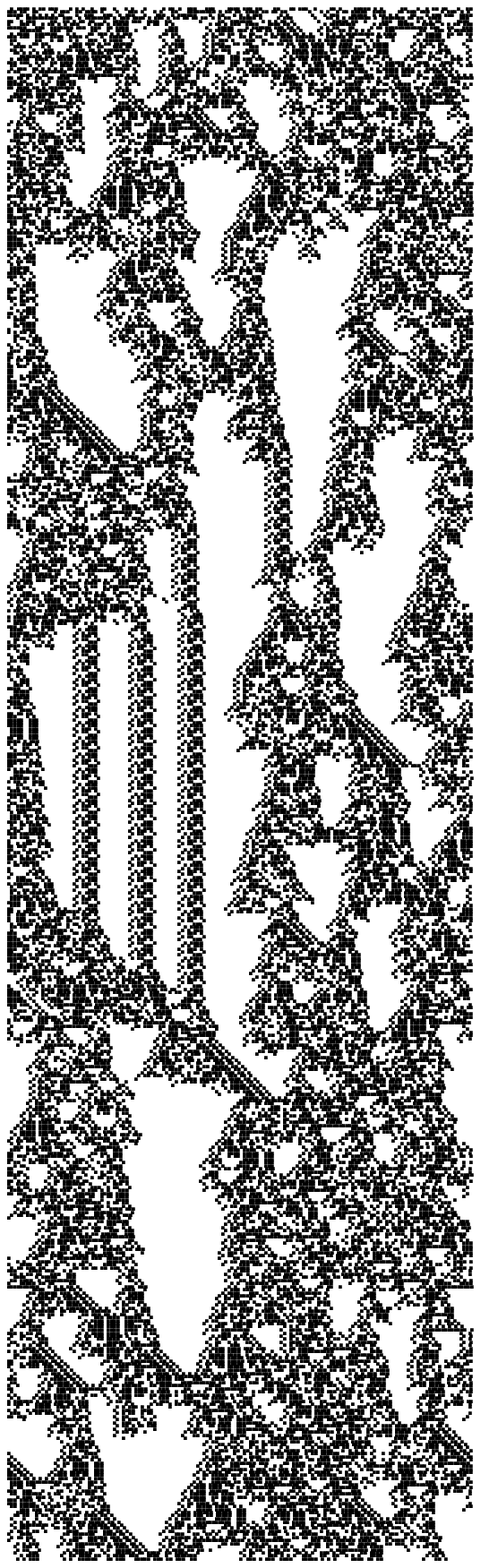}}}
\end{tabular}
\caption{Sample spatiotemporal patterns for the update rules of
Table~\ref{updtrules} in the cylindrical, $d=1$ cases with $150$ cells observed
for $500$ time steps. Each plot displays cell states horizontally for each time
step; time grows from top to bottom. The topmost row of plots corresponds to
$r_1=2$, the bottommost to $r_1=3$. Within each row, from left to right, the
update rules belong each to classes (i)--(iv).}
\label{patterns-cyld1}
\end{figure}

\begin{figure}
\fboxsep=0pt
\centering
\begin{tabular}{c@{\hspace{0.200in}}c@{\hspace{0.200in}}c@{\hspace{0.200in}}c}
\fbox{\scalebox{0.400}{\includegraphics{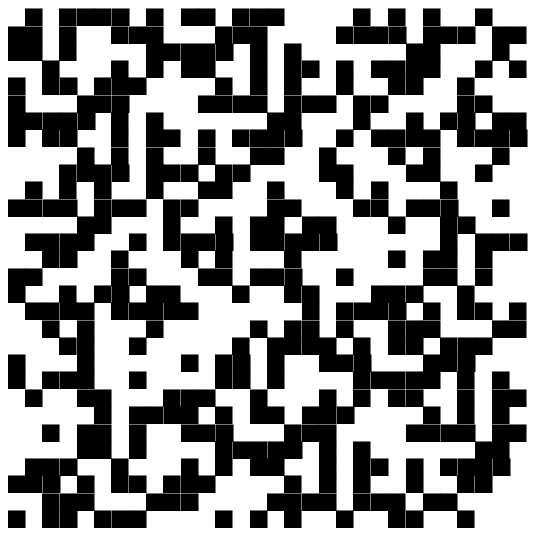}}}&
\fbox{\scalebox{0.400}{\includegraphics{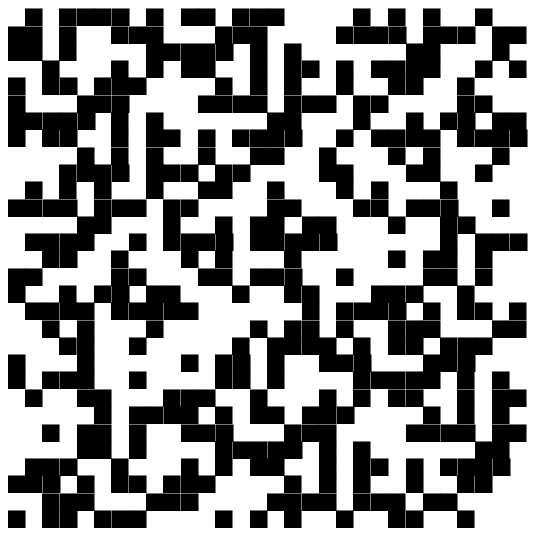}}}&
\fbox{\scalebox{0.400}{\includegraphics{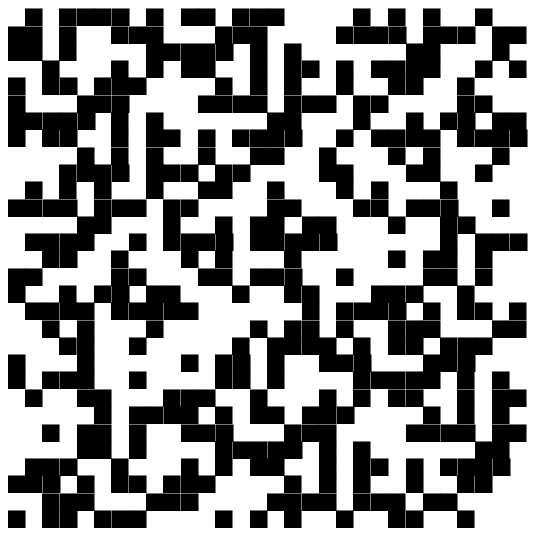}}}&
\fbox{\scalebox{0.400}{\includegraphics{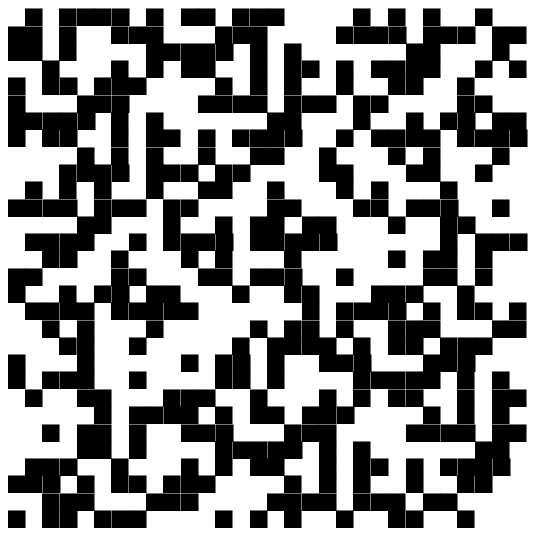}}}\\
\fbox{\scalebox{0.400}{\includegraphics{patc-30-v-ib.ps}}}&
\fbox{\scalebox{0.400}{\includegraphics{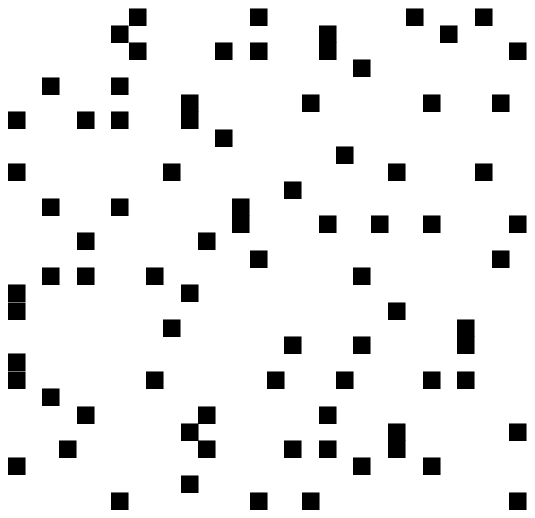}}}&
\fbox{\scalebox{0.400}{\includegraphics{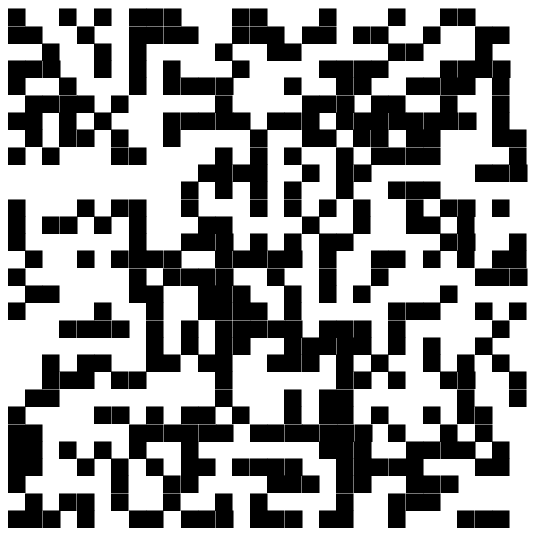}}}&
\fbox{\scalebox{0.400}{\includegraphics{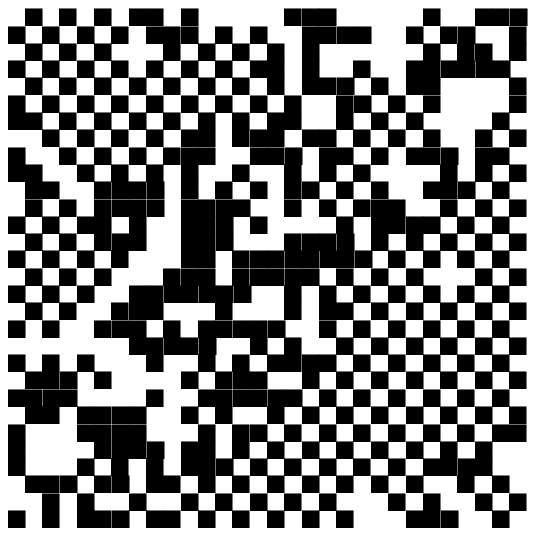}}}\\
\fbox{\scalebox{0.400}{\includegraphics{patc-30-v-ic.ps}}}&
\fbox{\scalebox{0.400}{\includegraphics{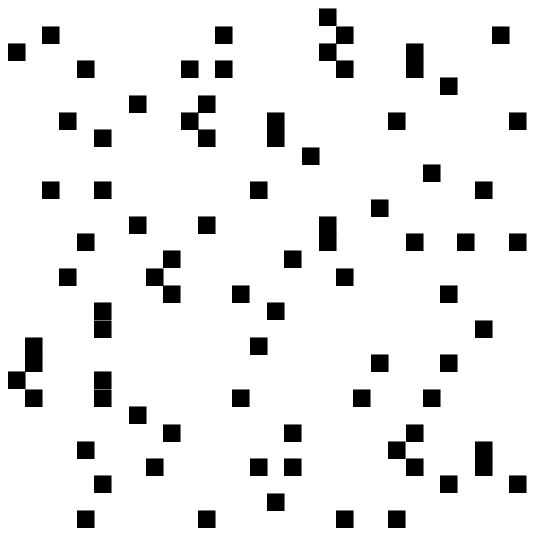}}}&
\fbox{\scalebox{0.400}{\includegraphics{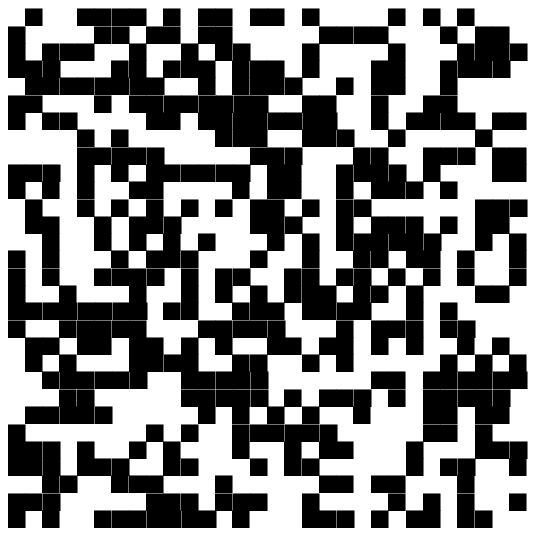}}}&
\fbox{\scalebox{0.400}{\includegraphics{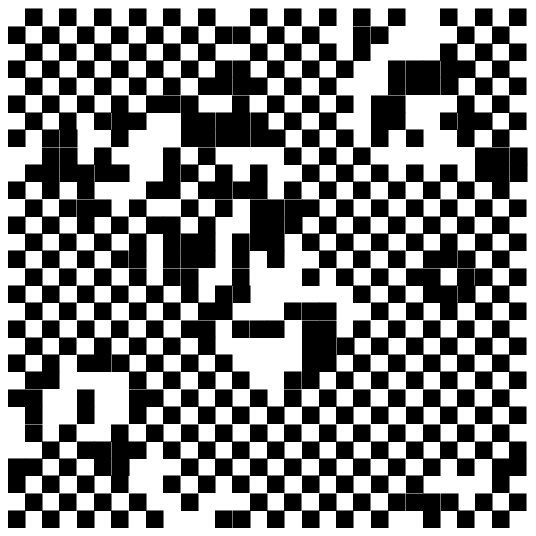}}}\\\\
\fbox{\scalebox{0.400}{\includegraphics{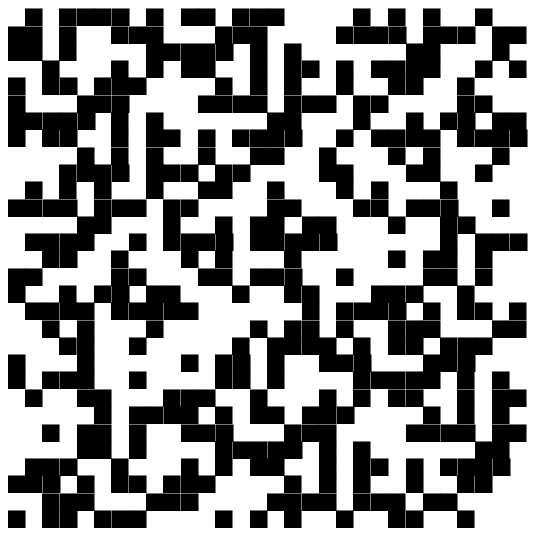}}}&
\fbox{\scalebox{0.400}{\includegraphics{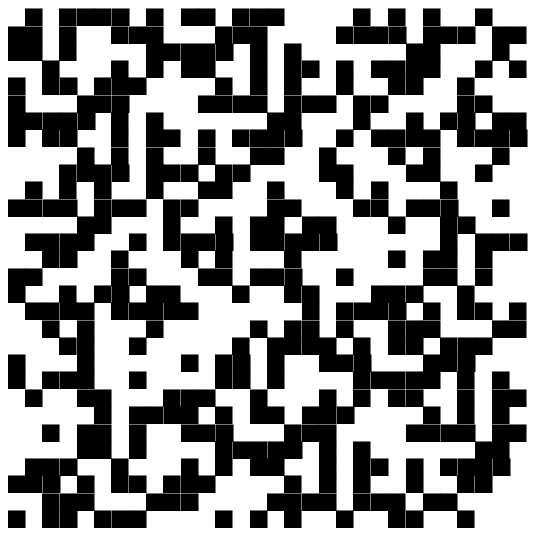}}}&
\fbox{\scalebox{0.400}{\includegraphics{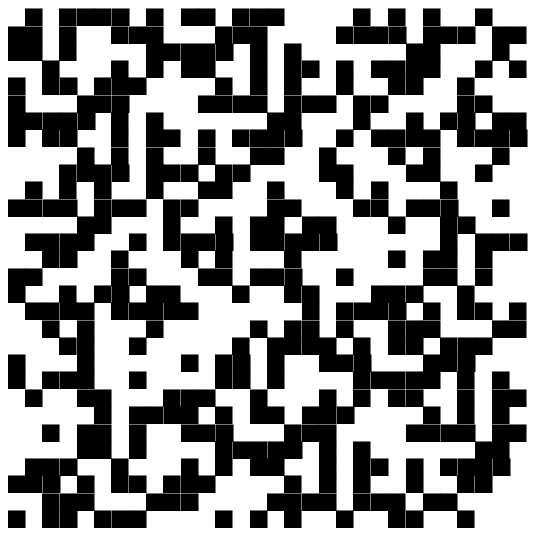}}}&
\fbox{\scalebox{0.400}{\includegraphics{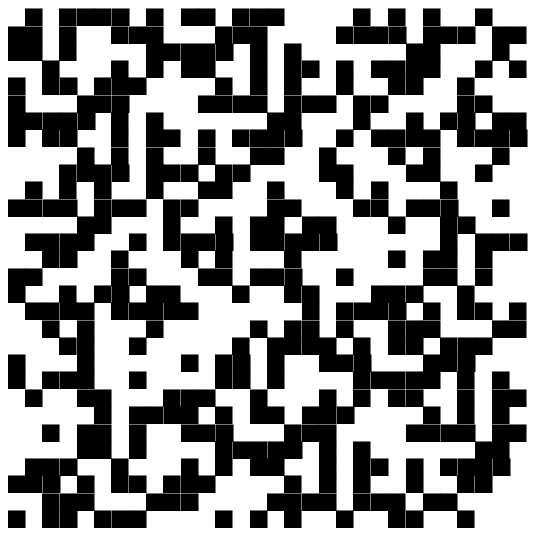}}}\\
\fbox{\scalebox{0.400}{\includegraphics{patc-30-m-ib.ps}}}&
\fbox{\scalebox{0.400}{\includegraphics{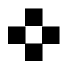}}}&
\fbox{\scalebox{0.400}{\includegraphics{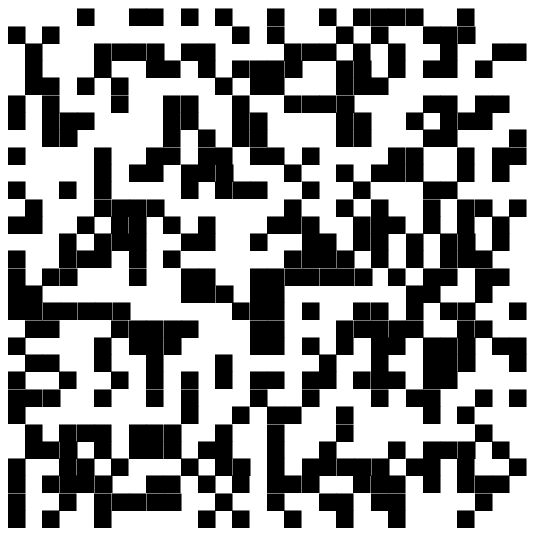}}}&
\fbox{\scalebox{0.400}{\includegraphics{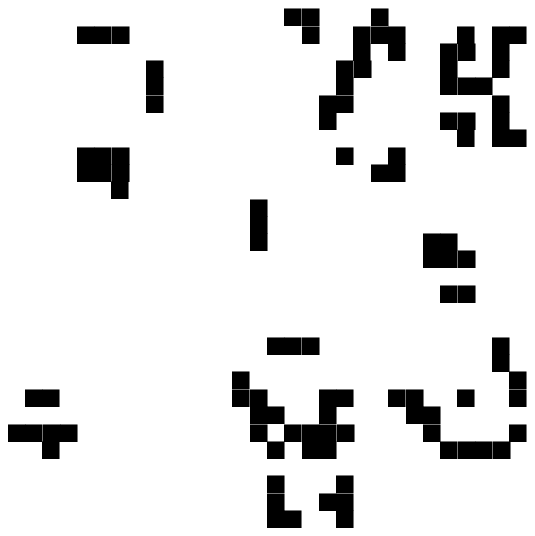}}}\\
\fbox{\scalebox{0.400}{\includegraphics{patc-30-m-ic.ps}}}&
\fbox{\scalebox{0.400}{\includegraphics{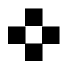}}}&
\fbox{\scalebox{0.400}{\includegraphics{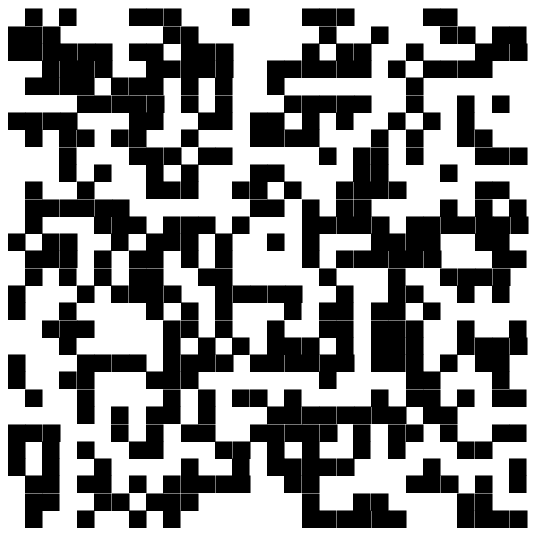}}}&
\fbox{\scalebox{0.400}{\includegraphics{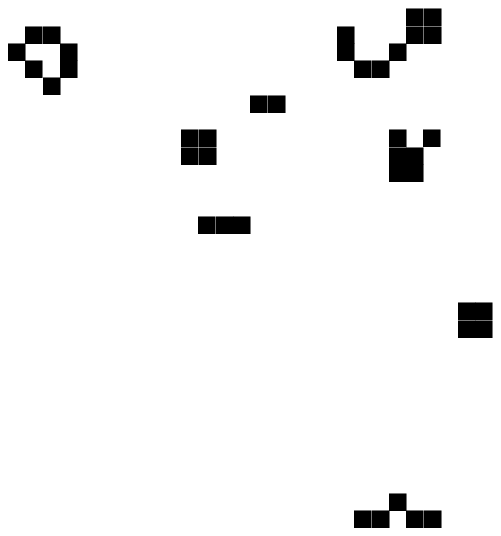}}}
\end{tabular}
\caption{Sample spatiotemporal patterns for the update rules of
Table~\ref{updtrules} in the cylindrical, $d=2$ cases with $30\times 30$
observed cells. Each plot displays a configuration during the evolution of the
automaton. The topmost three rows of plots are relative to the von Neumann
update rules, the bottommost three rows to the Moore update rules. Within each
triple of rows, the topmost row corresponds to $t=0$, the middle one to $t=125$,
and the bottommost to $t=250$. Within each row, from left to right, the update
rules belong each to classes (i)--(iv).}
\label{patterns-cyld2}
\end{figure}

A similar second set of illustrations depicts the evolution of the same cells,
but now under cylindrical boundaries, following our remarks in
Section~\ref{concl}. These are given in Figures~\ref{patterns-cyld1} and
\ref{patterns-cyld2}, respectively for the one- and two-dimensional cases. They
are related to the data shown in Table~\ref{cylindrical} only in principle,
because they correspond, for the sake of comparison, to initial configurations
that match those used for the infinite cases.

\clearpage
\subsection*{Acknowledgments}

The authors acknowledge partial support from CNPq, CAPES, and a FAPERJ BBP
grant.

\bibliography{caclass}
\bibliographystyle{elsart-num}

\end{document}